\documentclass[aps,prx,reprint,superscriptaddress,nofootinbib]{revtex4-2}
\usepackage{blindtext}
\usepackage{lipsum}
\usepackage{graphics}
\usepackage{amsmath}
\usepackage{graphicx}
\usepackage{graphics}
\usepackage{amssymb}
\usepackage{verbatim}
\usepackage{physics}
\usepackage{float}
\usepackage{dsfont}
\usepackage[dvipsnames]{xcolor}
\usepackage{framed}
\definecolor{shadecolor}{RGB}{224,224,224}

\makeatletter
\def\l@subsubsection#1#2{}
\makeatother

\usepackage[dvipsnames]{xcolor}
\usepackage[normalem]{ulem}
\usepackage{wasysym} % for "\varhexagon" macro
\usepackage[export]{adjustbox}

\makeatletter
\DeclareFontFamily{OMX}{MnSymbolE}{}
\DeclareSymbolFont{MnLargeSymbols}{OMX}{MnSymbolE}{m}{n}
\SetSymbolFont{MnLargeSymbols}{bold}{OMX}{MnSymbolE}{b}{n}
\DeclareFontShape{OMX}{MnSymbolE}{m}{n}{
    <-6>  MnSymbolE5
   <6-7>  MnSymbolE6
   <7-8>  MnSymbolE7
   <8-9>  MnSymbolE8
   <9-10> MnSymbolE9
  <10-12> MnSymbolE10
  <12->   MnSymbolE12
}{}
\DeclareFontShape{OMX}{MnSymbolE}{b}{n}{
    <-6>  MnSymbolE-Bold5
   <6-7>  MnSymbolE-Bold6
   <7-8>  MnSymbolE-Bold7
   <8-9>  MnSymbolE-Bold8
   <9-10> MnSymbolE-Bold9
  <10-12> MnSymbolE-Bold10
  <12->   MnSymbolE-Bold12
}{}

\let\llangle\@undefined
\let\rrangle\@undefined
\DeclareMathDelimiter{\llangle}{\mathopen}%
                     {MnLargeSymbols}{'164}{MnLargeSymbols}{'164}
\DeclareMathDelimiter{\rrangle}{\mathclose}%
                     {MnLargeSymbols}{'171}{MnLargeSymbols}{'171}
\makeatother

\usepackage{tikz}
\usepackage{tikz-cd}
\usetikzlibrary{arrows}
\usetikzlibrary{snakes}
\usetikzlibrary{intersections}
\usetikzlibrary{shapes.geometric}
\usetikzlibrary{decorations.pathmorphing, patterns,shapes}
\usetikzlibrary{decorations.markings}

\def\A{\mathds A}
\def\T{\mathds T}

\tikzset{
	partial ellipse/.style args={#1:#2:#3}{
		insert path={+ (#1:#3) arc (#1:#2:#3)}
	}
}

\tikzset{
	mid arrow/.style={postaction={decorate,decoration={
				markings,
				mark=at position .575 with {\arrow[#1]{stealth}}
	}}},
	near arrow/.style={postaction={decorate,decoration={
				markings,
				mark=at position .275 with {\arrow[#1]{stealth}}
	}}},
	far arrow/.style={postaction={decorate,decoration={
				markings,
				mark=at position .800 with {\arrow[#1]{stealth}}
	}}},
}

\pgfdeclarepatternformonly{south west lines}{\pgfqpoint{-0pt}{-0pt}}{\pgfqpoint{3pt}{3pt}}{\pgfqpoint{3pt}{3pt}}{
	\pgfsetlinewidth{0.4pt}
	\pgfpathmoveto{\pgfqpoint{0pt}{0pt}}
	\pgfpathlineto{\pgfqpoint{3pt}{3pt}}
	\pgfpathmoveto{\pgfqpoint{2.8pt}{-.2pt}}
	\pgfpathlineto{\pgfqpoint{3.2pt}{.2pt}}
	\pgfpathmoveto{\pgfqpoint{-.2pt}{2.8pt}}
	\pgfpathlineto{\pgfqpoint{.2pt}{3.2pt}}
	\pgfusepath{stroke}}

\newcommand{\phii}{\varphi}

\definecolor{orange(ryb)}{HTML}{FFA500}
\definecolor{lightorange(ryb)}{HTML}{FFB300}
\definecolor{dodgerblue}{HTML}{1E90FF}
\definecolor{lightdodgerblue}{HTML}{4dbff7}
\definecolor{crimson}{HTML}{FF4C4C}%{FB4324}%{E53214}%{DC143C}
\definecolor{pinkerton}{HTML}{EC368D}
\definecolor{forest}{HTML}{6DD189}
\definecolor{lightishgray}{HTML}{DFDFDF}
\definecolor{error-red}{HTML}{EFB2B6}

\usepackage[colorlinks=true,citecolor=dodgerblue,linkcolor=dodgerblue,urlcolor=dodgerblue,pdftitle={Glue}]{hyperref}

\begin{document}

\title{Classifying One-Dimensional Quantum States\\Prepared by a Single Round of Measurements}
\author{Rahul Sahay}
\affiliation{Department of Physics, Harvard University, Cambridge, Massachusetts 02138 USA}
\author{Ruben Verresen}
\affiliation{Department of Physics, Harvard University, Cambridge, Massachusetts 02138 USA}
\affiliation{Department of Physics, Massachusetts Institute of Technology, Cambridge, MA 02139, USA}

\date{\today}

\begin{abstract}
Measurements and feedback have emerged as powerful resources for creating many-body quantum states.
However, a detailed understanding has been restricted to fixed-point representatives of phases of matter.
Here, we go beyond this and characterize the patterns of many-body entanglement that can be deterministically created from measurement.
Focusing on one spatial dimension, a framework is developed for the case where a single round of measurements is the only entangling operation.
We show this creates matrix product states and identify necessary and sufficient tensor conditions for preparability, which uniquely determine the preparation protocol.
We use these conditions to both classify preparable quantum states and characterize their physical constraints.
In particular, we find a trade-off between the richness of the preparable entanglement spectrum and correlation functions, which leads to a no-go theorem for preparing certain quantum states.
More broadly, we connect properties of the preparation protocol to the resulting phase of matter, including trivial, symmetry-breaking, and symmetry-protected topological phases---for both uniform and modulated symmetries.
This work offers a resource-theoretic perspective on preparable quantum entanglement and shows how to systematically create states of matter, away from their fixed points, in quantum devices.
\end{abstract}

\maketitle

\tableofcontents

\section{Introduction}

Emergence in many-body quantum states can give rise to remarkable physical properties \cite{WenBook,NewSachdevBook}.
Indeed, decades of exploration have revealed a plethora of possible phases of matter, including symmetry-breaking states and more exotic `topological' states \cite{Wen_90, WenRMP}.
In one spatial dimension, a complete classification has even been achieved for ground states of local gapped Hamiltonians, including symmetry-protected topological (SPT) states with interesting boundary properties \cite{Su1979-rl,Haldane1983-ya,AKLT_original,Gu09,Schuch2011,Pollmann_2010,Turner2011-zi,fidkowski2011topological,Chen2011-et,son12,Chen2011-kz,Pollmann2012-lv,Chen2012-oa}.
However, knowing the physical \emph{properties} of a many-body quantum state is quite different from knowing how to \emph{create} it.
This has become of increasing relevance due to the rapid development of highly tunable quantum devices \cite{NISQ,PRXQuantum.2.017003} and raises the question: \emph{which states of matter can be created with the capabilities of a given platform?}
This question is even of fundamental importance, since from a resource-theoretic perspective, we learn something new about a state when we understand how to create it.

A conceptual and practical development in the efficient preparation of many-body quantum states is the use of measurements. 
Although this is a rather novel ingredient in the context of condensed matter physics---where one has typically focused on the unitary aspects of quantum physics---the use of measurement has a rich history in quantum information theory \cite{Teleportation,NielsenChuangBook}.
In particular, the idea of error correction \cite{gottesman1997stabilizer} naturally suggests that certain `stabilizer states' (including the famed GHZ and toric code states \cite{Kitaev_2003}) can be prepared by measuring an initial quantum state and correcting the non-deterministic outcomes with a unitary feedback circuit, conditioned on the measurement outcomes \cite{Briegel01,Raussendorf05,Aguado08,Bolt16}. 
In stark contrast to the case with only unitary circuits \cite{bravy2006LR, chen2010wen}, measurements allow one to create such states in a time (or `circuit depth') which is independent of system size\footnote{More precisely, the `quantum time' is finite, by which we mean the depth of the quantum operations in the circuit; classical information-gathering and -processing still has a linear overhead, but such ingredients tend to be cheap and are only limited by the speed of light.}.

\begin{figure}
    \centering
    \includegraphics[width = 247pt]{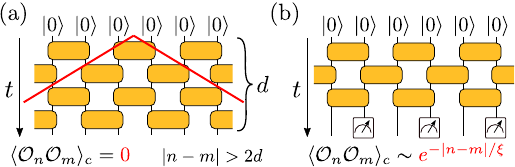}
    \caption{\textbf{Correlations from Unitaries and Measurements.} (a) Local unitary circuits can only create correlations at a distance that is proportional to their depth. Notably, connected correlation functions $\langle \mathcal{O}_n \mathcal{O}_m\rangle_c \equiv \langle \mathcal{O}_n \mathcal{O}_m\rangle - \langle \mathcal{O}_n \rangle \langle \mathcal{O}_m\rangle$ beyond a certain distance will be identically zero. (b) In contrast, the present work explores how and when finite-depth protocols with measurements and feedforward can create correlation functions whose asymptotic \textit{correlation length} $\xi$ is nonzero.}
    \label{fig:correlation_unitary_v_measurement}
\end{figure}

Recent years have seen rapid progress in developing new ways of preparing many-body states using measurement, both at the theory level \cite{piroli2021locc,LREfromSPT,verresen2021efficiently,lu2022shortcut,Bravyi22,lee2022decoding,hierarchy,zhu2023nishimori,smith2023aklt,shortestroute,gunn2023phases,li2023set,lu2023mixedstateLRO, piroli2024approximate,malz2024logdepth, li2023measuring, Wu_2023, friedman2023locality, friedman2023feedback, sukeno2024LGT,buhrman2023state, lee2024symmetry, hong2024quantum} as well as experimental implementations \cite{iqbal2023topological,foss2023advantage,chen2023nishimori,Iqbal2024nonabel,Bluvstein_2023}.
However, most works consider the deterministic preparation of \emph{fixed-point} states, which represent exotic states of matter but otherwise have rather featureless correlation functions and entanglement spectra.
Two notable exceptions in literature are the deterministic constant-depth preparation of the spin-1 AKLT state \cite{smith2023aklt} and of certain wavefunction deformations of the trivial and GHZ states (Appendix G of Ref.~\onlinecite{zhu2023nishimori}).
These constitute examples of states with non-trivial correlation length $0 < \xi < \infty$ which can nevertheless be \emph{deterministically} prepared in \emph{finite time} using measurement.
The existence of such examples is a priori surprising: the emergence of smooth and exponentially decaying correlations seems in tension with the discrete nature of quantum processes appearing in digital quantum devices.
Indeed, as is highlighted in Fig.~\ref{fig:correlation_unitary_v_measurement}, such smooth correlations cannot be created by the strictly local unitary circuits that appear in such devices.

Emboldened by these limited examples, we explore the question: \emph{what types of many-body entanglement and correlation functions can possibly arise from a measurement-based protocol?} In addition to the immediate practical relevance, an answer to this type of question can give insight into the nature of phases of matter and entanglement---similar to knowing how the \emph{inability} of preparing the toric code using finite-depth unitary circuits \cite{bravy2006LR} captures part of the essence of the toric code itself. If we hope to provide a crisp answer, we must provide a crisp list of ingredients of what we do (not) allow. To single out the power of measurement, we will explore a \emph{measurement-only} protocol: measurement will be the only entangling operation. Moreover, in this work we will allow only a single layer of measurement. Despite this restrictive setting, we will see that the resulting landscape is remarkably rich and structured, even in one spatial dimension.

While we provide a detailed description of our set-up in Sec.~\ref{sec:setup}, let us give a brief taste of what we consider:
\begin{shaded}
    \textbf{Ingredients (Informal)} We explore and characterize one-dimensional quantum states which are deterministically preparable with the following ingredients: 
    \begin{enumerate}
        \item[(1)] An initial product state of disentangled clusters (shown in blue)

        \item[(2)] Finite-range complete-basis measurements on a subset of sites (white)

        \item[(3)] On-site (i.e., non-entangling) unitary feedback conditioned on measurement outcomes (orange). 
    \end{enumerate}
    Schematically:
    \noindent
        \begin{equation*}
        \includegraphics[valign = c, scale = 0.9]{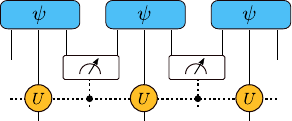}
    \end{equation*}
\end{shaded}
\noindent These ingredients are in part inspired by those used in Ref.~\onlinecite{smith2023aklt} for preparing the AKLT state, although we have minimized our assumptions (e.g., we do not presume that the measurement basis is fully-entangled, also called a fusion measurement \cite{bartolucci2021fusionbased}). Let us also stress that we demand deterministic state-preparation, which means we can correct for any possible measurement outcome. This is in contrast to works which consider approximate, probabilistic or partial preparation, whereby one prepares, e.g., mixed states \cite{lee2022decoding,lu2022shortcut,zhu2023nishimori,lu2023mixedstateLRO}  or (arbitrarily-good) approximations to the target state \cite{piroli2021locc,piroli2024approximate}.

In this work, we classify and detail the physical properties of states which can result from the above set of ingredients. 
A general insight that emerges is that such states can have very rich entanglement spectra and very rich correlation functions---\emph{but not at the same time}. 
Indeed, we find a trade-off between the two. This is crystallized by proving a no-go theorem for states which cannot be prepared in this way.

These and other results follow from a general framework we develop.
This is built on the observation that any state obtained from these ingredients must be a matrix product state (MPS) \cite{Fannes1992, Cirac2021MPSReview}. 
Using the MPS formalism, we are then able to convert the above list of ingredients into necessary and sufficient conditions on the MPS tensors which guarantee preparability of the state and constrain their form, enabling a classification.
Remarkably, if an MPS tensor satisfies these conditions, it also essentially specifies the preparation protocol.

The advantage of this approach is that once we have encoded these ingredients into conditions on the local MPS tensor, we can study what physical properties are implied by these conditions. This leads to a series of results, such as general connections between the feedback-protocol and the resulting phase of matter. To make these results broadly accessible, Sec.~\ref{sec:examples_and_prelude} introduces key ideas through examples and informally states some of our key results without requiring MPS-based language.

After the overview section (Sec.~\ref{sec:examples_and_prelude}), we go over the basic ingredients in Sec.~\ref{sec:setup}. We show how this implies an MPS description of the state, along with the local characterization which guarantees preparability, leading to a classification. In Sec.~\ref{sec:Topological} we explore the case where feedback requires non-local classical communication. We relate this to various physical quantities, such as a degenerate entanglement spectrum, and we even present full classifications in certain cases. This is followed by Sec.~\ref{sec:localvtopological}, which also studies the case with local error correction. The latter culminates in the aforementioned no-go theorem.

\section{Motivating Examples and Prelude of Results}\label{sec:examples_and_prelude}

Before delving into a general formalism for characterizing states preparable using the ingredients above, we begin with an informal prelude of our results.
Specifically, we discuss three examples of deterministically preparable states, each of which elucidates aspects of the general problem.
We subsequently highlight the essential features of these examples that generalize, stating informally and without proof the key physical ideas and results of this work with formal statements postponed to later sections.
Readers seeking a sense of what can be prepared with the ingredients above and a conceptual pr\'ecis of what constraints preparability places on quantum states can choose only to read this section.
Alternatively, those interested in the general formalism and not requiring motivating examples are encouraged to read only Sec.~\ref{subsec-Prelude} of this section, before going to the next section.

\begin{figure*}
    \centering
    \includegraphics[width = 494 pt]{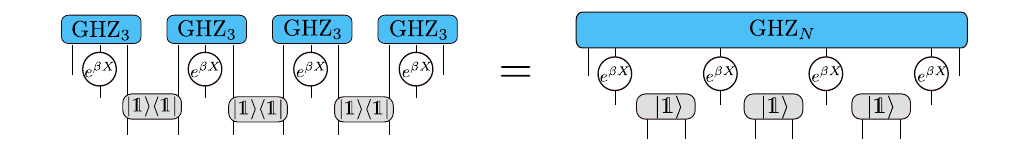}
    \caption{\textbf{Preparation of Beyond Fixed Point GHZ State.} Entangling measurements between disentangled clusters and tensor product unitary feedback can deterministially create quantum states away from their fixed points.
    This is exemplified in the above where, to create a deformed GHZ state $e^{\beta \sum_{x = 2}^{N - 1} X_x} \ket{\text{GHZ}_{N }}$ (shown on right), one first prepares disentangled three-qubit clusters $e^{\beta X}|\text{GHZ}_{3} \rangle$ (left).
    Such clusters can be prepared unitarily by first creating the three-qubit GHZ state and then time evolving with a particular Hamiltonian: $e^{-i tY_{l} Z_{c}} \ket{\text{GHZ}_3}$ with $l$ and $c$ labeling the left and center qubits respectively and $\tan(t) = \tanh(\beta)$.
    Subsequently, one measures nearest-neighbor qubits of adjacent clusters in the Bell basis.
    This ``glues'' the clusters together and, in the ideal case of every measurement outcome yielding $\ket{\mathds{1}} = \ket{00} + \ket{11}/\sqrt{2}$, prepares the deformed GHZ exactly.
    When the measurements outcomes are different, ``errors'' will exist atop the desired deformed GHZ state, which we show can be corrected with a tensor product of unitaries conditioned on the measurement outcomes.
    }. 
    \label{fig:prepGHZ}
\end{figure*}

\subsection{Example 1: Deformed GHZ State \label{subsec-deformedGHZ}}

For our first motivating example, we describe the preparation protocol for a parameterized class of long-range entangled wavefunctions, which spontaneously break a $\mathbb{Z}_2$ spin-flip symmetry.
The wavefunctions are given by 
\begin{align} \label{eq-ebetaXGHZ}
    \ket{\Psi(\beta)} &\propto \exp\left( \beta \sum_{x = 2}^{N -1} X_x \right) \ket{\text{GHZ}_N} \\
    &= \begin{tikzpicture}[scale = 1, baseline = {([yshift=-.5ex]current bounding box.center)}]
        \draw[rounded corners, fill = lightdodgerblue] (0, 0) rectangle (3.6, 0.4);
        \foreach \i in {0, ..., 6} {
        %\draw[rounded corners, fill = lightdodgerblue] (\i*1.3,0) rectangle (\i*1.3 + 1, 0.4) {};
        \draw[color = black] (0.1 + 0.57*\i, 0) -- (0.1 + 0.57*\i, -0.5);
        %\draw[color = black] (0.5 + 1.3*\i, 0) -- (0.5 + 1.3*\i, -0.5);
        };
        \foreach \i in {1, ..., 5} {
        \draw[fill = white] (0.1 + 0.57*\i, -0.25) circle (0.1);
        };
        \node at (1.8, 0.2) {\small $\text{GHZ}_N$};
     \end{tikzpicture}
\end{align}
that are deformed versions of the standard GHZ state beyond its fixed point (where boundary perturbations are omitted to simplify the discussion\footnote{This guarantees \emph{deterministic} state-preparation. Boundary perturbations still allow for statistical state-preparation, with a success probability \emph{independent of system size}.}).
Here, the white circles are the non-unitary time-evolution $e^{\beta X}$ ($\beta \in \mathbb R$) and $\ket{\text{GHZ}_N} \propto \left(\ket{0\cdots 0} + \ket{1 \cdots 1}\right)$ is the standard GHZ state defined on a one-dimensional chain of $N$ qubits and $X, Y, Z$ are the Pauli matrices. One can straightforwardly confirm this state retains its `cat state' nature for all $\beta < \infty$:
\begin{equation}
\small \prod_{x=1}^N X_x | \Psi(\beta)\rangle =| \Psi(\beta)\rangle,\; \langle \Psi(\beta) | Z_x Z_{y\neq x} | \Psi(\beta) \rangle = \sech(2\beta)^{2}.
\end{equation}
However, for $0 \neq \beta < \infty$ this state has non-trivial correlation lengths\footnote{This is evidenced by its symmetry string correlator $\prod_{x=a>1}^{b<N} X_x$ decaying as $e^{-|b-a|/\xi}$ with $\xi = -1/\ln \tanh(2\beta)$.}. Our preparation protocol will generalize the one in Ref.~\onlinecite{smith2023aklt}, which considered $\beta=0$.

In line with the ingredients presented in the introduction, we prepare this state by starting with a product state of $(N-2)$ decoupled three-qubit states\footnote{Each state can be unitarily obtained from $\ket{\text{GHZ}_3}$ as $e^{-i t Z_{l_x} Y_{c_x}} \ket{\text{GHZ}_3}$ with $\tan t = \tanh \beta$.}:
\begin{align} \label{eq-decoupled-cluster}
    \ket{\Psi_0(\beta)} &\propto \bigotimes_{x = 2}^{N-1} e^{\beta X_{c_x}} \ket{\text{GHZ}_3}_{l_x, c_x, r_x} \\ 
    &=\cdots \begin{tikzpicture}[scale = 1, baseline = {([yshift=-.5ex]current bounding box.center)}]
        \foreach \i in {0, ..., 2} {
        \draw[rounded corners, fill = lightdodgerblue] (\i*1.3,0) rectangle (\i*1.3 + 1, 0.4) {};
        \node at (0.5 + \i * 1.3, 0.2) {\small $\text{GHZ}_3$};
        \draw[color = black] (0.1 + 1.3*\i, 0) -- (0.1 + 1.3*\i, -0.4);
        \draw[color = black] (0.5 + 1.3*\i, 0) -- (0.5 + 1.3*\i, -0.4);
        \draw[color = black] (0.9 + 1.3*\i, 0) -- (0.9 + 1.3*\i, -0.4);
        % \draw[color = black] (1.4, 0) -- (1.4, -0.5);
        % \draw[color = black] (1.8, 0) -- (1.8, -0.65);
        % \draw[color = black] (2.2, 0) -- (2.2, -0.25);
        \draw[fill = white] (0.5 + 1.3*\i, -0.2) circle (0.1);
        %\draw[fill = white] (1.8, -0.3) circle (0.1);
        % \node at (1.15, -0.4) { \scriptsize $\bra{\mathds{1}}$};
        };
        %\node at (2.7, -0.3) {\small $Z$};
        %\node at (1.15 + 2.6, -0.4) { \scriptsize $\bra{ZX}$};
     \end{tikzpicture} \cdots
\end{align}
where $l_x, c_x, r_x$ label the center, left, and right qubit of the three qubit cluster.
Subsequently, we perform nearest-neighbor measurements between qubits $r_x$ and $l_{x + 1}$ on the above cluster in the Bell basis:
\begin{align}
    &\ket{\mathds{1}} = \frac{1}{\sqrt{2}}(\ket{00} + \ket{11})\quad \ket{X} = \frac{1}{\sqrt{2}}(\ket{01} + \ket{10}) \label{eq-Bell} \\
    &\ket{Z} = \frac{1}{\sqrt{2}}(\ket{00} - \ket{11})\quad \ket{ZX} = \frac{1}{\sqrt{2}} (\ket{10} - \ket{01}) \notag
\end{align}
which satisfy $\ket{V}_{12} = V_{2} \ket{\mathds{1}}_{12} = V_{1}^{\dagger} \ket{\mathds{1}}_{12}$.
To understand the result of this, let us consider the case where each of the measurement outcomes yield the $\ket{\mathds{1}}$ Bell state.
In this case, the post-measurement state on the unmeasured qubits is:
\begin{equation}
     \left( \bigotimes_{x = 1}^{N - 3} \bra{\mathds{1}}_{r_{x} l_{x + 1}} \right)\ket{\Psi_0(\beta)} \propto \cdots \begin{tikzpicture}[scale = 1, baseline = {([yshift=-.5ex]current bounding box.center)}]
        \draw[rounded corners, fill = lightdodgerblue] (0,0) rectangle (1, 0.4) {};
        \draw[rounded corners, fill = lightdodgerblue] (1.3,0) rectangle (2.3, 0.4) {};
        \node at (0.5, 0.2) {\small $\text{GHZ}_3$};
        \node at (1.8, 0.2) {\small $\text{GHZ}_3$};
        \draw[color = black] (0.1, 0) -- (0.1, -0.25);
        \draw[color = black] (0.5, 0) -- (0.5, -0.65);
        \draw[color = black] (0.9, 0) -- (0.9, -0.5);
        \draw[color = black] (1.4, 0) -- (1.4, -0.5);
        \draw[color = black] (1.8, 0) -- (1.8, -0.65);
        \draw[color = black] (2.2, 0) -- (2.2, -0.25);
        \draw[fill = white] (0.5, -0.3) circle (0.1);
        \draw[fill = white] (1.8, -0.3) circle (0.1);
        \draw[rounded corners, fill = lightishgray] (0.8,-0.25) rectangle (1.5, -0.55) {};
        \node at (1.15, -0.4) { \scriptsize $\bra{\mathds{1}}$};
     \end{tikzpicture} \cdots 
\end{equation}
where the white circles indicate the $e^{\beta X}$ operator appearing in the initial cluster and time runs downwards.
Note that, the projector onto the $\ket{\mathds{1}}$ state will ``glue'' the neighboring $\text{GHZ}_3$'s together to a many-body wavefunction.
Specifically, since the projector $\bra{\mathds{1}}$ enforces the qubits $r_x$ and $l_{x+1}$ to be in the same computational state and the $\ket{\text{GHZ}_3}_x$ state has the property that each of its three qubits are in the same computational state, the result of this ``gluing'' will be to produce a many-body GHZ state on the $N$ unmeasured qubits.
Further taking into account the $e^{\beta X}$ gates, we conclude that, in the case where all measurement outcomes yield $\ket{\mathds{1}}$, the resulting wavefunction is the desired target state $\ket{\Psi(\beta)}$ with no unitary  feedback necessary.

Now, let us consider what happens for an arbitrary measurement outcome.
For example, the circuit diagram for such an outcome may look like:
\begin{equation}
      \begin{tikzpicture}[scale = 1, baseline = {([yshift=-.5ex]current bounding box.center)}]
        \foreach \i in {0, ..., 2} {
        \draw[rounded corners, fill = lightdodgerblue] (\i*1.3,0) rectangle (\i*1.3 + 1, 0.4) {};
        \node at (0.5 + \i * 1.3, 0.2) {\small $\text{GHZ}_3$};
        \draw[color = black] (0.1 + 1.3*\i, 0) -- (0.1 + 1.3*\i, -0.5);
        \draw[color = black] (0.5 + 1.3*\i, 0) -- (0.5 + 1.3*\i, -0.65);
        \draw[color = black] (0.9 + 1.3*\i, 0) -- (0.9 + 1.3*\i, -0.5);
        % \draw[color = black] (1.4, 0) -- (1.4, -0.5);
        % \draw[color = black] (1.8, 0) -- (1.8, -0.65);
        % \draw[color = black] (2.2, 0) -- (2.2, -0.25);
        \draw[fill = white] (0.5 + 1.3*\i, -0.3) circle (0.1);
        %\draw[fill = white] (1.8, -0.3) circle (0.1);
        % \node at (1.15, -0.4) { \scriptsize $\bra{\mathds{1}}$};
        };
        % \foreach \i in {0, ..., 1} {
        %     \draw[rounded corners, fill = lightishgray] (0.8 + 1.3*\i,-0.25) rectangle (1.5 + 1.3*\i, -0.55) {};
        % }
        \draw[rounded corners, fill = error-red] (0.8 + 1.3*0,-0.25) rectangle (1.5 + 1.3*0, -0.55) {};
        \draw[rounded corners, fill = lightishgray] (0.8 + 1.3*1,-0.25) rectangle (1.5 + 1.3*1, -0.55) {};
        \node at (1.15, -0.4) { \scriptsize $\bra{X}$};
        \node at (1.15 + 1.3, -0.4) { \scriptsize $\bra{\mathds{1}}$};
        %\node at (1.15 + 2.6, -0.4) { \scriptsize $\bra{ZX}$};
     \end{tikzpicture}\ =\ \begin{tikzpicture}[scale = 1, baseline = {([yshift=-.5ex]current bounding box.center)}]
        \foreach \i in {0, ..., 2} {
        \draw[rounded corners, fill = lightdodgerblue] (\i*1.3,0) rectangle (\i*1.3 + 1, 0.4) {};
        \node at (0.5 + \i * 1.3, 0.2) {\small $\text{GHZ}_3$};
        \draw[color = black] (0.1 + 1.3*\i, 0) -- (0.1 + 1.3*\i, -0.5);
        \draw[color = black] (0.5 + 1.3*\i, 0) -- (0.5 + 1.3*\i, -0.65);
        \draw[color = black] (0.9 + 1.3*\i, 0) -- (0.9 + 1.3*\i, -0.5);
        % \draw[color = black] (1.4, 0) -- (1.4, -0.5);
        % \draw[color = black] (1.8, 0) -- (1.8, -0.65);
        % \draw[color = black] (2.2, 0) -- (2.2, -0.25);
        \draw[fill = white] (0.5 + 1.3*\i, -0.3) circle (0.1);
        %\draw[fill = white] (1.8, -0.3) circle (0.1);
        % \node at (1.15, -0.4) { \scriptsize $\bra{\mathds{1}}$};
        };
        \foreach \i in {0, ..., 1} {
            \draw[rounded corners, fill = lightishgray] (0.8 + 1.3*\i,-0.25 - 0.25) rectangle (1.5 + 1.3*\i, -0.55 -0.25) {};
        }
        \node at (1.15, -0.65) { \scriptsize $\bra{\mathds{1}}$};
        \node at (1.15 + 1.3, -0.65) { \scriptsize $\bra{\mathds{1}}$};
        \node at (1.4, -0.3) {\small $\textcolor{red}{X}$};
        %\node at (2.7, -0.3) {\small $Z$};
        %\node at (1.15 + 2.6, -0.4) { \scriptsize $\bra{ZX}$};
     \end{tikzpicture}
\end{equation}
Here, a measurement outcome of anything other than $\ket{\mathds{1}}$ will be referred to as an error, that needs to be corrected by acting a tensor product unitary at the physical level.
To see how a $\ket{X}$ measurement error is to be corrected, we remark upon a convenient property of the GHZ state.
Namely, $X_{l_x} \ket{\text{GHZ}_3}_x = X_{c_x} X_{r_x} \ket{\text{GHZ}_3}$.
Consequently, we have that: 
\begin{equation}
    \begin{tikzpicture}[scale = 1, baseline = {([yshift=-.5ex]current bounding box.center)}]
        \foreach \i in {0, ..., 2} {
        \draw[rounded corners, fill = lightdodgerblue] (\i*1.3,0) rectangle (\i*1.3 + 1, 0.4) {};
        \node at (0.5 + \i * 1.3, 0.2) {\small $\text{GHZ}_3$};
        \draw[color = black] (0.1 + 1.3*\i, 0) -- (0.1 + 1.3*\i, -0.5);
        \draw[color = black] (0.5 + 1.3*\i, 0) -- (0.5 + 1.3*\i, -0.65);
        \draw[color = black] (0.9 + 1.3*\i, 0) -- (0.9 + 1.3*\i, -0.5);
        % \draw[color = black] (1.4, 0) -- (1.4, -0.5);
        % \draw[color = black] (1.8, 0) -- (1.8, -0.65);
        % \draw[color = black] (2.2, 0) -- (2.2, -0.25);
        \draw[fill = white] (0.5 + 1.3*\i, -0.3) circle (0.1);
        %\draw[fill = white] (1.8, -0.3) circle (0.1);
        % \node at (1.15, -0.4) { \scriptsize $\bra{\mathds{1}}$};
        };
        \foreach \i in {0, ..., 1} {
            \draw[rounded corners, fill = lightishgray] (0.8 + 1.3*\i,-0.25 - 0.25) rectangle (1.5 + 1.3*\i, -0.55 -0.25) {};
        }
        \node at (1.15, -0.65) { \scriptsize $\bra{\mathds{1}}$};
        \node at (1.15 + 1.3, -0.65) { \scriptsize $\bra{\mathds{1}}$};
        \node at (1.4, -0.3) {\small $\textcolor{red}{X}$};
        %\node at (2.7, -0.3) {\small $Z$};
        \node at (2.4, -1.1) {\scriptsize $\ $};
        %\node at (1.15 + 2.6, -0.4) { \scriptsize $\bra{ZX}$};
     \end{tikzpicture} =     \begin{tikzpicture}[scale = 1, baseline = {([yshift=-.5ex]current bounding box.center)}]
        \foreach \i in {0, ..., 2} {
        \draw[rounded corners, fill = lightdodgerblue] (\i*1.3,0) rectangle (\i*1.3 + 1, 0.4) {};
        \node at (0.5 + \i * 1.3, 0.2) {\small $\text{GHZ}_3$};
        \draw[color = black] (0.1 + 1.3*\i, 0) -- (0.1 + 1.3*\i, -0.5);
        \draw[color = black] (0.5 + 1.3*\i, 0) -- (0.5 + 1.3*\i, -0.65);
        \draw[color = black] (0.9 + 1.3*\i, 0) -- (0.9 + 1.3*\i, -0.5);
        % \draw[color = black] (1.4, 0) -- (1.4, -0.5);
        % \draw[color = black] (1.8, 0) -- (1.8, -0.65);
        % \draw[color = black] (2.2, 0) -- (2.2, -0.25);
        \draw[fill = white] (0.5 + 1.3*\i, -0.3) circle (0.1);
        %\draw[fill = white] (1.8, -0.3) circle (0.1);
        % \node at (1.15, -0.4) { \scriptsize $\bra{\mathds{1}}$};
        };
        \foreach \i in {0, ..., 1} {
            \draw[rounded corners, fill = lightishgray] (0.8 + 1.3*\i,-0.25 - 0.25) rectangle (1.5 + 1.3*\i, -0.55 -0.25) {};
        }
        \node at (1.15, -0.65) { \scriptsize $\bra{\mathds{1}}$};
        \node at (1.15 + 1.3, -0.65) { \scriptsize $\bra{\mathds{1}}$};
        %\node at (0.9 + 2.6, -0.3) {\small $Z$};
        \node at (0.9 + 1.3, -0.3) {\small $\textcolor{red}{X}$};
        \node at (0.5 + 1.3, -0.6) {\small $\textcolor{red}{X}$};
        \draw[color = red, fill = none] (0.7 + 1.3, -0.9) rectangle (1.6 + 1.3, -0.1);
        \node at (2.4, -1.1) {\scriptsize $\textcolor{red}{=\bra{X}}$};
        %\node at (1.15 + 2.6, -0.4) { \scriptsize $\bra{ZX}$};
     \end{tikzpicture}
\end{equation}
where we used the fact that $X$ commutes through $e^{\beta X}$.
The above can be iterated and shows that measurement outcomes that result in an $\ket{X}$ can be corrected by ``sweeping'' the measurement error to the right by acting with a physical unitary string of $X$'s from the location of the error to the boundary.
Similarly, the GHZ state has the property that and $Z_{l_x} \ket{\text{GHZ}_3}_x = \mathds{1}_{c_x} Z_{r_x} \ket{\text{GHZ}_3}_x$ implying that $\ket{Z}$ measurement outcomes can be similarly swept to the right, with no required action at the physical level except for a local $Z$ correction applied at the boundary.
Finally, corrections of $\ket{ZX}$ proceed similarly by using a combination of the techniques for $X$ and $Z$.
This demonstrates that the quantum state in Eq.~\eqref{eq-ebetaXGHZ} can be prepared deterministically using the ingredients referenced in the introduction.

Let us take stock of the qualitative features of the measurement protocol and the physical properties of the resulting state.
We saw that correcting a measurement error involved ``sweeping'' it off to the boundary. This means that a single error is corrected by a physical unitary action which was spatially delocalized from the location of the error. We call such errors \textbf{topological} (indeed as we discuss later, they correspond to defects in either long-range entangled phases or in symmetry-protected topological phases).
Equivalently, the feedback-unitary applied at a given physical site was non-locally conditioned on the classical data of every measurement outcome to the left of it.

The fact that all errors are non-locally corrected requires that the post-measurement state has the ability to `teleport' information across the system. In Sec.~\ref{sec:setup} we will prove quite generally that this necessitates the post-measurement state to have a flat entanglement spectrum, congruent with intuition gained from measurement-based quantum computation \cite{Raussendorf2001-xm,stephen2017MBQCSPT, Wei_2021}. Indeed, $\ket{\Psi(\beta)}$ has a bipartite entanglement spectrum given by  $\Lambda^2 = (1/2, 1/2)$, similar to the GHZ state.
Furthermore, the state has the seemingly unrelated property that it has no observables with zero correlation length (which we define in Sec.~\ref{sec:localvtopological}). We will later show that this is no accident.
By considering two more examples, we will phenomenologically motivate a key result of our work that these properties of the state are direct consequences of the fact that each measurement error is non-locally corrected.

\subsection{Example 2: Deformed Cluster SPT State \label{subsec-deformedcluster}}

Our next motivating example is a class of wavefunctions in the non-trivial symmetry-protected topological (SPT) phase protected by a $\mathbb{Z}_2 \times \mathbb{Z}_2$ symmetry generated by spin-flips on even and odd sublattices.
The wavefunctions are given by:
\begin{equation} \label{eq-ebetaXcluster}
    \ket{\Psi(\beta)} \propto \exp\left( \beta \sum_{x = 2}^{N -1} X_x \right) \ket{\text{cluster}_N}
\end{equation}
which are now deformed versions of the $N$-qubit ``cluster'' state, defined as the state stabilized by $Z_x X_{x + 1} Z_{x + 2}$ in the bulk and $X_{1} Z_2$ and $Z_{N-1} X_N$ at the boundary~\cite{Briegel01,son12,geraedts2014exact,Santos15}.
Physically, such states are short-range entangled but share many features in common with the deformed GHZ state.
In particular, they similarly have a flat bipartite entanglement spectrum given by  $\Lambda^2 = (1/2, 1/2)$ and also have the property that every connected correlation function that is not identically zero has exponentially decaying tails (see Appendix~\ref{app:examples}).
We now show that these shared physical properties between the GHZ  are accompanied by a similarity in the correction of measurement errors in the preparation of $\ket{\Psi(\beta)}$.

To see this, we note that this state is prepared by first preparing the same product state of $(N - 2)$ decoupled clusters as in the first example [c.f. Eq.~\eqref{eq-decoupled-cluster}] but instead of measuring in the Bell basis, we measure in a basis related to it by a Hadamard gate $H$, acting on the first qubit, e.g. our basis is $\ket{V'} = H_{r_x} \ket{V} = H_{r_x} V_{l_{x + 1}} \ket{\mathds{1}}$.
The claim is then when all the measurement outcomes are $H_{r_x} \ket{\mathds{1}}$:
\begin{align} \label{eq-secretcluster}
     \left( \bigotimes_{x = 1}^{N - 3} \bra{\mathds{1}}_{r_{x} l_{x + 1}} H_{r_x} \right)&\ket{\Psi_0(\beta)}\propto \cdots \begin{tikzpicture}[scale = 1, baseline = {([yshift=-.5ex]current bounding box.center)}]
        \draw[rounded corners, fill = lightdodgerblue] (0,0) rectangle (1, 0.4) {};
        \draw[rounded corners, fill = lightdodgerblue] (1.3,0) rectangle (2.3, 0.4) {};
        \node at (0.5, 0.2) {\small $\text{GHZ}_3$};
        \node at (1.8, 0.2) {\small $\text{GHZ}_3$};
        \draw[color = black] (0.1, 0) -- (0.1, -0.25);
        \draw[color = black] (0.5, 0) -- (0.5, -0.65);
        \draw[color = black] (0.9, 0) -- (0.9, -0.5);
        \draw[color = black] (1.4, 0) -- (1.4, -0.5);
        \draw[color = black] (1.8, 0) -- (1.8, -0.65);
        \draw[color = black] (2.2, 0) -- (2.2, -0.25);
        \draw[fill = white] (0.5, -0.3) circle (0.1);
        \draw[fill = white] (1.8, -0.3) circle (0.1);
        \draw[rounded corners, fill = lightishgray] (0.8,-0.5) rectangle (1.5, -0.8) {};
        \node at (1.15, -0.65) { \scriptsize $\bra{\mathds{1}}$};
        %\node at (0.9, -0.25) { \small $H$};
        \draw[fill = lightorange(ryb)] (0.9, -0.3) circle (0.1);
     \end{tikzpicture} \cdots \\
     &=\begin{tikzpicture}[scale = 1, baseline = {([yshift=-.5ex]current bounding box.center)}]
        \draw[rounded corners, fill = lightdodgerblue] (0, 0) rectangle (3.6, 0.4);
        \foreach \i in {0, ..., 6} {
        %\draw[rounded corners, fill = lightdodgerblue] (\i*1.3,0) rectangle (\i*1.3 + 1, 0.4) {};
        \draw[color = black] (0.1 + 0.57*\i, 0) -- (0.1 + 0.57*\i, -0.5);
        %\draw[color = black] (0.5 + 1.3*\i, 0) -- (0.5 + 1.3*\i, -0.5);
        };
        \foreach \i in {1, ..., 5} {
        \draw[fill = white] (0.1 + 0.57*\i, -0.25) circle (0.1);
        };
        \node at (1.8, 0.2) {\small $\text{cluster}_N$};
        \draw[fill = lightorange(ryb)] (0.1, -0.25) circle (0.1);
        \draw[fill = lightorange(ryb)] (0.1 + 6*0.57, -0.25) circle (0.1);
     \end{tikzpicture}
\end{align}
%\rv{add orange dots on qubit 1 and N in Eq. 14}
where the orange circle is the Hadamard matrix.
The above can be easily verified by showing that, when $\beta = 0$, the state is stabilized by $Z_x X_{x + 1} Z_{x + 2}$ in the bulk and  $X_1 Z_2$ and $Z_{N-1} X_N$ at the boundary (see Appendix~\ref{app:examples}).
%The claim is that the above state is $e^{\beta X} \ket{ \text{cluster}}$.
%

For an arbitrary measurement outcome, we see that the correction indeed proceeds similarly to the first example, with:
\begin{equation}
      \begin{tikzpicture}[scale = 1, baseline = {([yshift=-.5ex]current bounding box.center)}]
        \foreach \i in {0, ..., 2} {
        \draw[rounded corners, fill = lightdodgerblue] (\i*1.3,0) rectangle (\i*1.3 + 1, 0.4) {};
        \node at (0.5 + \i * 1.3, 0.2) {\small $\text{GHZ}_3$};
        \draw[color = black] (0.1 + 1.3*\i, 0) -- (0.1 + 1.3*\i, -0.5);
        \draw[color = black] (0.5 + 1.3*\i, 0) -- (0.5 + 1.3*\i, -0.65);
        \draw[color = black] (0.9 + 1.3*\i, 0) -- (0.9 + 1.3*\i, -0.5);
        % \draw[color = black] (1.4, 0) -- (1.4, -0.5);
        % \draw[color = black] (1.8, 0) -- (1.8, -0.65);
        % \draw[color = black] (2.2, 0) -- (2.2, -0.25);
        \draw[fill = white] (0.5 + 1.3*\i, -0.3) circle (0.1);
        \draw[fill = white] (1.8, -0.3) circle (0.1);
        % \node at (1.15, -0.4) { \scriptsize $\bra{\mathds{1}}$};
        };
        \draw[rounded corners, fill = error-red] (0.8 + 1.3*0,-0.25 - 0.25) rectangle (1.5 + 1.3*0, -0.55 -0.25) {};
        \draw[rounded corners, fill = lightishgray] (0.8 + 1.3*1,-0.25 - 0.25) rectangle (1.5 + 1.3*1, -0.55 -0.25) {};
        \node at (1.15, -0.65) { \scriptsize $\bra{X}$};
        \node at (1.15 + 1.3, -0.65) { \scriptsize $\bra{\mathds{1}}$};
        %\node at (1.8, -0.65) {\small $X$};
        %\node at (0.5, -0.65) {\small $Z$};
        %\node at (3.1, -0.65) {\small $Z$};
        %\node at (0.9, -0.25) {\small $H$};
        %\node at (2.2, -0.25) {\small $H$};
        %\node at (3.5, -0.25) {\small $H$};
        \draw[fill = lightorange(ryb)] (0.9, -0.3) circle (0.1);
        \draw[fill = lightorange(ryb)] (2.2, -0.3) circle (0.1);
        \draw[fill = lightorange(ryb)] (3.5, -0.3) circle (0.1);
        %\node at (1.15 + 2.6, -0.4) { \scriptsize $\bra{ZX}$};
     \end{tikzpicture} =       \begin{tikzpicture}[scale = 1, baseline = {([yshift=-.5ex]current bounding box.center)}]
        \foreach \i in {0, ..., 2} {
        \draw[rounded corners, fill = lightdodgerblue] (\i*1.3,0) rectangle (\i*1.3 + 1, 0.4) {};
        \node at (0.5 + \i * 1.3, 0.2) {\small $\text{GHZ}_3$};
        \draw[color = black] (0.1 + 1.3*\i, 0) -- (0.1 + 1.3*\i, -0.5);
        \draw[color = black] (0.5 + 1.3*\i, 0) -- (0.5 + 1.3*\i, -0.65);
        \draw[color = black] (0.9 + 1.3*\i, 0) -- (0.9 + 1.3*\i, -0.5);
        % \draw[color = black] (1.4, 0) -- (1.4, -0.5);
        % \draw[color = black] (1.8, 0) -- (1.8, -0.65);
        % \draw[color = black] (2.2, 0) -- (2.2, -0.25);
        \draw[fill = white] (0.5 + 1.3*\i, -0.3) circle (0.1);
        \draw[fill = white] (1.8, -0.3) circle (0.1);
        % \node at (1.15, -0.4) { \scriptsize $\bra{\mathds{1}}$};
        };
        \foreach \i in {0, ..., 1} {
            \draw[rounded corners, fill = lightishgray] (0.8 + 1.3*\i,-0.25 - 0.25) rectangle (1.5 + 1.3*\i, -0.55 -0.25) {};
        }
        \node at (1.15, -0.65) { \scriptsize $\bra{\mathds{1}}$};
        \node at (1.15 + 1.3, -0.65) { \scriptsize $\bra{\mathds{1}}$};
        \node at (1.8, -0.65) {\small $\textcolor{red}{X}$};
        %\node at (0.5, -0.65) {\small $Z$};
        \node at (3.5, -0.65) {\small $\textcolor{red}{X}$};
        % \node at (0.9, -0.25) {\small $H$};
        % \node at (2.2, -0.25) {\small $H$};
        % \node at (3.5, -0.25) {\small $H$};
        \draw[fill = lightorange(ryb)] (0.9, -0.3) circle (0.1);
        \draw[fill = lightorange(ryb)] (2.2, -0.3) circle (0.1);
        \draw[fill = lightorange(ryb)] (3.5, -0.3) circle (0.1);
        %\node at (1.15 + 2.6, -0.4) { \scriptsize $\bra{ZX}$};
     \end{tikzpicture}
\end{equation}
following from the GHZ identities we have regularly used.
The above implies that measurement outcomes that result in an $\ket{X}$ at location $x$ can be corrected by ``sweeping'' the measurement error to the right by acting with the string $\prod_{y \geq 0 }X_{x + 2y}$ from the location of the error to the boundary.
A similar analysis shows that measurement outcomes that result in an $\ket{Z}$ or $\ket{ZX}$ at location $x$ can be corrected by ``sweeping'' the measurement error to the right by acting with strings $\prod_{y \geq 0 }X_{x + 2y + 1}$ or $\prod_{y \geq 0} X_{x + y}$ from the location of the error to the boundary.
Consequently, the quantum state in Eq.~\eqref{eq-ebetaXcluster} can also be prepared deterministically using the ingredients referenced in the introduction.
Moreover, similar to the GHZ, each type of measurement error was non-locally corrected.
We note that this preparation protocol resembles the one introduced in Ref.~\onlinecite{smith2023aklt} for preparing the AKLT state \cite{AKLT_original}, although here we have a tunable correlation length. Crucially, we now turn to an example whose preparation and physical properties are radically distinct.

\subsection{Example 3: Deformed Trivial State \label{subsec-deformedtrivial}}

We conclude by discussing one final representative class of examples, which have both qualitatively different physical properties accompanied by a qualitatively different preparation protocol.
The examples are given by\footnote{One can also include the term for $x=1$ at the expense of applying an additional $e^{\alpha X_{l_1}}$ factor on the first cluster in Eq.~\eqref{eq-decoupled-cluster-v2}.}:
\begin{equation}\label{eq-deformedtriv}
    \ket{\Psi(\beta)} \propto \exp\left( \beta \sum_{x = 2}^{N -1} Z_x Z_{x + 1} \right) \ket{+}^{\otimes N}
\end{equation}
which are short-range entangled wavefunctions in the trivial phase for all $\beta < \infty$ that asymptote to the GHZ state exactly at $\beta = \infty$.
In this case, the state is preparable by first preparing the following entangled clusters:
\begin{equation} \label{eq-decoupled-cluster-v2}
    \ket{\Psi_0(\beta)} \propto \bigotimes_{x = 1}^{N} e^{\alpha X_{r_x}} \ket{\text{GHZ}_3}_{l_x, c_x, r_x} 
\end{equation}
with $\alpha = \text{arctanh}(e^{-2\beta})$. Crucially, while this looks similar to Eq.~\eqref{eq-decoupled-cluster}, note that the imaginary time-evolution is now on the \emph{right} rather than the \emph{central} qubit of each cluster.

Subsequently, we again measure in the familiar Bell basis \eqref{eq-Bell}.
We show in Appendix~\ref{app:examples} (using tensor network state methods) that in the case where each of the measurement outcome yields the $\ket{\mathds{1}}$ Bell state, the post-meaurement state is exactly $\ket{\Psi(\beta)}$.
We turn to discussing the correctability properties of the state.

Similar to the GHZ and cluster state case we find that: 
\begin{equation}
    \begin{tikzpicture}[scale = 1, baseline = {([yshift=-.5ex]current bounding box.center)}]
        \foreach \i in {0, ..., 2} {
        \draw[rounded corners, fill = lightdodgerblue] (\i*1.3,0) rectangle (\i*1.3 + 1, 0.4) {};
        \node at (0.5 + \i * 1.3, 0.2) {\small $\text{GHZ}_3$};
        \draw[color = black] (0.1 + 1.3*\i, 0) -- (0.1 + 1.3*\i, -0.5);
        \draw[color = black] (0.5 + 1.3*\i, 0) -- (0.5 + 1.3*\i, -0.65);
        \draw[color = black] (0.9 + 1.3*\i, 0) -- (0.9 + 1.3*\i, -0.5);
        % \draw[color = black] (1.4, 0) -- (1.4, -0.5);
        % \draw[color = black] (1.8, 0) -- (1.8, -0.65);
        % \draw[color = black] (2.2, 0) -- (2.2, -0.25);
        \draw[fill = white] (0.9 + 1.3*\i, -0.3) circle (0.1);
        %\draw[fill = white] (1.8, -0.3) circle (0.1);
        % \node at (1.15, -0.4) { \scriptsize $\bra{\mathds{1}}$};
        };
        % \foreach \i in {0, ..., 1} {
        %     \draw[rounded corners, fill = lightishgray] (0.8 + 1.3*\i,-0.25 - 0.25) rectangle (1.5 + 1.3*\i, -0.55 -0.25) {};
        % }
        \draw[rounded corners, fill = error-red] (0.8 + 1.3*0,-0.25 - 0.25) rectangle (1.5 + 1.3*0, -0.55 -0.25) {};
        \draw[rounded corners, fill = lightishgray] (0.8 + 1.3*1,-0.25 - 0.25) rectangle (1.5 + 1.3*1, -0.55 -0.25) {};
        \node at (1.15, -0.65) { \scriptsize $\bra{X}$};
        \node at (1.15 + 1.3, -0.65) { \scriptsize $\bra{\mathds{1}}$};
        %\node at (1.4, -0.3) {\small $X$};
        %\node at (2.7, -0.3) {\small $Z$};
        %\node at (1.15 + 2.6, -0.4) { \scriptsize $\bra{ZX}$};
     \end{tikzpicture} =     \begin{tikzpicture}[scale = 1, baseline = {([yshift=-.5ex]current bounding box.center)}]
        \foreach \i in {0, ..., 2} {
        \draw[rounded corners, fill = lightdodgerblue] (\i*1.3,0) rectangle (\i*1.3 + 1, 0.4) {};
        \node at (0.5 + \i * 1.3, 0.2) {\small $\text{GHZ}_3$};
        \draw[color = black] (0.1 + 1.3*\i, 0) -- (0.1 + 1.3*\i, -0.5);
        \draw[color = black] (0.5 + 1.3*\i, 0) -- (0.5 + 1.3*\i, -0.65);
        \draw[color = black] (0.9 + 1.3*\i, 0) -- (0.9 + 1.3*\i, -0.5);
        % \draw[color = black] (1.4, 0) -- (1.4, -0.5);
        % \draw[color = black] (1.8, 0) -- (1.8, -0.65);
        % \draw[color = black] (2.2, 0) -- (2.2, -0.25);
        \draw[fill = white] (0.9 + 1.3*\i, -0.3) circle (0.1);
        %\draw[fill = white] (1.8, -0.3) circle (0.1);
        % \node at (1.15, -0.4) { \scriptsize $\bra{\mathds{1}}$};
        };
        \foreach \i in {0, ..., 1} {
            \draw[rounded corners, fill = lightishgray] (0.8 + 1.3*\i,-0.25 - 0.25) rectangle (1.5 + 1.3*\i, -0.55 -0.25) {};
        }
        \node at (1.15, -0.65) { \scriptsize $\bra{\mathds{1}}$};
        \node at (1.15 + 1.3, -0.65) { \scriptsize $\bra{\mathds{1}}$};
        \node at (1.8, -0.6) {\small $\textcolor{red}{X}$};
        \node at (3.1, -0.6) {\small $\textcolor{red}{X}$};
        \node at (3.5, -0.6) {\small $\textcolor{red}{X}$};
        %\node at (1.8, -0.6) {\small $X$};
        %\node at (2.7, -0.3) {\small $Z$};
        %\node at (1.15 + 2.6, -0.4) { \scriptsize $\bra{ZX}$};
     \end{tikzpicture}
\end{equation}
which is to say that $\ket{X}$ at location $x$ can be corrected by ``sweeping'' the measurement error to the right by acting with the string $\prod_{y \geq 0 }X_{x + y}$ from the location of the error to the boundary.
However, unlike the other two cases, a $\ket{Z}$ error is corrected as: 
\begin{equation}
    \begin{tikzpicture}[scale = 1, baseline = {([yshift=-.5ex]current bounding box.center)}]
        \foreach \i in {0, ..., 2} {
        \draw[rounded corners, fill = lightdodgerblue] (\i*1.3,0) rectangle (\i*1.3 + 1, 0.4) {};
        \node at (0.5 + \i * 1.3, 0.2) {\small $\text{GHZ}_3$};
        \draw[color = black] (0.1 + 1.3*\i, 0) -- (0.1 + 1.3*\i, -0.5);
        \draw[color = black] (0.5 + 1.3*\i, 0) -- (0.5 + 1.3*\i, -0.65);
        \draw[color = black] (0.9 + 1.3*\i, 0) -- (0.9 + 1.3*\i, -0.5);
        % \draw[color = black] (1.4, 0) -- (1.4, -0.5);
        % \draw[color = black] (1.8, 0) -- (1.8, -0.65);
        % \draw[color = black] (2.2, 0) -- (2.2, -0.25);
        \draw[fill = white] (0.9 + 1.3*\i, -0.3) circle (0.1);
        %\draw[fill = white] (1.8, -0.3) circle (0.1);
        % \node at (1.15, -0.4) { \scriptsize $\bra{\mathds{1}}$};
        };
        \foreach \i in {0, ..., 1} {
            \draw[rounded corners, fill = lightishgray] (0.8 + 1.3*\i,-0.25 - 0.25) rectangle (1.5 + 1.3*\i, -0.55 -0.25) {};
        }
                \draw[rounded corners, fill = error-red] (0.8 + 1.3*0,-0.25 - 0.25) rectangle (1.5 + 1.3*0, -0.55 -0.25) {};
        \draw[rounded corners, fill = lightishgray] (0.8 + 1.3*1,-0.25 - 0.25) rectangle (1.5 + 1.3*1, -0.55 -0.25) {};
        \node at (1.15, -0.65) { \scriptsize $\bra{Z}$};
        \node at (1.15 + 1.3, -0.65) { \scriptsize $\bra{\mathds{1}}$};
     \end{tikzpicture} =     \begin{tikzpicture}[scale = 1, baseline = {([yshift=-.5ex]current bounding box.center)}]
        \foreach \i in {0, ..., 2} {
        \draw[rounded corners, fill = lightdodgerblue] (\i*1.3,0) rectangle (\i*1.3 + 1, 0.4) {};
        \node at (0.5 + \i * 1.3, 0.2) {\small $\text{GHZ}_3$};
        \draw[color = black] (0.1 + 1.3*\i, 0) -- (0.1 + 1.3*\i, -0.5);
        \draw[color = black] (0.5 + 1.3*\i, 0) -- (0.5 + 1.3*\i, -0.65);
        \draw[color = black] (0.9 + 1.3*\i, 0) -- (0.9 + 1.3*\i, -0.5);
        \draw[fill = white] (0.9 + 1.3*\i, -0.3) circle (0.1);
        };
        \foreach \i in {0, ..., 1} {
            \draw[rounded corners, fill = lightishgray] (0.8 + 1.3*\i,-0.25 - 0.25) rectangle (1.5 + 1.3*\i, -0.55 -0.25) {};
        }
        \node at (1.15, -0.65) { \scriptsize $\bra{\mathds{1}}$};
        \node at (1.15 + 1.3, -0.65) { \scriptsize $\bra{\mathds{1}}$};
        \node at (1.8, -0.6) {\small $\textcolor{red}{Z}$};
        \node at (3.1, -0.6) {\small $\ $};
        \node at (3.5, -0.6) {\small $\ $};
     \end{tikzpicture}
\end{equation}
which is entirely local (i.e. no need for infinite strings or correction at the boundary). We call this a \textbf{local error}.

As stated earlier, the difference in the correction properties of the errors in this example are accompanied by different physical properties.
Unlike the other two examples, here the entanglement spectrum is \emph{not flat}: one can calculate that $\Lambda^2 = \left(\frac{1+\delta}{2},\frac{1-\delta}{2} \right)$ with $\delta^{-1} = \cosh(2\beta)$. Indeed, now there is no longer a need for being able to teleport arbitrary Pauli matrices.
Moreover, we prove in Appendix~\ref{app:examples} that, unlike the other two examples, there exists an operator with a strictly finite range of correlations that does not exponentially decay (i.e. its correlation length is zero)\footnote{In particular, if we define $\mathcal O^L_n = Z_{n+1} X_{n+1} e^{-2\beta Z_n Z_{n+1}}$ and $\mathcal O^R_n = X_{n} Z_n e^{-2\beta Z_n Z_{n+1}}$ then $\langle \mathcal O^L_{2m} \mathcal O^R_{2n} \rangle $ equals $\tanh(2\beta)$ if $n=m+1$ but vanishes for $n>m+1$.}, although other operators have nonzero correlation length $\xi \sim 1/|\ln \tanh \beta|$.

\subsection{Informal Prelude of Results} \label{subsec-Prelude}

With these motivating examples in hand, our results fall into two categories which we summarize below. In these cases we refer to states as `\textbf{gluable}' if they can be prepared with the above ingredients: decoupled clusters which are `glued' together with a single-round of complete-basis measurements. In the above examples, we saw errors could be swept in a single direction. We will generally require that measurement errors can be corrected by conditioning only on gates to the left. For a detailed discussion of the ingredients, we refer to Sec.~\ref{sec:setup}.

\subsubsection{Interplay Between Preparability, Entanglement, Correlations, and Order}

A key physical result of our work is a relationship between how a state is prepared using measurement, and the physical properties of the resulting state (in particular, its correlation lengths, entanglement spectrum, and phase of matter).
To give a taste of these general results, we highlight five informally stated theorems, whose precise statements are in the parenthetically referenced sections below: 
\begin{enumerate}
    \item[I.] \textbf{Theorem} (\textit{Non-Local Correction Constrains Entanglement Spectrum}) If a wavefunction is preparable by gluing and all measurement errors require non-local correction, then its bipartite entanglement spectrum $\Lambda^2$ is constrained to be flat (Sec.~\ref{sec:simplygluablegeneral}).

    \item[II.] \textbf{Theorem} (\textit{Local Correction Constrains Correlations}) If a wavefunction is preparable by gluing and there exists a measurement error that can be locally corrected, then there exists an operator with zero correlation length (Sec.~\ref{sec:localvtopological}).
\end{enumerate}

A corollary of the above theorems is a powerful no-go theorem:

\begin{enumerate}
    \item[III.] \textbf{Theorem} \textit{(No-Go Theorem)} Suppose $\ket{\Psi}$ is a wavefunction whose entanglement spectrum $\Lambda^2$ is not flat. If there does not exist an operator whose connected correlation function in $\ket{\Psi}$ has a zero correlation length, then $\ket{\Psi}$ is not preparable with the ingredients listed in the introduction (Sec.~\ref{sec:localvtopological}).
\end{enumerate}

Despite this powerful restriction on the class of preparable many-body wavefunctions, we prove that the landscape of preparable states are rich.
This is exemplified by the following ``go theorem'' 
\begin{enumerate}
    \item[IV.] \textbf{Theorem} \textit{(Go Theorem)} Let $\ket{\Psi}$ be in a non-trivial SPT phase for abelian on-site symmetries. If it has minimal entanglement (i.e., the bipartition of an infinite chain has the smallest Schmidt rank allowed by the non-trivial SPT phase), then it is gluable with the above ingredients.
\end{enumerate}
Note that our second example (Sec.~\ref{subsec-deformedcluster}) is a special case of this general theorem, for the abelian group $\mathbb Z_2 \times \mathbb Z_2$ and Schmidt spectrum $\Lambda^2 = \left( \frac{1}{2} , \frac{1}{2} \right)$.

We also have a converse for this theorem, which arises as a special case of a classification of gluable states:
\begin{enumerate}
    \item[V.] \textbf{Theorem} \textit{(Classification)} We have a full classification of gluable quantum states (shown in Eq.~\eqref{eq-commutant}).
    This moreover allows one to construct tuneable ``phase diagrams'' of preparable states (e.g. see Eq.~\eqref{eq-Aclassification}), which we explore in more detail in a companion work \cite{sahay2024finite}.

    As a converse to the go theorem, our classification tells us that if a state is (i) short-range entangled; (ii) all errors are non-locally corrected; (iii) all errors commute up to phase (which means that correcting one error does not affect other errors); and (iv) the errors satisfy a technical homogeneity condition, then the state must be a minimally-entangled state for a non-trivial abelian SPT phase (Sec.~\ref{subsec:simplygluableSPT}).
\end{enumerate}
Such a classification highlights that precise answers can be found to the question of which states can be prepared with certain measurement-based resources. Once one is in possession of this closed-form solution of gluable states, their physical properties can be gleaned. In fact, one can use this to even \emph{engineer} preparable states with certain desired properties; the latter is illustrated in our companion work \cite{sahay2024finite}.

\subsubsection{Connection to Matrix Product States and Required Resources for Preparability}

Our second class of results connects preparable states to matrix product states (MPSs) \cite{Fannes1992, Cirac2021MPSReview} and further constrains the possible ways in which one could create interesting correlated states.
These class of results inform one ``where to look'' if one wants to prepare a given quantum many-body state from the ingredients above.

\begin{enumerate}
    \item[I.] \textbf{Theorem} Every gluable quantum state is a matrix product state.

    \item[II.] \textbf{Theorem} \textit{(Requirements on Initial Clusters)} The initial entangled clusters are guaranteed to be the matrix product state tensors in right-canonical form up to a local unitary acting individually on the ``physical'' qudit of the cluster and each ``virtual'' qudit of the cluster.

    \item[III.] \textbf{Theorem} \textit{(Requirements on Measurements)} The measurement basis used for preparation are guaranteed to be maximally entangled.
\end{enumerate}

The above are all formally stated in Sec.~\ref{sec:setup}. In addition, we note that we identify \emph{necessary and sufficient conditions} on the local matrix product state tensor to guarantee that a given quantum state is gluable with our ingredients (Eq.~\eqref{eq-VTTV}). Having provided an informal prelude of our results, we now introduce the formalism from which we can prove these and other results.

\section{General Setup and Formalism \label{sec:setup}}

In this section, we begin our more formal treatment of determining which quantum states are preparable using the ingredients present in the introduction.
To do so, we start by stating the assumed ingredients in the introduction more formally, thereby defining the notion of gluable quantum states.
We subsequently show that such states are necessarily matrix product states and then state a theorem that the clusters used to prepare them are necessarily the tensors of the matrix product state in canonical form (up to local unitary actions on the qudits making up the cluster).
These tensors will be shown to satisfy a strict set of criteria, that will pave the way to a general mathematical formalism that enables proving the remaining results of this work.

\subsection{Formalizing Setup and Reduction to Matrix Product States \label{subsec:setup}}

\begin{figure*}
    \centering
    \includegraphics[width = 494pt]{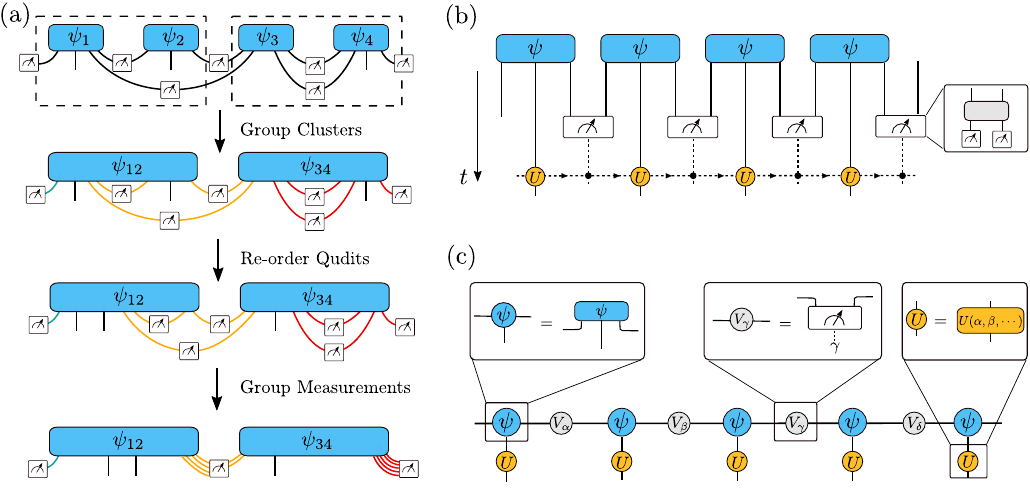}
    \caption{\textbf{Setup and Reduction to Matrix Product States.} We address the question of which quantum states can be prepared with an initial product state of entangled clusters in a one-dimensional geometry, a single round of finite-range complete basis measurement with non-overlapping support, and left-conditioned tensor product unitary feedback. 
    (a) While a generic protocol allowed within these set of ingredients may look unwieldy [top row], we first show that it can always be viewed as nearest neighbor measurements between disentangled clusters.
    To do so, we use a set of three moves: grouping clusters together [top to second row], re-ordering measured qudits [second to third row], and grouping measurements together [third to last row].
    A detailed discussion of this is provided in Appendix~\ref{app:reduction}. 
    A consequence is that the generic case is the setup shown in panel (b).
    Here, we remark that classical feedback for the measurement outcomes is shown using dotted lines coming from the measurements and informing the unitary action provided on the qudit clusters.
    The requirement of left-conditioned measurement is shown graphically with arrows moving to the left amongst the unitaries, depicting the flow of classical information used to create the feedback.
    (c) By reshaping the elements of panel (b), we are naturally led to an MPS representation of the state, with the clusters, measurement basis states, and unitaries forming the MPS tensors.
    We show in Theorem 1 that the unentangled clusters $\psi$ required to prepare the MPS are always the MPS tensors---up to a unitary basis change on each qudit of the cluster.
    }
    \label{fig:reduction_to_MPS}
\end{figure*}

Let us start by defining the notion of a `right-gluable' many-body quantum state (with a corresponding notion of `left-gluable' after making the appropriate changes):

\begin{shaded}
    \textbf{Definition} (Right-Gluable Quantum States) We say that a many-body quantum state is right-gluable if it can be prepared deterministically with:
    \begin{enumerate}
        \item[(1)] an initial product state of clusters of qudits arranged in a one-dimensional geometry;

        \item[(2)] a single round of finite-range, complete basis measurements with non-overlapping qudit support;

        \item[(3)] left-conditioned tensor product unitary feedback on unmeasured qubits.
        Here,  the tensor product structure is defined relative to the clusters, i.e. unitaries can act arbitrarily within clusters but not between them.
        Moreover, left-conditioning is a technical restriction we impose, requiring the correction unitary at a given unmeasured qubit only depends on measurements between clusters to the left of it. We thus only need to propagate classical information from left to right.
    \end{enumerate}

    We will also assume a minimality condition, which physically means that we do not use more measurements than necessary.
    Mathematically, this means that for any bipartition between the clusters, the Schmidt rank of the target state equals the Schmidt rank of the measurement projectors.
    See the discussion below Eq.~\eqref{eq-minimality} for a simpler rephrasing.
    
\end{shaded}

The above ingredients appear rather general and could enable measurement patterns of the form shown in Fig.~\ref{fig:reduction_to_MPS} (a, top row), which look rather different than the measurements in the examples in Sec.~\ref{sec:examples_and_prelude} that were nearest-neighbor and had non-overlapping geometric support.
Nevertheless, we show in Appendix~\ref{app:reduction} that more elaborate measurement schemes such as the one shown in the figure are, in fact, equivalent to the restrictive scheme used in the examples and shown in Fig.~\ref{fig:reduction_to_MPS}(b).
The technique for demonstrating equivalence is depicted for the readers benefit in Fig.~\ref{fig:reduction_to_MPS}(a).
We will henceforth, without loss of generality, work with a nearest-neighbor protocol.
Moreover, since we only consider right-gluable quantum states in this work, we will sometimes refer to them as gluable states for brevity.

A further convenient observation is that we can always presume that at least one of the measurement outcomes gives the desired state---without requiring correction. Indeed, consider any particular choice of measurement outcomes $\{\alpha_i\}$ where $i$ labels the location and $\alpha_i$ the outcome: by virtue of the state being gluable, we know there exists a tensor product unitary $U_{\{\alpha_i\}}$ which corrects the post-measurement state and produces the desired state. This means if we had simply applied this (non-entangling) unitary $U_{\{\alpha_i\}}$ to the initial clusters, then this particular measurement outcome would immediately give the desired result.

While the above are conventions which do not restrict the generality of our results, we will introduce a physical restriction for convenience: we consider translation-invariant set-ups, although many of our results hold more generally. More concretely, the initial clusters in Fig.~\ref{fig:reduction_to_MPS}(b) will be taken to be identical, and the measurement bases will be invariant under translation by one cluster. Of course we will \emph{not} presume the measurement outcomes to be translation-invariant, nor the feedback step.

\subsubsection{Entangled Clusters and Minimality Condition}

We now relate the outcomes of such measurement protocols (shown in Fig.~\ref{fig:reduction_to_MPS}(b)) to tensor network states (shown in Fig.~\ref{fig:reduction_to_MPS}(c)).
Let us observe that each cluster can be re-expressed graphically in a manner that looks more like a matrix product state tensor: 
\begin{equation} \label{eq-minimality}
    \begin{tikzpicture}[scale = 1, baseline = {([yshift=-.5ex]current bounding box.center)}]
        \draw[rounded corners, fill = lightdodgerblue] (0*1.3,0) rectangle (0*1.3 + 1, 0.4) {};
        \node at (0.5 + 0 * 1.3, 0.2) {\small $\psi$};
        \draw[color = black] (0.1 + 1.3*0, 0) -- (0.1 + 1.3*0, -0.4);
        \draw[color = black] (0.5 + 1.3*0, 0) -- (0.5 + 1.3*0, -0.4);
        \draw[color = black] (0.9 + 1.3*0, 0) -- (0.9 + 1.3*0, -0.4);
        \node at (0.14, -0.55) {\small $\ell_x$};
        \node at (0.54, -0.55) {\small $x$};
        \node at (0.94, -0.55) {\small $r_x$};
         \end{tikzpicture}\quad \longrightarrow \quad  \begin{tikzpicture}[scale = 1, baseline = {([yshift=-.5ex]current bounding box.center)}] 
        \draw[color = black] (-1, 0) -- (1, 0);
        \draw[color = black] (0, 0) -- (0, -0.8);
        \draw[fill = lightdodgerblue] (0,0) circle (0.3);
        \node at (0,0) {\small $\psi$};
        \node at (-1.1, 0) {\small $\ell_x$};
        \node at (0, -0.9) {\small $x$};
        \node at (1.15, 0) {\small $r_x$};
        %\node at (-1.3, 0) {\small $V^{[n]}$};
        %\node at (0, -1.2) {\small $\ $};
 %       \node at (-1.4, 0) {\small $\{V_{\alpha} \}$ };
    \end{tikzpicture} 
\end{equation}
where the `virtual' legs labeled $\ell_x$ and $r_x$ are assumed to have local Hilbert space dimension $\chi$, and the `physical' leg labeled $x$ has dimension $d$.
The aforementioned ``minimality'' condition is then equivalent to requiring that $\chi =\text{rank}(\Lambda)$, where $\Lambda$ is the bipartite entanglement spectrum of the state. I.e., the resulting matrix product state has no zeros in the entanglement spectrum.

\subsubsection{Measurements and Error Bases}

Similar to the clusters, it will also be useful to adopt a graphical notation for the different measurement outcomes.
In particular, let us suppose that we are measuring in a basis of states $\mathcal{V} = \{\ket{V_{\alpha}}\}$ with $\alpha = 1, \cdots, \chi^2$.
The projection operator onto each state can be graphically depicted as:
\begin{equation} \label{eq-V}
    \begin{tikzpicture}[scale = 1.5, baseline = {([yshift=-.5ex]current bounding box.center)}]
        \draw[rounded corners, fill = lightishgray] (0,-0.25) rectangle (0.7, -0.55) {};
        \draw[color = black] (0.1, -0.25) -- (0.1, 0);
        \draw[color = black] (0.6, -0.25) -- (0.6, 0);
        \node at (0.35, -0.4) {\small $\bra{V_{\alpha}}$};
    \end{tikzpicture}\quad \longrightarrow \quad \frac{1}{\sqrt{\chi}} \begin{tikzpicture}[scale = 1, baseline = {([yshift=-.5ex]current bounding box.center)}] 
        \draw[color = black] (-1, 0) -- (1, 0);
        %\draw[color = black] (0, 0) -- (0, -0.8);
        \draw[fill = lightishgray] (0,0) circle (0.3);
        \node at (0,0) {\small $V_{\alpha}$};
        %\node at (-1.3, 0) {\small $V^{[n]}$};
        %\node at (0, -1.2) {\small $\ $};
 %       \node at (-1.4, 0) {\small $\{V_{\alpha} \}$ };
    \end{tikzpicture}
\end{equation}
where the factor of $\chi^{-1/2}$ is chosen for later convenience.
In this language, the states in the measurement outcome are re-interpreted as operators $V_{\alpha}$, which means that the usual orthonormality of the measurement basis (i.e., $\langle V_\alpha|V_\beta\rangle = \delta_{ab}$) is equivalent to the following trace-orthonormality condition: 
\begin{equation}
    \text{tr}(V_{\alpha}^{\dagger} V_{\beta}) = \chi\, \delta_{\alpha \beta}.
\end{equation}
I.e., the basis $\mathcal{V}$ can be re-interpreted as an orthonormal basis for the space of $\chi \times \chi$ matrices.
In the error correction literature, such operators have been called an \textbf{error basis} \cite{knill1996group, klappenecker2005monomiality, klappenecker2002beyond}.
In fact, such literature tends to study the more restrictive notion of a \textbf{unitary error basis}, where each $V_\alpha$ is a unitary operator.
This is equivalent to the measurement basis being maximally entangled.
We emphasize that we do \emph{not impose} this condition (since it is a priori not essential for preparing states); however, we will soon prove that unitarity is a \emph{consequence} of our protocol.
As an example, in Sec.~\ref{sec:examples_and_prelude} we considered the case $\chi=2$, where the Pauli matrices defined an error basis $\{\mathds{1}, X, Z, ZX\}$; since these operators are all unitary (equivalently, the basis of Bell states \eqref{eq-Bell} is maximally entangled), it is indeed a unitary error basis.

Finally, as a technical remark, it will often be useful in proofs to use the following graphical notation for the full error basis:
\begin{equation}
    V_{\alpha} = \begin{tikzpicture}[scale = 1, baseline = {([yshift=-.5ex]current bounding box.center)}] 
        \draw[color = black] (-0.8, 0) -- (0.8, 0);
        \draw[color = black] (0, 0) -- (0, -0.8);
        \draw[fill = lightishgray] (0,0) circle (0.3);
        \node at (-0.0,0) {\small $\mathcal{V}$};
        \node at (-0.9, 0) {\small $\chi$};
        \node at (0.9, 0) {\small $\chi$};
        \node at (0.25, -0.6) {\small $\chi^2$};
        %\node at (1.2, -0.05) {\small $v_i$};
        % \node at (0.0, -1.2) {\small $\ $};
 %       \node at (-1.4, 0) {\small $\{V_{\alpha} \}$ };
    \end{tikzpicture}
\end{equation}
where the leg labels indicate dimensionality of the leg and the downwards facing leg is the ``$\alpha$'' leg.
Note that the trace orthonormality of the basis is equivalent to:
\begin{equation} 
    \text{tr}(V_{\alpha}^{\dagger} V_{\beta}) =\  \begin{tikzpicture} [scale = 1, baseline = {([yshift=-.5ex]current bounding box.center)}] 
    \draw[color = black] (-0.8, 0) -- (0.8, 0); %  top horizontal bar
    \draw[color = black] (0, 0) -- (0, 0.5); % vertical bar
    \node at (0.25, 0.5) {\small $\alpha$};
    \node at (0.25, -1.5) {\small $\beta$};
    \draw[fill = lightishgray] (0,0) circle (0.3); % circle
    % \draw[fill = white] (0,0.5) circle (0.14);
    %% Text
    \node at (0.0,0) {\small $\bar{\mathcal{V}}$};
    % \node at (0.0 ,0.5) {\small $\hat{s}$};
    %
    %
    \draw[color = black] (-0.8, 0) -- (-0.8, -1);
    \draw[color = black] (0.8, 0) -- (0.8, -1);
    \draw[color = black] (-0.8, -1) -- (0.8, -1); %  top horizontal bar
    \draw[color = black] (0, -1) -- (0, -0.5-1); % vertical bar
    \draw[fill = lightishgray] (0,-1) circle (0.3); % circle
    % \draw[fill = white] (0,-0.5-1) circle (0.14);
    % \draw[rounded corners, fill = lightdodgerblue] (-1, -0.7) rectangle (1, -0.3);
    %% Text
    \node at (0.,-1) {\small $\mathcal{V}$};
    \end{tikzpicture}\ =\  \chi \  \begin{tikzpicture}[scale = 0.8, baseline = {([yshift=-.5ex]current bounding box.center)}] 
        \draw[color = black] (0,0) -- (0, -1); 
    \end{tikzpicture}\ = \chi\, \delta_{\alpha \beta} .
\end{equation}
Moreover, we remark that the graphical notation here reveals an intriguing fact: $\mathcal{V}^{\dagger}$, when viewed as a $\chi^2 \times \chi^2$ matrix in the vertical direction of the above diagram, is the inverse of $\mathcal{V}$ and is hence unitary (we stress that this does \emph{not}   presume or imply that $V_\alpha$ is unitary as a $\chi \times \chi$ matrix). Therefore, we also obtain the following relation:
\begin{equation} \label{eq-niceidentity}
    \frac{1}{\chi} \begin{tikzpicture} [scale = 0.9, baseline = {([yshift=-.5ex]current bounding box.center)}] 
    \draw[color = black] (-0.8, 0) -- (0.8, 0); %  top horizontal bar
    \draw[color = black] (-0.8, -1.2) -- (0.8, -1.2); %  bottom horizontal bar
    \draw[color = black] (0, 0) -- (0, -1.2); % vertical bar
    \draw[fill = lightishgray] (0,0) circle (0.3); % circle
    \draw[fill = lightishgray] (0,-1.2) circle (0.3); % circle
    \draw[color = black] (-0.8, 0) -- (-0.8, 0.2);
    \draw[color = black] (0.8, 0) -- (0.8, 0.2);
    \draw[color = black] (-0.8, -1.2) -- (-0.8, -1.4);
    \draw[color = black] (0.8, -1.2) -- (0.8, -1.4);
    %% Text
    \node at (-0.0,0) {\small $\mathcal{V}$};
    \node at (-0.0,-1.2) {\small $\bar{\mathcal{V}}$};
    \end{tikzpicture}\ \ \  =\ \ \  \begin{tikzpicture} [scale = 0.8, baseline = {([yshift=-.5ex]current bounding box.center)}] 
        \draw[color = black] (-0.8, 0.4) -- (-0.8, -1.6);
        \draw[color = black] (0.8, 0.4) -- (0.8, -1.6);
    \end{tikzpicture}
\end{equation}
which will be invaluable in several proofs.

\subsubsection{Connection to Matrix Product States}

The result of using this graphical notation is summarized in Fig.~\ref{fig:reduction_to_MPS}(c), which manifestly reveals the connection to matrix product states.
We make this connection stronger by showing that the resulting state actually has a very particular and useful form:
\begin{shaded}
    \textbf{Theorem 1} (Resources for Gluable Quantum States) Suppose that $\ket{\Psi}$ is a translation-invariant right-gluable quantum state.
    Then the following are true: 
    \begin{enumerate}
        \item[(1)] $\ket{\Psi}$ has an exact matrix product state description

        \item[(2)] the unentangled clusters in the state preparation protocol $\ket{\psi}$ are its matrix product state tensors in canonical form (labeled $A$) up to a tensor product of unitaries acting on each qudit of the cluster.
        In other words, 
        \begin{equation} \label{eq-uLuPuRA}
            \begin{tikzpicture}[scale = 1, baseline = {([yshift=-.5ex]current bounding box.center)}] 
            \draw[color = black] (-1, 0) -- (1, 0);
            \draw[color = black] (0, 0) -- (0, -0.8);
            \draw[fill = lightdodgerblue] (0,0) circle (0.3);
            \node at (-0.0,0) {\small $\psi$};
            %\node at (-1.3, 0) {\small $u_L$ };
            %\node at (1.5, 0) {\small $u_R$ }; %\\
            \node at (0, -1.0) {\small $\ $};
            %\node at (-1.3, 0) {\small $V^{[n]}$};
            %\node at (0, -1.2) {\small $\ $};
     %       \node at (-1.4, 0) {\small $\{V_{\alpha} \}$ };
        \end{tikzpicture} = \begin{tikzpicture}[scale = 1, baseline = {([yshift=-.5ex]current bounding box.center)}] 
            \draw[color = black] (-1, 0) -- (1, 0);
            \draw[color = black] (0, 0) -- (0, -0.8);
            \draw[fill = lightdodgerblue] (0,0) circle (0.3);
            \node at (-0.03,0) {\small $A$};
            \node at (-1.3, 0) {\small $u_L$ };
            \node at (1.5, 0) {\small $u_R$ };
            \node at (0, -1.0) {\small $u_P$};
        \end{tikzpicture}
    \end{equation}
    where $A$ is the MPS in right-canonical form and $u_L, u_R,$ and $u_P$ are unitaries (which are determined by the feedback protocol and measurement basis).

    \item[(3)] The measurement basis $\mathcal{V}$ used for the preparation is maximally entangled.
    Equivalently, viewed as operators, the measurement basis is a unitary error basis.
    \end{enumerate}

\end{shaded}

Let us first remark on some technical points.
In the above theorem, by right-canonical form, we mean that the Kronecker delta is a dominant right-eigenvector of the transfer matrix with unit eigenvalue \cite{perezgarcia2006MPSrep, Cirac2021MPSReview, tenpy, Bridgeman_2017}. In tensor network notation, this becomes:
\begin{equation}
    \begin{tikzpicture} [scale = 1, baseline = {([yshift=-.5ex]current bounding box.center)}] 
    
    \draw[color = black] (-0.8, 0) -- (0.8, 0); %  top horizontal bar
    \draw[color = black] (-0.8, -1.2) -- (0.8, -1.2); %  bottom horizontal bar

    %\draw[color = black] (-1, 0) -- (-1.4, 0); %  top horizontal bar
    %\draw[color = black] (-1, -1.2) -- (-1.4, -1.2); %  bottom horizontal bar
    \draw[color = black] (0, 0) -- (0, -1.2); % vertical bar
     \draw[color = black] (0.8, 0) -- (0.8, -1.2);
    % \draw[color = black] (-1, 0) -- (-1, -1.2);
    \draw[fill = lightdodgerblue] (0,0) circle (0.3); % circle
    \draw[fill = lightdodgerblue] (0,-1.2) circle (0.3); % circle
    %% Text
    \node at (-0.0,0) {\small $A$};
    \node at (-0.0,-1.2) {\small $\bar{A}$};
    %
    % \node at (-1, -1.2) {\small $\bar{V}_{\alpha}$};
    % \node at (1.05, -1.2) {\small $\bar{W}_{\alpha} $ };
    % \node at (-1, 0) {\small ${V}_{\alpha}$};
    % \node at (1.05, 0) {\small $W_{\alpha} $ };
    \end{tikzpicture} \quad = \quad   \begin{tikzpicture} [scale = 1, baseline = {([yshift=-.5ex]current bounding box.center)}] 
        \draw[color = black] (0, 0) -- (0.6, 0);
        \draw[color = black] (0, -1.2) -- (0.6, -1.2);
        \draw[color = black] (0.6, 0) -- (0.6, -1.2);
    \end{tikzpicture}
\end{equation}
Such a form is highly convenient, as it allows to reduce many physical quantities of the state to local expressions.
In addition, this form has the property that the remaining `gauge' degree of freedom $A_i \to W A_i W^\dagger$ (where $W$ is unitary) can be used to also make the dominant left eigenvector diagonal, which has the physical interpretation of the Schmidt spectrum $\Lambda^2$ of a bipartition of the state.
While the full proof of the above theorem is provided in detail in Appendix~\ref{app:resourcetheorem}, it logically relies on first proving Theorem $2$, which we will state momentarily.

The above theorem is powerful in that it heavily constrains ``where to look'' if one wants to prepare a given quantum state with a measurement-only protocol.
Indeed, the target state essentially fully specifies the clusters that can be used for the preparation protocol, and the measurement basis required for preparation is constrained to be maximally entangled (these are also called fusion measurements \cite{bartolucci2021fusionbased}).
In other words, the ingredients we utilize tell us it can never help to attempt to create a state by gluing with, e.g., non-maximally-entangled measurement bases.
To highlight the non-triviality of this theorem, we point out that if we had dropped the left-conditioning requirement, there \emph{do} exist non-maximally-entangled bases for gluing certain quantum states, as shown in Sec.~\ref{subsec-examples-again}.
We thus see that the deceptively simple ingredient of left-conditioning (i.e., classical information only travels to the right) helps to build a remarkably structured framework, which we will use to uncover the physical properties (not) preparable via measurement.

The above theorem leads to a very useful corollary:
\begin{shaded}
    \textbf{Corollary 1} If $\ket{\Psi}$ is a right-gluable quantum state, then the initial clusters of its preparation protocol can always be chosen to be the matrix product state tensors in canonical form and one of the measurement outcomes $V_1 \in \mathcal{V}$ can be chosen to be the identity.
\end{shaded}
The last claim follows from Theorem 1 stating that $V_\alpha$ is unitary. That means we can always use $\tilde V_\alpha = V_\alpha V_1^\dagger$ as an alternative error basis, which will still be correctable and a unitary error basis, which now satisfies $\tilde V_1 = \mathds{1}$.

Another corollary of Theorem 1 is that all measurement outcomes are equally likely (see Appendix~\ref{app-equalprob} for a proof). This is not important for chains with open boundary conditions (such as the examples in Sec.~\ref{sec:examples_and_prelude}), since the state is deterministically preparable, i.e., we can correct any possible measurement outcome. However, it can be used to infer that the state is efficiently probabilistically prepaparable even with periodic boundary conditions. For instance, in Sec.~\ref{subsec-uniform} we explore additional conditions on the $V_\alpha$ operators which ensure that classical post-processing is very simple, and in such cases we can infer that the probability of successfully preparing the state on periodic boundary conditions is lower bounded by $\frac{1}{\chi^2}$ (in particular, it is independent of system size).

With these results in mind, we now work to find appropriate conditions on the MPS tensor and the measurement basis which enable correctability.

\subsection{Local Tensor Criteria for Gluable States}

We now discuss the properties of both the matrix product state tensor and the measurement basis that are required for the state to be gluable.
To do so, we start by recognizing that the corollary above implies that measurement outcomes $V_{\alpha} \neq V_1 \in \mathcal{V}$ can be viewed as virtual errors in the target matrix product state.
Correctability then reduces to asking the question as to when such virtual errors ``push through to unitaries''.
Recall that by definition of gluable states we know that each wrong measurement outcome can be corrected by some physical unitary operator to the right of the location of the error:
\begin{equation}
\begin{tikzpicture}[scale = 0.65, baseline = {([yshift=-.5ex]current bounding box.center)}] 
    \node at (-1.25, 0) {\small $V_{\alpha}$};
    \draw[color = black] (-1.0,0) -- (4, 0);
    \node at (4.4, 0) {\small $\cdots$};
    \foreach \i in {0, ..., 2}{
        \draw[color = black] (1.5*\i,0) -- (1.5*\i, -1);
        \draw[fill = lightdodgerblue] (1.5*\i,0) circle (0.3);
        \node at (1.5*\i, 0) {\small $A$};
        \node at (1.5*\i, -1.6) {$\ $};
    } 
\end{tikzpicture} = \begin{tikzpicture}[scale = 0.65, baseline = {([yshift=-.5ex]current bounding box.center)}] 
    %\node at (-1.25, 0) {\small $V_{\alpha}$};
    \draw[color = black] (-1.0,0) -- (4, 0);
    \node at (4.4, 0) {\small $\cdots$};
    \foreach \i in {0, ..., 2}{
        \draw[color = black] (1.5*\i,0) -- (1.5*\i, -1);
        \draw[fill = lightdodgerblue] (1.5*\i,0) circle (0.3);
        \node at (1.5*\i, 0) {\small $A$};
        \node at (1.5*\i, -1.3) {\small $U^{[\i]}_{\alpha}$};
    } 
\end{tikzpicture}
\end{equation}
This suggests that we should be able to push the error through each matrix product state tensor \emph{individually}.
In order to make this notion precise, let us first introduce the general concept of pushable operators of a matrix product state:

\begin{shaded} 
    \textbf{Definition} (Pushable Operators) Let $A$ be a translationally-invariant matrix product state tensor.
    We say that an operator $V^{[0]}$ is right-pushable with respect to $A$ if there exists a sequence of operators $V^{[n \in \mathbb{N}]}, U^{[n \in \mathbb{N}]}$ such that:
    \begin{equation} \label{eq-pushable}
        \begin{tikzpicture}[scale = 1, baseline = {([yshift=-.5ex]current bounding box.center)}] 
        \draw[color = black] (-1, 0) -- (1, 0);
        \draw[color = black] (0, 0) -- (0, -0.8);
        \draw[fill =  lightdodgerblue] (0,0) circle (0.3);
        \node at (-0.03,0) {\small $A$};
        \node at (-1.3, 0) {\small $V^{[n]}$};
        \node at (0, -1.2) {\small $\ $};
 %       \node at (-1.4, 0) {\small $\{V_{\alpha} \}$ };
    \end{tikzpicture} = \begin{tikzpicture}[scale = 1, baseline = {([yshift=-.5ex]current bounding box.center)}] 
        \draw[color = black] (-1, 0) -- (1, 0);
        \draw[color = black] (0, 0) -- (0, -0.8);
        \draw[fill = lightdodgerblue] (0,0) circle (0.3);
        \node at (-0.03,0) {\small $A$};
        %\node at (-1.4, 0) {\small $\{V_{\alpha} \}$ };
        \node at (1.5, 0) {\small $V^{[n + 1]}$ }; \\
        \node at (0.1, -1.0) {\small $U^{[n]}$};
    \end{tikzpicture}
    \end{equation}
    where $U^{[n]}$ is unitary.
    A similar definition can be applied for left-pushable operators.
\end{shaded}
\noindent
The notion of pushable operators provides the bedrock of a formalism for understanding which matrix product states can be created deterministically from measurements and provides a condition on the matrix product state tensor required for gluability. Indeed, by formalizing the above intuition, we prove in the Appendix~\ref{app:pushable} that the errors in gluable states are pushable operators. More generally:
\begin{shaded}
    \textbf{Theorem 2} (Tensor Characterization for Right-Gluable States) 
    A translation-invariant state $\ket{\Psi}$ is right-gluable if and only if its matrix product state representation $A$ admits an error basis of right-pushable operators.
    
    This condition is equivalent to the existence of $\chi^2$ trace-orthonormal operators $\{V_\alpha\}$ such that
\begin{equation} \label{eq-VTTV}
        \begin{tikzpicture} [scale = 1, baseline = {([yshift=-.5ex]current bounding box.center)}] 
    
    \draw[color = black] (-0.6, 0) -- (0.6, 0); %  top horizontal bar
    \draw[color = black] (-0.6, -1.2) -- (0.6, -1.2); %  bottom horizontal bar
    \draw[color = black] (0, 0) -- (0, -1.2); % vertical bar
     %\draw[color = black] (-0.8, 0) -- (-0.8, -1.2);
    % \draw[color = black] (-1, 0) -- (-1, -1.2);
    \draw[fill = lightdodgerblue] (0,0) circle (0.3); % circle
    \draw[fill = lightdodgerblue] (0,-1.2) circle (0.3); % circle
    %% Text
    \node at (-0.0,0) {\small $A$};
    \node at (-0.0,-1.2) {\small $\bar{A}$};
    %\draw[fill = lightishgray] (-1,0) circle (0.4);
    \node at (-1, 0) {\small $V_{\alpha}^{[n]}$};
    %\draw[fill = lightishgray] (-1,-1.2) circle (0.4);
    \node at (-1, -1.2) {\small $\bar{V}_{\alpha}^{[n]}$};
    \end{tikzpicture}  \; \; = \; \;
        \begin{tikzpicture} [scale = 1, baseline = {([yshift=-.5ex]current bounding box.center)}] 
    
    \draw[color = black] (-0.6, 0) -- (0.6, 0); %  top horizontal bar
    \draw[color = black] (-0.6, -1.2) -- (0.6, -1.2); 
    \draw[color = black] (0, 0) -- (0, -1.2); % vertical bar
     %\draw[color = black] (-0.8, 0) -- (-0.8, -1.2);
    % \draw[color = black] (-1, 0) -- (-1, -1.2);
    \draw[fill = lightdodgerblue] (0,0) circle (0.3); % circle
    \draw[fill = lightdodgerblue] (0,-1.2) circle (0.3); % circle
    %\draw[fill = lightishgray] (1.05,0) circle (0.5);
    \node at (1.05, 0) {\small $V^{[n+1]}_{\alpha}$};
    %\draw[fill = lightishgray] (1.05,-1.2) circle (0.5);
    \node at (1.05, -1.2) {\small $\bar{V}^{[n+1]}_{\alpha}$};
    %% Text
    \node at (-0.0,0) {\small $A$};
    \node at (-0.0,-1.2) {\small $\bar{A}$};
    \end{tikzpicture} 
\end{equation}
where $\chi$ is the bond dimension of the MPS tensor, $n$ is any non-negative integer, and $V_\alpha^{[0]} = V_\alpha$.
\end{shaded}

This local condition on the MPS tensor is crucial to proving Theorem 1. To give a taste of this, let us note that Eq.~\eqref{eq-VTTV} gives us a local condition on the transfer matrix. This is rather powerful, since the transfer matrix contains physical properties of interest; in particular its spectrum encodes correlation lengths and its eigenvectors encode entanglement properties.
If we combine this local criterion with Eq.~\eqref{eq-niceidentity} (trace-orthonormality of the basis), we obtain
\begin{equation}
        \begin{tikzpicture} [scale = 1, baseline = {([yshift=-.5ex]current bounding box.center)}] 
    
    \draw[color = black] (-0.8, 0) -- (0.8, 0); %  top horizontal bar
    \draw[color = black] (-0.8, -1.2) -- (0.8, -1.2); %  bottom horizontal bar
    \draw[color = black] (-0.8, 0) -- (-0.8, -1.2);
    \draw[color = black] (-1, 0) -- (-1, -1.2);
    \draw[color = black] (-1, 0) -- (-1.4, 0);
    \draw[color = black] (-1, -1.2) -- (-1.4, -1.2);
    %\draw[color = black] (-1, 0) -- (-1.4, 0); %  top horizontal bar
    %\draw[color = black] (-1, -1.2) -- (-1.4, -1.2); %  bottom horizontal bar
    \draw[color = black] (0, 0) -- (0, -1.2); % vertical bar
     %\draw[color = black] (-0.8, 0) -- (-0.8, -1.2);
    % \draw[color = black] (-1, 0) -- (-1, -1.2);
    \draw[fill = lightdodgerblue] (0,0) circle (0.3); % circle
    \draw[fill = lightdodgerblue] (0,-1.2) circle (0.3); % circle
    %% Text
    \node at (-0.0,0) {\small $A$};
    \node at (-0.0,-1.2) {\small $\bar{A}$};
    %
    % \node at (-1, -1.2) {\small $\bar{V}_{\alpha}$};
    % \node at (1.05, -1.2) {\small $\bar{W}_{\alpha} $ };
    % \node at (-1, 0) {\small ${V}_{\alpha}$};
    % \node at (1.05, 0) {\small $W_{\alpha} $ };
    \end{tikzpicture} = \chi \sum_{\alpha} 
        \begin{tikzpicture} [scale = 1, baseline = {([yshift=-.5ex]current bounding box.center)}] 
    
    \draw[color = black] (-0.8, 0) -- (1.4, 0); %  top horizontal bar
    \draw[color = black] (-0.8, -1.2) -- (1.4, -1.2); %  bottom horizontal bar

    %\draw[color = black] (-1, 0) -- (-1.4, 0); %  top horizontal bar
    %\draw[color = black] (-1, -1.2) -- (-1.4, -1.2); %  bottom horizontal bar
    \draw[color = black] (0, 0) -- (0, -1.2); % vertical bar
     %\draw[color = black] (-0.8, 0) -- (-0.8, -1.2);
    % \draw[color = black] (-1, 0) -- (-1, -1.2);
    \draw[fill = lightdodgerblue] (0,0) circle (0.3); % circle
    \draw[fill = lightdodgerblue] (0,-1.2) circle (0.3); % circle
    \draw[fill = lightishgray] (0.9,0) circle (0.3);
    \node at (0.9, 0) {\small $V^{[1]}_{\alpha}$};
    \draw[fill = lightishgray] (0.9,-1.2) circle (0.3);
    \node at (0.9, -1.2) {\small $\bar{V}^{[1]}_{\alpha}$};
    %% Text
    \node at (-0.0,0) {\small $A$};
    \node at (-0.0,-1.2) {\small $\bar{A}$};
    %
    % \node at (-1, -1.2) {\small $\bar{V}_{\alpha}$};
    % \node at (1.05, -1.2) {\small $\bar{W}_{\alpha} $ };
    % \node at (-1, 0) {\small ${V}_{\alpha}$};
    % \node at (1.05, 0) {\small $W_{\alpha} $ };
    \end{tikzpicture} 
\end{equation}
Here we see the Kronecker delta appear. By further manipulating this local condition, one can in fact prove that the Kronecker delta is a dominant right eigenvector of the MPS transfer matrix, thereby proving we are in canonical form (property (2) in Theorem 1). While the details are in the Appendix, we refer the interested reader to Sec.~\ref{sec:simplygluablegeneral} where we use the same ingredients in a somewhat simpler setting.

Finally, we note that having proven that we are in canonical form also allows us to deduce the last property in Theorem 1, namely that the error basis is unitary. This is because an MPS in canonical form has a more natural relationship between `virtual' and `physical' legs, such that $V_\alpha$ pushing through to a physical unitary operator can be used to prove that, indeed, $V_\alpha$ itself must be unitary; we again refer to the Appendix for a proof.

\subsection{Classification of Gluable Quantum States}

Beyond its utility in proving the resource theorem, the tensor characterization of Theorem 2 can even be leveraged to prove a classification of all right-gluable quantum states.
To do so, it will be convenient to first note that by performing a change of basis on each physical site, any matrix product tensor can be brought into what we will call the \textbf{definite form}. More precisely, there always exists a physical isometry\footnote{\label{footnoteboi}In practice one can think of this as a unitary operator. Indeed, if $d \leq \chi^2$, we can embed the physical site into a $\chi^2$-dimensional qudit, and the isometry can be extended into a unitary. If $d > \chi^2$, we can always truncate the physical dimension since the MPS can only give nonzero weight to a $\chi^2$-dimensional subspace.} $W$ such that
\begin{equation} \label{eq-definiteform}
    \begin{tikzpicture}[scale = 1, baseline = {([yshift=-.5ex]current bounding box.center)}] 
    \draw[color = black] (-1, 0) -- (1, 0);
    \draw[color = black] (0, 0) -- (0, -0.8);
    \draw[color = black] (-0.05, -0.8) -- (-0.05, -1.4);
    \draw[color = black] (0.05, -0.8) -- (0.05, -1.4);
    \draw[fill = lightdodgerblue] (0,0) circle (0.3);
    \node at (-0.03,0) {\small $A$};
    \draw[fill = lightorange(ryb)] (0, -0.8) circle (0.25);
    \node at (0, -0.8) {\small $W$};
    %\node at (-1.4, 0) {\small $\{V_{\alpha} \}$ };
   % \node at (1.3, 0) {\small ${V}_{\alpha}$ }; \\
\end{tikzpicture} \quad = \quad \begin{tikzpicture}[scale = 1, baseline = {([yshift=-.5ex]current bounding box.center)}] 
    \draw[color = black] (-1, 0) -- (-0.1, 0);
    \draw[color = black] (1, 0) -- (0.1, 0);
    \draw[color = black] (0.1, 0) -- (0.1, -1);
    \draw[color = black] (-0.1, 0) -- (-0.1, -1);
    %\draw[color = black] (0, -0.4) -- (0, -1.0);
    \draw[fill = lightdodgerblue] (0,-0.5) circle (0.3) node{$\A$};
\end{tikzpicture}   
\end{equation}
where $\A$ is a \emph{positive semi-definite (PSD) operator in the vertical direction} (i.e., as a map between virtual and physical legs). In the special case $\A = \mathds{1}_{\chi^2}$, this just gives a product state of Bell pairs, and Eq.~\eqref{eq-definiteform} makes manifest that any MPS can be thought of as gluing together Bell pairs with imaginary time-evolution or projectors (depending on whether $\A$ has strictly positive or zero eigenvalues). Indeed, since $\A$ is PSD, we can write it as $\A = \sum_{i=1}^{{\rm rank}(\A)} e^{E_i} \ket{v_i}\bra{v_i}$ where $\{ \ket{v_i} \}$ is an orthogonal set and $E_i \in \mathbb R$. We stress that being in definite form is independent from (and compatible with) being in canonical form, since the former involves a physical change of basis and the latter a virtual action. When both properties hold, we can say the MPS is in definite canonical form.

There are two ways of seeing that the isometry in Eq.~\eqref{eq-definiteform} exists. Let us first give a conceptual high-level argument, which uses the transfer matrix. Although this is often regarded as a matrix in the horizontal direction, here we will want to use it as a matrix in the vertical direction, i.e., from the virtual bonds of the `ket' layer to the virtual bonds of the `bra' layer. To avoid confusion we will denote it as a $\T$ when interpreted in this vertical direction, reserving $T$ for the horizontal case. Then we can write $\T = A^\dagger A$, where we similarly interpret $A$ as a matrix from virtual to physical bonds. This manifestly shows that $\T$ is a PSD operator in this vertical direction. Hence, we can define its square root $\A\equiv \sqrt{\T}$, which is also PSD. If we interpret this as a matrix product state tensor, we observe it generates the same transfer matrix: $\A^\dagger \A = \A^2 = \T = A^\dagger A$.  This proves Eq.~\eqref{eq-definiteform}, since it is known that the transfer matrix fully characterizes an MPS tensor, up to a physical isometry \cite{Cirac2021MPSReview}.

An alternative constructive proof is to simply take a polar decomposition of $A$, again regarded as a matrix from virtual legs to the physical leg:
\begin{equation}\label{eq-polardecomp}
    \begin{tikzpicture}[scale = 1, baseline = {([yshift=-.5ex]current bounding box.center)}] 
    \draw[color = black] (-1, 0) -- (1, 0);
    \draw[color = black] (0, 0) -- (0, -0.8);
    \draw[fill = lightdodgerblue] (0,0) circle (0.3);
    \node at (-0.03,0) {\small $A$};
    %\node at (-1.4, 0) {\small $\{V_{\alpha} \}$ };
   % \node at (1.3, 0) {\small ${V}_{\alpha}$ }; \\
\end{tikzpicture}\ \ =\ \     \begin{tikzpicture}[scale = 1, baseline = {([yshift=-.5ex]current bounding box.center)}] 
    \draw[color = black] (-1, 0) -- (1, 0);
    \draw[color = black] (0.0, 0) -- (0.0, -0.9);
    %\draw[color = black] (0.05, 0) -- (0.05, -0.9);
    \draw[fill = lightorange(ryb)] (0,0) circle (0.25);
    \node at (0,0) {\scriptsize $W^\dagger$};
    \draw[fill = white] (0,-0.6) circle (0.2);
    \node at (0,-0.6) {\small $Q$}; 
    %\node at (-1.4, 0) {\small $\{V_{\alpha} \}$ };
   % \node at (1.3, 0) {\small ${V}_{\alpha}$ }; \\
\end{tikzpicture} 
\end{equation}
where $W$ is an isometry and $Q$ is PSD. We note that such a polar decomposition was a key ingredient in Ref.~\onlinecite{Schuch2011} for classifying one-dimensional phases of matter.
Here we use it to remark that simply applying $W$ to both sides of Eq.~\eqref{eq-polardecomp} gives us Eq.~\eqref{eq-definiteform} with $\A = W Q W^\dagger$.

We can now state and prove the classification of gluable states:
\begin{shaded}
    \textbf{Theorem 3} (Classification of Right-Gluable States) Let $A$ be a matrix product state tensor associated with quantum state $\ket{\Psi}$.
    Then, $A$ admits an error basis of right-pushable operators $\mathcal{V}^{[0]}$ that pushes through to virtual $\mathcal{V}^{[n] \in \mathbb{N}}$ (equivalently, by virtue of Theorem 2, $\ket{\Psi}$ is right-gluable) if and only if for all $n$,
    \begin{equation} \label{eq-commutant}
        A \in \mathcal{C}(\{ V_{\alpha}^{[n]} \otimes (V_{\alpha}^{[n + 1]})^{\dagger}\})
    \end{equation}
    for an appropriate choice of on-site change of basis implemented by an isometry\footnotemark, where we view $A$ as a matrix from virtual to physical bonds and where $\mathcal{C}(S)$ is the commutant algebra of the set $S$ (i.e. the set of $\chi^2 \times \chi^2$ matrices that commute with all elements of $S$).
\end{shaded}
\footnotetext[\value{footnote}]{Explicitly, this holds if one takes $A$ to definite form (see Eq.~\eqref{eq-definiteform}).}
%See footnote \ref{footnoteboi}.}

\textit{Proof.} By virtue of the preceding discussion, it is sufficient to prove this for $A$ being in definite form, i.e.., $A = \A$ in Eq.~\eqref{eq-definiteform}. As discussed, we can then equate $\A = \sqrt{\T}$, where $\T$ is the transfer matrix interpreted as an operator going in the vertical direction. Since $\A$ is PSD, Eq.~\eqref{eq-commutant} is equivalent\footnote{Indeed, $M \A^k = \A^k M$ is equivalent to $M$ preserving the eigenspaces of $\A^k$. Crucially, since $\A$ is PSD, the eigenspaces of $\A^k$ are independent of $k \neq 0$.} to $\A^2 = \T \in \mathcal{C}(\{ V_{\alpha}^{[n]} \otimes (V_{\alpha}^{[n + 1]})^{\dagger}\})$. This is in turn equivalent to Eq.~\eqref{eq-VTTV} in Theorem 2 if $V_\alpha^{[m]}$ is unitary for any $m$. This indeed guaranteed by Theorem 1 for $m=0$, but the unitarity proof in fact applies for any $m$ (see Appendix~\ref{app:resourcetheorem}).

\hspace{0.44 \textwidth} $\blacksquare$

The above theorem completely characterizes the space of all right-gluable quantum states.
Indeed, for a given error basis and a sequence of operators $\mathcal{V}^{[n \in \mathbb{N}]}$ which one demands the basis pushes through to, the space of translationally-invariant gluable quantum states can be explicitly computed from the commutant. 
As as an example, for $\mathcal{V}^{[n]} \equiv \mathcal{V} = \{\mathds{1}, X, Z, ZX\}$ for all $n$, the commutant consists of all linear combinations of $V_{\alpha} \otimes V_{\alpha}^{\dagger}$, with $V_{\alpha} \in \mathcal{V}$.
Hence, for this error basis, the space of all gluable matrix product states are given by (up to a physical isometry): 
\begin{equation} \label{eq-pauliclass}
    A= \mu_1 \mathds{1} \otimes \mathds{1} + \mu_x X \otimes X + \mu_y Y\otimes Y + \mu_z Z \otimes Z,
\end{equation}
where $A$ is viewed as a matrix from the virtual legs to the physical legs and $\mu_{\alpha} \in \mathbb C$ can be arbitrarily chosen (although they can always be made real and non-negative by an isometry). For a more general unitary error basis, the commutant can be explicitly calculated as a finite linear algebra problem by sequentially going through $\alpha = 1,2,\cdots,\chi^2$ and identifying the invariant subspaces by simply diagonalizing $V_\alpha^{[n]}$.

\subsection{Types of Measurement Errors: Local and Topological}

Both the local tensor criteria and classification result above heavily relied on the the notion of pushable operators.
This  motivates us to phenomenologically characterize gluable quantum states by the manner by which measurement errors push through the matrix product state.
Indeed, inspired by the examples in Sec.~\ref{sec:examples_and_prelude}, we distinguish two types of pushable operators.

\begin{shaded}
    \textbf{Definition} (Topological and Local Errors) Let $V$ be a right-pushable operator with respect to $A$ associated with a sequence $V^{[n \in \mathbb{N}]}$ (where $V = V^{[0]}$).
    We can define two types of these operators: 
    \begin{itemize}
        \item We say that $V$ is a \textbf{local error}, and equivalently can be locally corrected, if the sequence of operators $(V^{[n]}, U^{[n]})$ \textit{can be chosen} to trivialize at some $n$ (i.e. there exists $m$ such that $V^{[n > m]}, U^{[n > m]} = \mathds{1}$).
    
        \item Conversely, we say that $V$ is a \textbf{topological error} if the sequence \textit{can be chosen} to never trivialize.
        In such a case, $V$ requires non-local correction, where one using string operators to pair up and annihilates topological errors (see below for examples).
    \end{itemize}
    Let us remark that local errors and topological errors are not mutually distinct.
    In other words, some errors could be both topological or local\footnotemark.
\end{shaded}
%\stepcounter[-1]{footnote}
\footnotetext[\value{footnote}]{We already saw an example of this in Sec.~\ref{sec:examples_and_prelude}: the GHZ state appeared as the $\beta=0$ limit in Sec.~\ref{subsec-deformedGHZ}, where the $Z$ error was topological, but it also appears as the $\beta = \infty$ limit in Sec.~\ref{subsec-deformedtrivial}, where $Z$ was a local error.}

Given these two types of errors, the remainder of the manuscript will be devoted to deriving physical properties of gluable quantum states from the properties of their measurement errors.
However, before this, we show our results in action concretely via the motivating examples introduced in Sec.~\ref{sec:examples_and_prelude}.

\subsection{Illustration of Results with Previous Examples \label{subsec-examples-again}}

\textbf{Deformed GHZ.} We start by illustrating our abstract results for the case of the deformed GHZ state which we encountered in Sec.~\ref{subsec-deformedGHZ}.
Note that the deformed GHZ state Eq.~\eqref{eq-ebetaXGHZ} is an exact bond dimension two tensor network, whose tensors in right canonical form are given by: 
\begin{equation}
    \begin{tikzpicture}[scale = 1, baseline = {([yshift=-.5ex]current bounding box.center)}] 
        \draw[color = black] (-0.8, 0) -- (0.8, 0);
        \draw[color = black] (0, 0) -- (0, -0.8);
        \draw[fill = lightdodgerblue] (0,0) circle (0.3);
        \node at (-0.0,0) {\small $A$};
        %\node at (-0.9, 0) {\small $i$};
        %\node at (0.9, 0) {\small $j$};
        %\node at (0.25, -0.6) {\small $k$};
        %\node at (1.2, -0.05) {\small $v_i$};
        % \node at (0.0, -1.2) {\small $\ $};
 %       \node at (-1.4, 0) {\small $\{V_{\alpha} \}$ };
    \end{tikzpicture} =  \begin{tikzpicture}[scale = 1, baseline = {([yshift=-.5ex]current bounding box.center)}] 
        \draw[color = black] (-0.8, 0) -- (0.8, 0);
        \draw[color = black] (0, 0) -- (0, -0.8);
        \draw[fill = white] (0, -0.45) circle (0.15); 
        %\draw[fill = lightdodgerblue] (0,0) circle (0.3);
        %\node at (-0.0,0) {\small $A$};
        %\node at (-0.9, 0) {\small $i$};
        %\node at (0.9, 0) {\small $j$};
        %\node at (0.25, -0.6) {\small $k$};
        %\node at (1.2, -0.05) {\small $v_i$};
        % \node at (0.0, -1.2) {\small $\ $};
 %       \node at (-1.4, 0) {\small $\{V_{\alpha} \}$ };
    \end{tikzpicture} \qquad \begin{tikzpicture}[scale = 1, baseline = {([yshift=-.5ex]current bounding box.center)}]
        \draw[fill = white] (0, -0.45) circle (0.15);
    \end{tikzpicture}\ =\  e^{\beta X}
\end{equation}
where the $T$-junction tensor above is the three-leg Kronecker delta tensor $\delta_{ijk}$.
Note that the tensor above, when viewed as a state via the reshaping of Eq.~\eqref{eq-minimality}, is precisely initial cluster in the preparation protocol for the deformed GHZ.
Explicitly: 
\begin{equation}
    \begin{tikzpicture}[scale = 1, baseline = {([yshift=-.5ex]current bounding box.center)}] 
        \draw[color = black] (-0.5, 0) -- (0.5, 0);
        \draw[color = black] (-0.5, 0) -- (-0.5, -0.8);
        \draw[color = black] (0.5, 0) -- (0.5, -0.8);
        \draw[color = black] (0, 0) -- (0, -0.8);
        \draw[fill = white] (0, -0.45) circle (0.15); 
        %\draw[fill = lightdodgerblue] (0,0) circle (0.3);
        %\node at (-0.0,0) {\small $A$};
        %\node at (-0.9, 0) {\small $i$};
        %\node at (0.9, 0) {\small $j$};
        %\node at (0.25, -0.6) {\small $k$};
        %\node at (1.2, -0.05) {\small $v_i$};
        % \node at (0.0, -1.2) {\small $\ $};
 %       \node at (-1.4, 0) {\small $\{V_{\alpha} \}$ };
    \end{tikzpicture}\ \ \propto\ \ \begin{tikzpicture}[scale = 1, baseline = {([yshift=-.5ex]current bounding box.center)}] 
        \draw[rounded corners, fill = lightdodgerblue] (0*1.3,0) rectangle (0*1.3 + 1, 0.4) {};
        \node at (0.5 + 0 * 1.3, 0.2) {\small $\text{GHZ}_3$};
        \draw[color = black] (0.1 + 1.3*0, 0) -- (0.1 + 1.3*0, -0.4);
        \draw[color = black] (0.5 + 1.3*0, 0) -- (0.5 + 1.3*0, -0.4);
        \draw[color = black] (0.9 + 1.3*0, 0) -- (0.9 + 1.3*0, -0.4);
        % \draw[color = black] (1.4, 0) -- (1.4, -0.5);
        % \draw[color = black] (1.8, 0) -- (1.8, -0.65);
        % \draw[color = black] (2.2, 0) -- (2.2, -0.25);
        \draw[fill = white] (0.5 + 1.3*0, -0.2) circle (0.1);
    \end{tikzpicture}
\end{equation}
In other words, we have that Eq.~\eqref{eq-uLuPuRA} holds with $u_L = u_R = u_P = \mathds{1}$.

Moreover, the measurements that we performed were in a maximally entangled basis---consistent with Theorem 1---and, in the operator language, they correspond to the Pauli matrices, which are unitary as predicted.
In particular, the Bell state projector $\bra{V}$ (with $V \in \{\mathds{1}, X, Z, ZX\}$) is given by: 
\begin{equation}
    \begin{tikzpicture}[scale = 1.5, baseline = {([yshift=-.5ex]current bounding box.center)}]
        \draw[rounded corners, fill = lightishgray] (0,-0.25) rectangle (0.7, -0.55) {};
        \draw[color = black] (0.1, -0.25) -- (0.1, 0);
        \draw[color = black] (0.6, -0.25) -- (0.6, 0);
        \node at (0.35, -0.4) {\small $\bra{V}$};
    \end{tikzpicture}\quad \longrightarrow \quad \frac{1}{\sqrt{2}} \begin{tikzpicture}[scale = 1, baseline = {([yshift=-.5ex]current bounding box.center)}] 
        \draw[color = black] (-1, 0) -- (1, 0);
        %\draw[color = black] (0, 0) -- (0, -0.8);
        \draw[fill = lightishgray] (0,0) circle (0.3);
        \node at (0,0) {\small $V_{\alpha}$};
        %\node at (-1.3, 0) {\small $V^{[n]}$};
        %\node at (0, -1.2) {\small $\ $};
 %       \node at (-1.4, 0) {\small $\{V_{\alpha} \}$ };
    \end{tikzpicture}
\end{equation}
where for the particular case of getting the identity, we have that: 
\begin{equation}
    \begin{tikzpicture}[scale = 1, baseline = {([yshift=-.5ex]current bounding box.center)}] 
        \draw[color = black] (-1, 0) -- (1, 0);
        %\draw[color = black] (0, 0) -- (0, -0.8);
        \draw[fill = lightishgray] (0,0) circle (0.3);
        \node at (0,0) {\small $\mathds{1}$};
        %\node at (-1.3, 0) {\small $V^{[n]}$};
        %\node at (0, -1.2) {\small $\ $};
 %       \node at (-1.4, 0) {\small $\{V_{\alpha} \}$ };
    \end{tikzpicture}\ =\ \begin{tikzpicture}[scale = 1, baseline = {([yshift=-.5ex]current bounding box.center)}] 
        \draw[color = black] (-1, 0) -- (1, 0);
    \end{tikzpicture}
\end{equation}
where the straight line is the two-leg Kronecker delta $\delta_{ij}$.
Notice that then the perfect outcome of Fig.~\ref{fig:prepGHZ} becomes: 
\begin{equation}
\begin{tikzpicture}[scale = 0.8, baseline = {([yshift=-.5ex]current bounding box.center)}] 
    \draw[color = black] (-1.5,0) -- (7.5 -2, 0);
    \foreach \i in {0, ..., 2}{
        \draw[color = black] (2*\i,0) -- (2*\i, -1);
        \draw[fill = lightdodgerblue] (2*\i,0) circle (0.3);
        \node at (2*\i, 0) {\small $A$};
    }
    \draw[color = black] (-1.5, 0) -- (-1.5, -1);
    \draw[color = black] (5.5, 0) -- (5.5, -1);
\end{tikzpicture}
\end{equation}
which is precisely the matrix product state for the deformed GHZ.
Moreover, the above measurement basis coincides with the basis of right-pushable operators for the matrix product state tensor.
In fact, each of these are topological errors:
    \begin{align}
        \begin{tikzpicture}[scale = 1, baseline = {([yshift=-.5ex]current bounding box.center)}] 
        \draw[color = black] (-1, 0) -- (1, 0);
        \draw[color = black] (0, 0) -- (0, -0.8);
        \draw[fill =  lightdodgerblue] (0,0) circle (0.3);
        \node at (-0.03,0) {\small $A$};
        \node at (-1.2, 0) {\small $X$};
        \node at (0, -1.2) {\small $\ $};
 %       \node at (-1.4, 0) {\small $\{V_{\alpha} \}$ };
    \end{tikzpicture} &= \begin{tikzpicture}[scale = 1, baseline = {([yshift=-.5ex]current bounding box.center)}] 
        \draw[color = black] (-1, 0) -- (1, 0);
        \draw[color = black] (0, 0) -- (0, -0.8);
        \draw[fill = lightdodgerblue] (0,0) circle (0.3);
        \node at (-0.03,0) {\small $A$};
        %\node at (-1.4, 0) {\small $\{V_{\alpha} \}$ };
        \node at (1.2, 0) {\small $X$ };
        \node at (0, -1.0) {\small $X$};
    \end{tikzpicture} \\ 
    \begin{tikzpicture}[scale = 1, baseline = {([yshift=-.5ex]current bounding box.center)}] 
        \draw[color = black] (-1, 0) -- (1, 0);
        \draw[color = black] (0, 0) -- (0, -0.8);
        \draw[fill =  lightdodgerblue] (0,0) circle (0.3);
        \node at (-0.03,0) {\small $A$};
        \node at (-1.2, 0) {\small $Z$};
        \node at (0, -1.2) {\small $\ $};
 %       \node at (-1.4, 0) {\small $\{V_{\alpha} \}$ };
    \end{tikzpicture} &= \begin{tikzpicture}[scale = 1, baseline = {([yshift=-.5ex]current bounding box.center)}] 
        \draw[color = black] (-1, 0) -- (1, 0);
        \draw[color = black] (0, 0) -- (0, -0.8);
        \draw[fill = lightdodgerblue] (0,0) circle (0.3);
        \node at (-0.03,0) {\small $A$};
        %\node at (-1.4, 0) {\small $\{V_{\alpha} \}$ };
        \node at (1.2, 0) {\small $Z$ };
        \node at (0, -1.0) {\small $\ $};
    \end{tikzpicture}
\end{align}
We interpret this as an instantiation of Theorem 2.

Lastly, we can also understand this example in the language of the classification in Theorem~3.
Specifically, one can show that if we bring $A$ to definite form \eqref{eq-definiteform}, then
\begin{equation} \label{eq-deformedGHZdefinite}
\A = \exp\left(\beta X \otimes X\right) \left( \frac{1 + Z \otimes Z}{2} \right).
\end{equation}
As this commutes with $V_{\alpha} \otimes V_{\alpha}^{\dagger}$ for all $V_{\alpha} \in \{\mathds{1}, X, Z, ZX\}$, this is definitionally in the commutant algebra $\mathcal{C}(\{V_{\alpha} \otimes V_{\alpha}^{\dagger}\})$, in line with Eq.~\eqref{eq-commutant} in the classification.
In fact, it is a special case of Eq.~\eqref{eq-pauliclass} with 
\begin{equation}
\mu_1 = \mu_z = \frac{\cosh \beta}{2}, \quad \mu_x = - \mu_y = \frac{\sinh \beta}{2}.
\end{equation}
While Eq.~\eqref{eq-deformedGHZdefinite} can be derived via a polar decomposition of $A$, an alternative is to first compute the transfer matrix of $A$ and write it as a matrix in the virtual direction: $\T = \frac{1+Z \otimes Z}{2} \; e^{2 \beta X\otimes X} \; \frac{1+Z \otimes Z}{2}$. Then $\A = \sqrt{\T}$.

%Specifically, let us remark that, using the polar decomposition of Eq.~\eqref{eq-polardecomp}, we have that $Q' = e^{\beta X}$ and $W_{i, jk} = \delta_{ijk}$ (i.e. the kronecker delta tensor).
%
%Hence, in definite form (defined in Eq.~\eqref{eq-simplifiedisometric}), the tensor is given by: $\mathbb{A} = W^{\dagger}Q'W = $.
%

\textbf{Deformed Cluster State.} The deformed cluster state (Sec.~\ref{subsec-deformedcluster}) is similarly a bond-dimension two matrix product state whose tensors in canonical form are given by: 
\begin{equation} \label{eq-deformedclusterMPS}
    \begin{tikzpicture}[scale = 1, baseline = {([yshift=-.5ex]current bounding box.center)}] 
        \draw[color = black] (-0.8, 0) -- (0.8, 0);
        \draw[color = black] (0, 0) -- (0, -0.8);
        \draw[fill = lightdodgerblue] (0,0) circle (0.3);
        \node at (-0.0,0) {\small $A$};
        %\node at (-0.9, 0) {\small $i$};
        %\node at (0.9, 0) {\small $j$};
        %\node at (0.25, -0.6) {\small $k$};
        %\node at (1.2, -0.05) {\small $v_i$};
        % \node at (0.0, -1.2) {\small $\ $};
 %       \node at (-1.4, 0) {\small $\{V_{\alpha} \}$ };
    \end{tikzpicture} =  \begin{tikzpicture}[scale = 1, baseline = {([yshift=-.5ex]current bounding box.center)}] 
        \draw[color = black] (-0.8, 0) -- (0.8, 0);
        \draw[color = black] (0, 0) -- (0, -0.8);
        \draw[fill = white] (0, -0.45) circle (0.15);
        \draw[fill = lightorange(ryb)] (0.4, 0) circle (0.15); 
        %\draw[fill = lightdodgerblue] (0,0) circle (0.3);
        %\node at (-0.0,0) {\small $A$};
        %\node at (-0.9, 0) {\small $i$};
        %\node at (0.9, 0) {\small $j$};
        %\node at (0.25, -0.6) {\small $k$};
        %\node at (1.2, -0.05) {\small $v_i$};
        % \node at (0.0, -1.2) {\small $\ $};
 %       \node at (-1.4, 0) {\small $\{V_{\alpha} \}$ };
    \end{tikzpicture}
\end{equation}
where the orange circle is once again the Hadamard and the white circle $e^{\beta X}$.
Note that in this case, we have that:
\begin{equation}
    \begin{tikzpicture}[scale = 1, baseline = {([yshift=-.5ex]current bounding box.center)}] 
        \draw[rounded corners, fill = lightdodgerblue] (0*1.3,0) rectangle (0*1.3 + 1, 0.4) {};
        \node at (0.5 + 0 * 1.3, 0.2) {\small $\psi$};
        \draw[color = black] (0.1 + 1.3*0, 0) -- (0.1 + 1.3*0, -0.4);
        \draw[color = black] (0.5 + 1.3*0, 0) -- (0.5 + 1.3*0, -0.4);
        \draw[color = black] (0.9 + 1.3*0, 0) -- (0.9 + 1.3*0, -0.4);
        % \draw[color = black] (1.4, 0) -- (1.4, -0.5);
        % \draw[color = black] (1.8, 0) -- (1.8, -0.65);
        % \draw[color = black] (2.2, 0) -- (2.2, -0.25);
        %\draw[fill = white] (0.5 + 1.3*0, -0.2) circle (0.1);
    \end{tikzpicture}\ \  =\  \ \begin{tikzpicture}[scale = 1, baseline = {([yshift=-.5ex]current bounding box.center)}] 
        \draw[rounded corners, fill = lightdodgerblue] (0*1.3,0) rectangle (0*1.3 + 1, 0.4) {};
        \node at (0.5 + 0 * 1.3, 0.2) {\small $\text{GHZ}_3$};
        \draw[color = black] (0.1 + 1.3*0, 0) -- (0.1 + 1.3*0, -0.4);
        \draw[color = black] (0.5 + 1.3*0, 0) -- (0.5 + 1.3*0, -0.4);
        \draw[color = black] (0.9 + 1.3*0, 0) -- (0.9 + 1.3*0, -0.4);
        % \draw[color = black] (1.4, 0) -- (1.4, -0.5);
        % \draw[color = black] (1.8, 0) -- (1.8, -0.65);
        % \draw[color = black] (2.2, 0) -- (2.2, -0.25);
        \draw[fill = white] (0.5 + 1.3*0, -0.2) circle (0.1);
    \end{tikzpicture}\ \ \propto\ \  \begin{tikzpicture}[scale = 1, baseline = {([yshift=-.5ex]current bounding box.center)}] 
        \draw[rounded corners, fill = lightdodgerblue] (0*1.3,0) rectangle (0*1.3 + 1, 0.4) {};
        \node at (0.5 + 0 * 1.3, 0.2) {\small $A$};
        \draw[color = black] (0.1 + 1.3*0, 0) -- (0.1 + 1.3*0, -0.4);
        \draw[color = black] (0.5 + 1.3*0, 0) -- (0.5 + 1.3*0, -0.4);
        \draw[color = black] (0.9 + 1.3*0, 0) -- (0.9 + 1.3*0, -0.4);
        % \draw[color = black] (1.4, 0) -- (1.4, -0.5);
        % \draw[color = black] (1.8, 0) -- (1.8, -0.65);
        % \draw[color = black] (2.2, 0) -- (2.2, -0.25);
        \draw[fill = lightorange(ryb)] (0.9 + 1.3*0, -0.2) circle (0.1);
    \end{tikzpicture}
\end{equation}
which means that $u_R = H$ in Eq.~\eqref{eq-uLuPuRA}, with $u_L, u_P = \mathds{1}$.
Similarly, here the measurement basis is given by $\{H, HX, HZ, HZX\}$, which is maximally entangled when viewed as quantum states and unitary when viewed as operators, thereby again providing an example of Theorem 1.
Consequently, one can check that the resulting tensor network will precisely be the deformed cluster state when all measurement outcomes yield $H$.

Note that, in this case, we glued using a measurement error basis which did \emph{not} contain the identity operator.
Relatedly, the initial $\ket{\psi}$ state is \emph{not} simply the MPS tensor of the target state (see Eq.~\eqref{eq-deformedclusterMPS}).
Moreover, the measurement basis is \textit{not} the basis of pushable operators. 
Instead, the latter basis is given by $\mathcal{V}^{[0]} = \{\mathds{1}, X, Z, ZX\}$, which consists of topological errors satisfying:
    \begin{align} \label{eq-clusterpush}
        \begin{tikzpicture}[scale = 1, baseline = {([yshift=-.5ex]current bounding box.center)}] 
        \draw[color = black] (-1, 0) -- (1, 0);
        \draw[color = black] (0, 0) -- (0, -0.8);
        \draw[fill =  lightdodgerblue] (0,0) circle (0.3);
        \node at (-0.03,0) {\small $A$};
        \node at (-1.2, 0) {\small $X$};
        \node at (0, -1.2) {\small $\ $};
 %       \node at (-1.4, 0) {\small $\{V_{\alpha} \}$ };
    \end{tikzpicture} &= \begin{tikzpicture}[scale = 1, baseline = {([yshift=-.5ex]current bounding box.center)}] 
        \draw[color = black] (-1, 0) -- (1, 0);
        \draw[color = black] (0, 0) -- (0, -0.8);
        \draw[fill = lightdodgerblue] (0,0) circle (0.3);
        \node at (-0.03,0) {\small $A$};
        %\node at (-1.4, 0) {\small $\{V_{\alpha} \}$ };
        \node at (1.2, 0) {\small $Z$ };
        \node at (0, -1.0) {\small $X$};
    \end{tikzpicture} \\ 
    \begin{tikzpicture}[scale = 1, baseline = {([yshift=-.5ex]current bounding box.center)}] 
        \draw[color = black] (-1, 0) -- (1, 0);
        \draw[color = black] (0, 0) -- (0, -0.8);
        \draw[fill =  lightdodgerblue] (0,0) circle (0.3);
        \node at (-0.03,0) {\small $A$};
        \node at (-1.2, 0) {\small $Z$};
        \node at (0, -1.2) {\small $\ $};
 %       \node at (-1.4, 0) {\small $\{V_{\alpha} \}$ };
    \end{tikzpicture} &= \begin{tikzpicture}[scale = 1, baseline = {([yshift=-.5ex]current bounding box.center)}] 
        \draw[color = black] (-1, 0) -- (1, 0);
        \draw[color = black] (0, 0) -- (0, -0.8);
        \draw[fill = lightdodgerblue] (0,0) circle (0.3);
        \node at (-0.03,0) {\small $A$};
        %\node at (-1.4, 0) {\small $\{V_{\alpha} \}$ };
        \node at (1.2, 0) {\small $X$ };
        \node at (0, -1.0) {\small $\ $};
    \end{tikzpicture}
\end{align}
implying that the ordered sets $\mathcal{V}^{[n \in 2\mathbb{N} + 1]} = \{\mathds{1}, Z, X, XZ\}$ and $\mathcal{V}^{[n \in 2\mathbb{N}]} = \mathcal{V}^{[0]}$.
The above naturally leads to a crucial observation that we can prepare the same state by using the MPS tensor as a starting ingredient and measuring in the Bell basis (by simply moving the Hadamard matrix from the error basis into $\ket{\psi}$).
This illustrates our general result stated in Corollary 1.
The fact that we can always restrict to such a simpler scenario without loss of generality is a consequence of Theorem 1 ensuring that the error basis of a right-gluable state is always unitary. This additional structure is key to proving the general results in the next sections.

We conclude our discussion of this example by understanding the gluability of the above in the language of the classification of Theorem 3.
If we bring $A$ to definite form, we obtain the same as Eq.~\eqref{eq-deformedGHZdefinite} up to conjugation by a single Hadamard, i.e., $\mathbb{A} = \exp\left(\beta X \otimes Z\right) \left(\frac{1 + Z \otimes X}{2}\right)$.
Note that $\mathbb{A}$ commutes with every element of the set $\{V^{[n]}_{\alpha} \otimes (V^{[n + 1]}_{\alpha})^{\dagger} \} = \{\mathds{1} \otimes \mathds{1}, X \otimes Z, Z \otimes X, ZX \otimes ZX \}$ for all $n$.
As such, definitionally, $\mathbb{A} \in \mathcal{C}(\{ V_{\alpha}^{[n]} \otimes (V_{\alpha}^{[n + 1]})^{\dagger}\})$.

\textbf{Deformed Trivial.} Finally, the deformed trivial state we considered (Sec.~\ref{subsec-deformedtrivial}) also has an exact tensor network description, with tensors in right-canonical form given by: 
\begin{equation} \label{eq-Adeformedtriv}
    \begin{tikzpicture}[scale = 1, baseline = {([yshift=-.5ex]current bounding box.center)}] 
        \draw[color = black] (-0.8, 0) -- (0.8, 0);
        \draw[color = black] (0, 0) -- (0, -0.8);
        \draw[fill = lightdodgerblue] (0,0) circle (0.3);
        \node at (-0.0,0) {\small $A$};
        %\node at (-0.9, 0) {\small $i$};
        %\node at (0.9, 0) {\small $j$};
        %\node at (0.25, -0.6) {\small $k$};
        %\node at (1.2, -0.05) {\small $v_i$};
        % \node at (0.0, -1.2) {\small $\ $};
 %       \node at (-1.4, 0) {\small $\{V_{\alpha} \}$ };
    \end{tikzpicture} =  \begin{tikzpicture}[scale = 1, baseline = {([yshift=-.5ex]current bounding box.center)}] 
        \draw[color = black] (-0.8, 0) -- (0.8, 0);
        \draw[color = black] (0, 0) -- (0, -0.8);
        %\draw[fill = white] (0, -0.45) circle (0.15); 
        \draw[fill = white] (0.4, 0) circle (0.15); 
        %\draw[fill = lightdodgerblue] (0,0) circle (0.3);
        %\node at (-0.0,0) {\small $A$};
        %\node at (-0.9, 0) {\small $i$};
        %\node at (0.9, 0) {\small $j$};
        %\node at (0.25, -0.6) {\small $k$};
        %\node at (1.2, -0.05) {\small $v_i$};
        % \node at (0.0, -1.2) {\small $\ $};
 %       \node at (-1.4, 0) {\small $\{V_{\alpha} \}$ };
    \end{tikzpicture}
\end{equation}
where the white circle is $e^{\alpha X}$ with $\tanh \alpha = e^{-2\beta}$.

In this case, the clusters used for preparation: 
\begin{equation} \label{eq-trivial_tensor_clusters}
    \begin{tikzpicture}[scale = 1, baseline = {([yshift=-.5ex]current bounding box.center)}] 
        \draw[color = black] (-0.5, 0) -- (0.5, 0);
        \draw[color = black] (-0.5, 0) -- (-0.5, -0.8);
        \draw[color = black] (0.5, 0) -- (0.5, -0.8);
        \draw[color = black] (0, 0) -- (0, -0.8);
        \draw[fill = white] (0.5, -0.45) circle (0.15); 
        %\draw[fill = lightdodgerblue] (0,0) circle (0.3);
        %\node at (-0.0,0) {\small $A$};
        %\node at (-0.9, 0) {\small $i$};
        %\node at (0.9, 0) {\small $j$};
        %\node at (0.25, -0.6) {\small $k$};
        %\node at (1.2, -0.05) {\small $v_i$};
        % \node at (0.0, -1.2) {\small $\ $};
 %       \node at (-1.4, 0) {\small $\{V_{\alpha} \}$ };
    \end{tikzpicture}\ \ \propto\ \ \begin{tikzpicture}[scale = 1, baseline = {([yshift=-.5ex]current bounding box.center)}] 
        \draw[rounded corners, fill = lightdodgerblue] (0*1.3,0) rectangle (0*1.3 + 1, 0.4) {};
        \node at (0.5 + 0 * 1.3, 0.2) {\small $\text{GHZ}_3$};
        \draw[color = black] (0.1 + 1.3*0, 0) -- (0.1 + 1.3*0, -0.4);
        \draw[color = black] (0.5 + 1.3*0, 0) -- (0.5 + 1.3*0, -0.4);
        \draw[color = black] (0.9 + 1.3*0, 0) -- (0.9 + 1.3*0, -0.4);
        % \draw[color = black] (1.4, 0) -- (1.4, -0.5);
        % \draw[color = black] (1.8, 0) -- (1.8, -0.65);
        % \draw[color = black] (2.2, 0) -- (2.2, -0.25);
        \draw[fill = white] (0.9 + 1.3*0, -0.2) circle (0.1);
    \end{tikzpicture}
\end{equation}
are precisely the matrix product state tensors with $u_L, u_R, u_P = \mathds{1}$ in the parlance of Eq.~\eqref{eq-uLuPuRA}.
Similar to the GHZ example, the measurement basis is the Bell basis, which now makes manifest why a perfect measurement $\ket{\mathds{1}}$ produces the target state with no need for correction.
Note that $X$ errors and $Z$ errors are both pushable, but now only $X$ is topological, with $Z$ being local.
In particular: 
    \begin{align} 
        \begin{tikzpicture}[scale = 1, baseline = {([yshift=-.5ex]current bounding box.center)}] 
        \draw[color = black] (-1, 0) -- (1, 0);
        \draw[color = black] (0, 0) -- (0, -0.8);
        \draw[fill =  lightdodgerblue] (0,0) circle (0.3);
        \node at (-0.03,0) {\small $A$};
        \node at (-1.2, 0) {\small $X$};
        \node at (0, -1.2) {\small $\ $};
 %       \node at (-1.4, 0) {\small $\{V_{\alpha} \}$ };
    \end{tikzpicture} &= \begin{tikzpicture}[scale = 1, baseline = {([yshift=-.5ex]current bounding box.center)}] 
        \draw[color = black] (-1, 0) -- (1, 0);
        \draw[color = black] (0, 0) -- (0, -0.8);
        \draw[fill = lightdodgerblue] (0,0) circle (0.3);
        \node at (-0.03,0) {\small $A$};
        %\node at (-1.4, 0) {\small $\{V_{\alpha} \}$ };
        \node at (1.2, 0) {\small $X$ };
        \node at (0, -1.0) {\small $X$};
    \end{tikzpicture} \label{eq:XrightTriv} \\ 
    \begin{tikzpicture}[scale = 1, baseline = {([yshift=-.5ex]current bounding box.center)}] 
        \draw[color = black] (-1, 0) -- (1, 0);
        \draw[color = black] (0, 0) -- (0, -0.8);
        \draw[fill =  lightdodgerblue] (0,0) circle (0.3);
        \node at (-0.03,0) {\small $A$};
        \node at (-1.2, 0) {\small $Z$};
        \node at (0, -1.2) {\small $\ $};
 %       \node at (-1.4, 0) {\small $\{V_{\alpha} \}$ };
    \end{tikzpicture} &= \begin{tikzpicture}[scale = 1, baseline = {([yshift=-.5ex]current bounding box.center)}] 
        \draw[color = black] (-1, 0) -- (1, 0);
        \draw[color = black] (0, 0) -- (0, -0.8);
        \draw[fill = lightdodgerblue] (0,0) circle (0.3);
        \node at (-0.03,0) {\small $A$};
        %\node at (-1.4, 0) {\small $\{V_{\alpha} \}$ };
        \node at (1.2, 0) {\small $\ $ };
        \node at (0, -1.0) {\small $Z$};
    \end{tikzpicture} \label{eq:ZrightTriv}
\end{align}
Hence, we would write that the error basis of pushable operators is $\mathcal{V}^{[0]} = \{\mathds{1}, X, Z, ZX\}$ and they push through to $\mathcal{V}^{[n \geq 1 \in \mathbb{N}]} = \{\mathds{1}, X, \mathds{1}, X\}$.

We conclude by once again describing how the above example fits into our classification.
To compute the definite form of $A$, we first write the transfer matrix as a matrix in the vertical direction:
\begin{equation}
    \T = e^{\alpha \mathds 1 \otimes X} \frac{1+ Z \otimes Z}{2} e^{\alpha \mathds 1 \otimes X} = \frac{ e^{2\alpha \mathds 1 \otimes X} + Z \otimes Z}{2}.
\end{equation}
One can straightforwardly show that $\T^2 \propto \T$, and hence $\A = \sqrt{\T} \propto \T$.
We thus see that $\A$ commutes with $\{V_{\alpha}^{[0]} \otimes V_{\alpha}^{[1]}\} = \{\mathds{1} \otimes \mathds{1}, X \otimes X, Z \otimes \mathds{1}, ZX \otimes X\}$ and $\{V_{\alpha}^{[n \geq 1]} \otimes V_{\alpha}^{[n + 1 \geq 2]}\} = \{\mathds{1} \otimes \mathds{1}, X \otimes X, \mathds{1} \otimes \mathds{1}, X \otimes X\}$.
Hence, we have that $\mathbb{A} \in \mathcal{C}(\{V_{\alpha}^{[n]} \otimes V_{\alpha}^{[n + 1]}\})$ for all $n$, consistent with Eq.~\eqref{eq-commutant} of Theorem 3.

\textbf{Deformed Trivial as a Non-Example.} Instead of Eq.~\eqref{eq-Adeformedtriv} we could consider the MPS tensor
\begin{equation}
    \begin{tikzpicture}[scale = 1, baseline = {([yshift=-.5ex]current bounding box.center)}] 
        \draw[color = black] (-0.8, 0) -- (0.8, 0);
        \draw[color = black] (0, 0) -- (0, -0.8);
        \draw[fill = lightdodgerblue] (0,0) circle (0.3);
        \node at (-0.0,0) {\small $A$};
        %\node at (-0.9, 0) {\small $i$};
        %\node at (0.9, 0) {\small $j$};
        %\node at (0.25, -0.6) {\small $k$};
        %\node at (1.2, -0.05) {\small $v_i$};
        % \node at (0.0, -1.2) {\small $\ $};
 %       \node at (-1.4, 0) {\small $\{V_{\alpha} \}$ };
    \end{tikzpicture} =  \begin{tikzpicture}[scale = 1, baseline = {([yshift=-.5ex]current bounding box.center)}] 
        \draw[color = black] (-0.8, 0) -- (0.8, 0);
        \draw[color = black] (0, 0) -- (0, -0.8);
        %\draw[fill = white] (0, -0.45) circle (0.15); 
        \draw[fill = white] (0.4, 0) circle (0.15);
        \draw[fill = white] (-0.4, 0) circle (0.15);
        %\draw[fill = lightdodgerblue] (0,0) circle (0.3);
        %\node at (-0.0,0) {\small $A$};
        %\node at (-0.9, 0) {\small $i$};
        %\node at (0.9, 0) {\small $j$};
        %\node at (0.25, -0.6) {\small $k$};
        %\node at (1.2, -0.05) {\small $v_i$};
        % \node at (0.0, -1.2) {\small $\ $};
 %       \node at (-1.4, 0) {\small $\{V_{\alpha} \}$ };
    \end{tikzpicture}
\end{equation}
where the white circle is now $e^{\alpha X/2}$ with $\tanh \alpha = e^{-2\beta}$. This clearly generates the same deformed trivial state. However, this MPS is \emph{not} in canonical form. Hence, according to our resource theorem (Theorem 1), we cannot use this as a wavefunction for our decoupled clusters. However, if we give up on our requirement that feedback is only left-conditioned, we can also glue these clusters. Remarkably, the error basis used for gluing will now be \emph{non-unitary}: $\{\mathds{1}, X, Ze^{\alpha X}, e^{\alpha X} Y \}$. This basis is now no longer maximally entangled, but we can still correct the errors: e.g., while $X$ pushes through to the right as in Eq.~\eqref{eq:XrightTriv}, but $Ze^{\alpha X}$ can only be corrected by (locally) pushing to the \emph{left}. This example highlights that a considerable amount of structure can be lost as soon as classical information needs to propagate \emph{both left and right}. It remains to be seen whether a similarly complete framework can be worked out for such cases; however, it is equally unclear whether such bi-directional classical propagation provides a distinct advantage over simply using left-conditioning.

\section{Topological Errors: Constraints on Entanglement and Order} \label{sec:Topological}

Having set up the basic formalism for understanding the structure of gluable quantum states in the previous section, here we consider constraints on the physical properties of such states when all measurement errors are topological or, equivalently, are non-locally corrected.
More precisely, we call a state \textbf{topologically gluable} if the unitary error basis $\mathcal V = \{ V_\alpha\}$ used for gluing remains an error basis\footnote{We note that this condition is automatically implied if $\{V_\alpha^{[0]}\}$ are all topological errors and, e.g., the transfer matrix $T$ has full rank (see Appendix~\ref{app-nogo} for a proof) or the state is uniformly gluable (Sec.~\ref{subsec-uniform}).} upon pushing through to $\mathcal V^{[n]} = \{ V^{[n]}_\alpha \}$ in Eq.~\eqref{eq-pushable}; this is in fact automatically a unitary error basis similar to the proof of Theorem 1. This is in contrast to states where we use local feedback, where necessarily $V^{[n]}_\alpha$ must trivialize for certain $\alpha$.

The section is organized by first proving a basic but powerful constraint on the entanglement structure for gluable matrix product states of this form.
After this, we explore the connection between topologically gluable matrix product states and symmetry, enabling us to provide a convenient parameterization of all such states in a particular setting.
We conclude by discussing the relationship between short-range entangled topologically gluable quantum states and symmetry protected topological phases with both uniform and modulated symmetries.

\subsection{Generalities of Topologically Gluable States} \label{sec:simplygluablegeneral}

Let us start by providing the most basic constraint on topologically gluable matrix product states.
This arises from the ability of pushing each error operator to a far-away region without losing its information (in particular, in finite chains this means the error gets pushed to the edge, as in the examples in Sec.~\ref{subsec-deformedGHZ} and Sec.~\ref{subsec-deformedcluster}). This is reminiscent of measurement-based quantum computation \cite{Raussendorf2001-xm,Wei_2021, stephen2017MBQCSPT}, where a flat (i.e., fully degenerate) entanglement spectrum is necessary to be able to `teleport' operators in this way. We show that this intuition indeed generalizes to our set-up:
\begin{shaded}
    \textbf{Theorem 4} (Exclusively Topological Errors Constrain Entanglement) Any topologically gluable matrix product state has a flat bipartite entanglement spectrum, i.e. the Schmidt spectrum $\Lambda^2$ is $\chi$-fold degenerate where $\chi$ is the bond dimension, i.e., $\Lambda^2 = \left( \frac{1}{\chi}, \frac{1}{\chi}, \cdots , \frac{1}{\chi} \right)$.
\end{shaded}

\textit{Proof.} To prove the claim, we first recall that Theorem 1 guarantees we are in canonical form. More precisely, the Kronecker delta (i.e., the Choi state for the identity operator) is a dominant right eigenvector. Since its corresponding left eigenvector is the Schmidt spectrum $\Lambda^2$, it now suffices to prove that its left eigenvector is also the Kronecker delta.
To do so, let us denote the transfer matrix of the matrix product state as: 
\begin{equation}
        \begin{tikzpicture} [scale = 1, baseline = {([yshift=-.5ex]current bounding box.center)}] 
    
    \draw[color = black] (-0.8, 0) -- (0.8, 0); %  top horizontal bar
    \draw[color = black] (-0.8, -1.2) -- (0.8, -1.2); %  bottom horizontal bar

    %\draw[color = black] (-1, 0) -- (-1.4, 0); %  top horizontal bar
    %\draw[color = black] (-1, -1.2) -- (-1.4, -1.2); %  bottom horizontal bar
    \draw[color = black] (0, 0) -- (0, -1.2); % vertical bar
     %\draw[color = black] (-0.8, 0) -- (-0.8, -1.2);
    % \draw[color = black] (-1, 0) -- (-1, -1.2);
    \draw[fill = lightdodgerblue] (0,0) circle (0.3); % circle
    \draw[fill = lightdodgerblue] (0,-1.2) circle (0.3); % circle
    %% Text
    \node at (-0.0,0) {\small $A$};
    \node at (-0.0,-1.2) {\small $\bar{A}$};
    %
    % \node at (-1, -1.2) {\small $\bar{V}_{\alpha}$};
    % \node at (1.05, -1.2) {\small $\bar{W}_{\alpha} $ };
    % \node at (-1, 0) {\small ${V}_{\alpha}$};
    % \node at (1.05, 0) {\small $W_{\alpha} $ };
    \end{tikzpicture}  =  \begin{tikzpicture} [scale = 1, baseline = {([yshift=-.5ex]current bounding box.center)}] 
        \draw[color = black] (-1.0, 0) -- (1.0, 0);
        \draw[color = black] (-1.0, -1.2) -- (1.0, -1.2);
        \draw[rounded corners, fill = lightdodgerblue] (-0.4, -1.4) rectangle (0.4, 0.2) {};
        \node at (0, -0.6) { $T$};
       % \draw[color = black] (0.9, 0) -- (0.9, -1.2);
        % \draw[fill = white] (0.9,0) circle (0.3);
        % \node at (0.9, 0) {\small $V^{[n]}$};
        % \draw[fill = white] (0.9,-1.2) circle (0.3);
        % \node at (0.9, -1.2) {\small $\bar{V}^{[n]}$};
     \end{tikzpicture}
\end{equation}
and note that for topologically gluable states, we have that:
\begin{equation}
\begin{tikzpicture} [scale = 0.95, baseline = {([yshift=-.5ex]current bounding box.center)}] 
        \draw[color = black] (-1.4, 0) -- (1.0, 0);
        \draw[color = black] (-1.4, -1.2) -- (1.0, -1.2);
        \draw[rounded corners, fill = lightdodgerblue] (-0.4, -1.4) rectangle (0.4, 0.2) {};
        \node at (0, -0.6) { $T^n$};
        %\draw[color = black] (-0.9, 0) -- (-0.9, -1.2);
        \draw[fill = lightishgray] (-0.9,0) circle (0.3);
        \node at (-0.9, 0) {\small $V^{[0]}_{\alpha}$};
        \draw[fill = lightishgray] (-0.9,-1.2) circle (0.3);
        \node at (-0.9, -1.2) {\small $\bar{V}^{[0]}_{\alpha}$};
     \end{tikzpicture} = 
    \begin{tikzpicture} [scale = 0.95, baseline = {([yshift=-.5ex]current bounding box.center)}] 
        \draw[color = black] (-1.0, 0) -- (1.4, 0);
        \draw[color = black] (-1.0, -1.2) -- (1.4, -1.2);
        \draw[rounded corners, fill = lightdodgerblue] (-0.4, -1.4) rectangle (0.4, 0.2) {};
        \node at (0, -0.6) { $T^n$};
       % \draw[color = black] (0.9, 0) -- (0.9, -1.2);
        \draw[fill = lightishgray] (0.9,0) circle (0.3);
        \node at (0.9, 0) {\small $V^{[n]}_{\alpha}$};
        \draw[fill = lightishgray] (0.9,-1.2) circle (0.3);
        \node at (0.9, -1.2) {\small $\bar{V}^{[n]}_{\alpha}$};
     \end{tikzpicture}
\end{equation}
for all $\alpha$, with $\mathcal{V}^{[n]} = \{V^{[n]}_{\alpha}\}$ forming an error basis for all $n$.
At this point, we can sum over $\alpha$ and use the diagrammatic identity introduced in Eq.~\eqref{eq-niceidentity} to show that, for all $n$: 
\begin{equation}
     \begin{tikzpicture} [scale = 1, baseline = {([yshift=-.5ex]current bounding box.center)}] 
    \draw[color = black] (-1.0, 0) -- (1.0, 0);
    \draw[color = black] (-1.0, 0) -- (-1.0, -1.2);
    \draw[color = black] (-1.2, 0) -- (-1.2, -1.2);
    \draw[color = black] (-1.2, 0) -- (-1.6, 0);
    \draw[color = black] (-1.2, -1.2) -- (-1.6, -1.2);
    \draw[color = black] (-1.0, -1.2) -- (1.0, -1.2);
    \draw[rounded corners, fill = lightdodgerblue] (-0.4, -1.4) rectangle (0.4, 0.2) {};
    \node at (0, -0.6) { $T^n$};
 \end{tikzpicture} =      \begin{tikzpicture} [scale = 1, baseline = {([yshift=-.5ex]current bounding box.center)}] 
    \draw[color = black] (-1.0, 0) -- (1.0, 0);
    \draw[color = black] (1.0, 0) -- (1.0, -1.2);
    \draw[color = black] (1.2, 0) -- (1.2, -1.2);
    \draw[color = black] (1.2, 0) -- (1.6, 0);
    \draw[color = black] (1.2, -1.2) -- (1.6, -1.2);
    \draw[color = black] (-1.0, -1.2) -- (1.0, -1.2);
    \draw[rounded corners, fill = lightdodgerblue] (-0.4, -1.4) rectangle (0.4, 0.2) {};
    \node at (0, -0.6) { $T^n$};
 \end{tikzpicture}
\end{equation}
Now, we can formally take the limit as $n \to \infty$ and using the fact that we are in canonical form, we have that:
\begin{equation}
    \sum_{\beta} \begin{tikzpicture}[scale = 1, baseline = {([yshift=-.5ex]current bounding box.center)}] 
        %\draw[color = black] (-1.0, 0) -- (-1.0, -1.2);
    \draw[color = black] (-1.2, 0) -- (-1.2, -1.2);
    \draw[color = black] (-1.2, 0) -- (-1.6, 0);
    \draw[color = black] (-1.2, -1.2) -- (-1.6, -1.2);
    % Box
    \draw[color = black] (-1.0, 0) -- (-1.0, -1.2);
    \draw[color = black] (0.0, 0) -- (0.0, -1.2);
    \draw[color = black] (-1.0, 0) -- (0.0, 0.0);
    \draw[color = black] (-1.0, -1.2) -- (0.0, -1.2);
    \draw[color = black] (0.6, 0) -- (0.6, -1.2);
    \draw[color = black] (0.6, 0) -- (1.0, 0);
    \draw[color = black] (0.6, -1.2) -- (1.0, -1.2);
    \draw[fill = white] (0.6, -0.6) circle (0.25);
    \node at (0.6, -0.6) {\small $L_{\beta}$};
    \draw[fill = white] (0.0, -0.6) circle (0.25);
    \node at (0.0, -0.6) {\small $R_{\beta}$};
    \end{tikzpicture} = \sum_{\gamma}    \begin{tikzpicture}[scale = 1, baseline = {([yshift=-.5ex]current bounding box.center)}] 
        %\draw[color = black] (-1.0, 0) -- (-1.0, -1.2);
    \draw[color = black] (-1.2 - 0.4, 0) -- (-1.2 - 0.4, -1.2);
    \draw[color = black] (-1.2 - 0.4, 0) -- (-1.6 - 0.4, 0);
    \draw[color = black] (-1.2 - 0.4, -1.2) -- (-1.6 - 0.4, -1.2);
    \draw[fill = white] (-1.6, -0.6) circle (0.25);
    \node at (-1.6, -0.6) {\small $R_{\gamma}$};
    % Box
    \draw[color = black] (-1.0, 0) -- (-1.0, -1.2);
    \draw[color = black] (0.0, 0) -- (0.0, -1.2);
    \draw[color = black] (-1.0, 0) -- (0.0, 0.0);
    \draw[color = black] (-1.0, -1.2) -- (0.0, -1.2);
    \draw[fill = white] (-1.0, -0.6) circle (0.25);
    \node at (-1.0, -0.6) {\small $L_{\gamma}$};
    \draw[color = black] (0.4 -0.2, 0) -- (0.4 -0.2, -1.2);
    \draw[color = black] (0.4 -0.2, 0) -- (0.8 -0.2, 0);
    \draw[color = black] (0.4 -0.2, -1.2) -- (0.8 -0.2, -1.2);
    % \draw[fill = white] (0.4, -0.6) circle (0.25);
    % \node at (0.4, -0.6) {\small $L_0$};
    \end{tikzpicture}
\end{equation}
where the sums over $\beta$ and $\gamma$ are over the dominant eigenspace of the transfer matrix (that is $1$-dimension for short-range entangled states) but higher dimensional for long-range entangled states.
Now, dotting on the left of both diagrams with the identity, yields: 
\begin{equation}
    \sum_{\beta} \chi\, \text{tr}(R_{\beta}) \begin{tikzpicture}[scale = 1, baseline = {([yshift=-.5ex]current bounding box.center)}] 
        \draw[color = black] (0.6, 0) -- (0.6, -1.2);
        \draw[color = black] (0.6, 0) -- (1.0, 0);
        \draw[color = black] (0.6, -1.2) -- (1.0, -1.2);
        \draw[fill = white] (0.6, -0.6) circle (0.25);
        \node at (0.6, -0.6) {\small $L_{\beta}$};
    \end{tikzpicture} = \left(\sum_{\gamma} \text{tr}(R_{\gamma}) \text{tr}(L_{\gamma}) \right)
    \begin{tikzpicture}[scale = 1, baseline = {([yshift=-.5ex]current bounding box.center)}] 
        \draw[color = black] (0.4 -0.2, 0) -- (0.4 -0.2, -1.2);
    \draw[color = black] (0.4 -0.2, 0) -- (0.8 -0.2, 0);
    \draw[color = black] (0.4 -0.2, -1.2) -- (0.8 -0.2, -1.2);
    \end{tikzpicture}
\end{equation}

Since the transfer matrix is gaurenteed to have at least one positive right eigenvector (since we are in right canonical form, that eigenvector is just the identity), the left hand side of the above equation is non-zero and hence the term in parenthesis on the right hand side is not zero.
As a direct consequence, the identity is in the span of the left eigenspace. 
Finally, since this has nonzero overlap with the Kronecker delta, it is the left eigenvector associated to the aforementioned dominant right eigenvector, proving that the entanglement spectrum of the transfer matrix is flat.

\hspace{0.44 \textwidth} $\blacksquare$

As a remark, we use a similar style of reasoning to prove canonical form in the first place (claimed in Theorem 1) as shown in Appendix~\ref{app:resourcetheorem}.

The flatness of the entanglement spectrum already suggests possible connections between topologically gluable quantum states and order.
Indeed, constraints on degeneracy of the entanglement spectrum regularly appear for non-trivial one-dimensional quantum states, ranging from long-range entangled states to symmetry-protected topological phases.
In what follows, we make this connection more precise by first formalizing the connection between topological errors and symmetry, which will then reveal how topological errors can constrain the order of gluable quantum states.
This reflects the two examples we saw in Sec.~\ref{subsec-deformedGHZ}~and~\ref{subsec-deformedcluster}, which consider deformed long-range entanglement and SPT order respectively.

\subsection{Topological Errors and Symmetry \label{subsec-uniform}}

In discussing the relationship between topological errors and symmetry, it is convenient to distinguish between two types of topologically gluable matrix product states.
In particular, we will say that a matrix product state is \textbf{uniform topologically gluable} if the elements of its error basis $\mathcal{V}$ satisfy a uniform push-through rule: 
\begin{equation} \label{eq-simplyglue}
    \begin{tikzpicture}[scale = 1, baseline = {([yshift=-.5ex]current bounding box.center)}] 
        \draw[color = black] (-1, 0) -- (1, 0);
        \draw[color = black] (0, 0) -- (0, -0.8);
        \draw[fill =  lightdodgerblue] (0,0) circle (0.3);
        \node at (-0.03,0) {\small $A$};
        \node at (-1.2, 0) {\small $V_{\alpha}$};
        \node at (0, -1.2) {\small $\ $};
 %       \node at (-1.4, 0) {\small $\{V_{\alpha} \}$ };
    \end{tikzpicture} = \begin{tikzpicture}[scale = 1, baseline = {([yshift=-.5ex]current bounding box.center)}] 
        \draw[color = black] (-1, 0) -- (1, 0);
        \draw[color = black] (0, 0) -- (0, -0.8);
        \draw[fill = lightdodgerblue] (0,0) circle (0.3);
        \node at (-0.03,0) {\small $A$};
        %\node at (-1.4, 0) {\small $\{V_{\alpha} \}$ };
        \node at (1.2, 0) {\small $V_{\alpha}$ };
        \node at (0.05, -1.0) {\small $U_{\alpha}$};
    \end{tikzpicture}
\end{equation}
for all $V_{\alpha} \in \mathcal{V}$, i.e. $V^{[n \in \mathbb{N}]}_{\alpha} = V_{\alpha}$ in Eq.~\eqref{eq-pushable} (in principle this only needs to hold up to a phase factor, which we omit for presentation purposes).
An example of such uniform states include our deformed GHZ state (Sec.~\ref{subsec-deformedGHZ}), but also the deformed SPT example (Sec.~\ref{subsec-deformedcluster}) if we block into a unit cell of two sites.
Conversely, in the case where a topologically gluable matrix product state is not uniform, we will call it \textbf{modulated}, an example of which is the deformed SPT without blocking (Sec.~\ref{subsec-deformedcluster}). As we will discuss in Sec.~\ref{subsec-modulated}, such cases are related to the notion of modulated symmetries, which are just starting to be explored in many-body quantum systems \cite{khemani2020localization, sala2020ergodicity, sala2024modulated, gromov2019modulated, han2024dipole, lam2024classification, watanabe_2023_SSB, watanabe2023ground, gui2023effective}.

Uniform topologically gluable states are associated with uniform symmetries, where the symmetry action does not vary from site to site of the chain. This is clear from rewriting Eq.~\eqref{eq-simplyglue} as:
\begin{equation} \label{eq-symsimpglue}
    \begin{tikzpicture}[scale = 1, baseline = {([yshift=-.5ex]current bounding box.center)}] 
        \draw[color = black] (-1, 0) -- (1, 0);
        \draw[color = black] (0, 0) -- (0, -0.8);
        \draw[fill =  lightdodgerblue] (0,0) circle (0.3);
        \node at (-0.03,0) {\small $A$};
        \node at (-1.2, 0) {\small $V_{\alpha}$};
        \node at (1.2, 0) {\small $V^{\dagger}_{\alpha}$ };
        \node at (0, -1.2) {\small $\ $};
 %       \node at (-1.4, 0) {\small $\{V_{\alpha} \}$ };
    \end{tikzpicture} = \begin{tikzpicture}[scale = 1, baseline = {([yshift=-.5ex]current bounding box.center)}] 
        \draw[color = black] (-1, 0) -- (1, 0);
        \draw[color = black] (0, 0) -- (0, -0.8);
        \draw[fill = lightdodgerblue] (0,0) circle (0.3);
        \node at (-0.03,0) {\small $A$};
        %\node at (-1.4, 0) {\small $\{V_{\alpha} \}$ };
        %\node at (1.2, 0) {\small $V_{\alpha}$ };
        \node at (0.05, -1.0) {\small $U_{\alpha}$};
    \end{tikzpicture}
\end{equation}
This is the usual symmetry condition for a matrix product state \cite{Cirac2021MPSReview}, indeed, it implies $\prod_n (U_\alpha)_n$ leaves the state invariant.
Hence, in the case where $U_{\alpha}$ is non-trivial operator, it gives a physical uniform symmetry of the matrix product state.
Conversely, when $U_{\alpha} = \mathds{1}$, we have that $V_{\alpha}$ forms a virtual symmetry of the matrix product state, whose presence indicates that the state is long-range entangled\footnote{This follows from the fact that for each eigenvector of the transfer matrix, $V_\alpha$ can be used to construct a second eigenvector with the same eigenvalue. We thus have a degenerate largest eigenvalue, which together with our minimality condition implies a physical diverging correlation length.}.
This is exemplified by the $Z$ errors in the deformed GHZ state.

The relationship between pushable measurement errors and symmetry naturally inspires taking an algebraic lens on measurement errors.
In particular, one can consider the case where the measurement errors commute up to a phase, which we will call \textbf{abelian errors} (indeed they give rise to abelian symmetries).
Cases with abelian errors have the useful property that correcting a measurement error does not affect other errors, since pushing one error along the matrix product will not transform the errors it passes along the way.
This property places rather strong constraints on the gluable quantum state and indeed, it is possible to provide a convenient explicit parameterization of matrix product states in these circumstances:

\begin{shaded}
    \textbf{Theorem 5} \textit{(Parameterization of All Uniform Topologically Gluable Matrix Product States with Abelian Errors)} Suppose that $A$ is a uniform topologically gluable matrix product state tensor with an error basis $\mathcal{V} = \{V_\alpha\}$, whose elements all commute up to a phase. Without loss of generality, we can choose one of them to contain the identity\footnotemark.
    Then, up to an isometry acting at the physical level, $A$ is a is a matrix product state tensor with a physical dimension of $\chi^2$ and virtual dimension of $\chi$ given by:
    \begin{equation} \label{eq-Aclassification}
        \begin{tikzpicture} [scale = 1, baseline = {([yshift=-.5ex]current bounding box.center)}] 
        \draw[color = black] (-0.8, 0) -- (0.8, 0);
        \draw[color = black] (0, 0) -- (0, -0.8);
        \draw[fill = lightdodgerblue] (0,0) circle (0.3);
        \node at (-0.0,0) {\small $A$};
        %\node at (-0.9, 0) {\small $i$};
        %\node at (0.9, 0) {\small $j$};
        %\node at (0.25, -0.6) {\small $k$};
        %\node at (1.2, -0.05) {\small $v_i$};
        \node at (0.2, -0.75) {\small $\alpha$};
        \end{tikzpicture} = t_{\alpha}\ \  \begin{tikzpicture} [scale = 1, baseline = {([yshift=-.5ex]current bounding box.center)}]
            \draw[color = black] (-0.8, 0) -- (0.8, 0);
            \draw[fill = white] (0,0) circle (0.3);
            \node at (0,0) {\small $V_\alpha$};
            \node at (0.0, -0.79) {\small $\ $};
        \end{tikzpicture}
    \end{equation}
where $t_{\alpha=1,2,\cdots,\chi^2}$ are arbitrary complex numbers such that $\sum_{\alpha} |t_\alpha|^2 = 1$ to ensure normalization.
\end{shaded}
\footnotetext[\value{footnote}]{One can always update the error basis as $\tilde V_\alpha = V_\alpha V_1^\dagger$, which has the property $\tilde V_1 = \mathds{1}$.}

\textit{Proof (abridged).} Since $\{V_\alpha\}_\alpha$ is a unitary error basis, we know $\{V_\alpha \otimes \bar V_\beta \}_{\alpha,\beta}$ is a basis for $\chi^2 \times \chi^2$-dimensional matrices. Hence, we can write the transfer matrix as:
\begin{equation} \label{eq-Tdecomposed}
    \begin{tikzpicture} [scale = 1, baseline = {([yshift=-.5ex]current bounding box.center)}] 
    \draw[color = black] (-0.8, 0) -- (0.8, 0); %  top horizontal bar
    \draw[color = black] (-0.8, -1.2) -- (0.8, -1.2); %  bottom horizontal bar
    \draw[color = black] (0, 0) -- (0, -1.2); % vertical bar
    \draw[fill = lightdodgerblue] (0,0) circle (0.3); % circle
    \draw[fill = lightdodgerblue] (0,-1.2) circle (0.3); % circle
    %% Text
    \node at (-0.03,0) {\small $A$};
    \node at (-0.03,-1.2) {\small $\bar{A}$};
    \end{tikzpicture} = \sum_{\alpha,\beta=1}^{\chi} \lambda_{\alpha,\beta} \begin{tikzpicture} [scale = 1, baseline = {([yshift=-.5ex]current bounding box.center)}] 
    \draw[color = black] (-0.8, 0) -- (0.8, 0); %  top horizontal bar
    \draw[color = black] (-0.8, -1.2) -- (0.8, -1.2); %  bottom horizontal bar
    %\draw[color = black] (0, 0) -- (0, -1.2); % vertical bar
    \draw[fill = white] (0,0) circle (0.25); % circle
    \draw[fill = white] (0,-1.2) circle (0.25); % circle
    %% Text
    \node at (0,0) {\small $V_\alpha$};
    \node at (0,-1.2) {\small $\bar{V}_\beta$};
    \end{tikzpicture}
\end{equation}
for certain coefficients $\lambda_{\alpha,\beta} \in \mathbb C$. The local characterization of gluability (Eq.~\eqref{eq-VTTV}) in the case of uniform symmetries tells us that for any $\gamma =1,2,\cdots,\chi$ we have
\begin{equation} \label{eq-AVAV}
     \begin{tikzpicture} [scale = 1, baseline = {([yshift=-.5ex]current bounding box.center)}] 
    \draw[color = black] (-0.8, 0) -- (0.8, 0); %  top horizontal bar
    \draw[color = black] (-0.8, -1.2) -- (0.8, -1.2); %  bottom horizontal bar
    \draw[color = black] (0, 0) -- (0, -1.2); % vertical bar
    \draw[fill = lightdodgerblue] (0,0) circle (0.3); % circle
    \draw[fill = lightdodgerblue] (0,-1.2) circle (0.3); % circle
    %% Text
    \node at (-0.03,0) {\small $A$};
    \node at (-0.03,-1.2) {\small $\bar{A}$};
    \end{tikzpicture} = \begin{tikzpicture} [scale = 1, baseline = {([yshift=-.5ex]current bounding box.center)}] 
    \draw[color = black] (-0.8, 0) -- (0.8, 0); %  top horizontal bar
    \draw[color = black] (-0.8, -1.2) -- (0.8, -1.2); %  bottom horizontal bar
    \draw[color = black] (0, 0) -- (0, -1.2); % vertical bar
    \draw[fill = lightdodgerblue] (0,0) circle (0.3); % circle
    \draw[fill = lightdodgerblue] (0,-1.2) circle (0.3); % circle
    %% Text
    \node at (-0.03,0) {\small $A$};
    \node at (-1,0) {\small $V_\gamma$};
    \node at (1,0) {\small $V^{\dagger}_\gamma$};
    \node at (-1,-1.2) {\small $\bar{V}_\gamma$};
    \node at (1,-1.2) {\small $V^{\mathsf{T}}_\gamma$};
    \node at (-0.03,-1.2) {\small $\bar{A}$};
    \end{tikzpicture} 
\end{equation}
So far, these equations hold for any right-gluable MPS. However, we will now argue that for the case of \emph{abelian} errors, only diagonal terms (i.e., where $\alpha = \beta$) can survive in Eq.~\eqref{eq-Tdecomposed}. Since the errors are abelian, we can write $V_\gamma V_\delta = \chi_{\gamma,\delta} V_\delta V_\gamma$ for $\chi_{\gamma,\delta} \in U(1)$. Then Eq.~\eqref{eq-AVAV} tells us
\begin{equation}
\lambda_{\alpha,\beta} = \frac{\chi_{\gamma,\alpha}}{\chi_{\gamma,\beta}} \lambda_{\alpha,\beta}.
\end{equation}
Hence, $\lambda_{\alpha,\beta} \neq 0$ implies that $\chi_{\gamma,\alpha} = \chi_{\gamma,\beta}$ for all $\gamma$. One can argue from the completeness of the unitary error basis that this implies $\alpha=\beta$. From this one can deduce Eq.~\eqref{eq-Aclassification}; we refer to Appendix~\ref{app-classification} for details.

\hspace{0.44 \textwidth} $\blacksquare$

It turns out that uniform topologically gluable states with abelian errors have additional properties which can be very convenient. In particular, in Appendix~\ref{app-classification} we show that abelian errors automatically have a group-like stucture. For instance, $V_\alpha V_\beta$ and $V_\alpha^\dagger$ are automatically part of the unitary error basis. Such structure can help with the classical post-processing step. For instance, it means that instead of correcting for a single $V_\alpha$ error with a semi-infinite string, one can pair up $V_\alpha$ with other nearby elements (such as $V_\alpha^\dagger$) with finite string operators. Algebraically, this structure is known as a \textbf{nice error basis}, and these have been classified for small degrees \cite{klappenecker2002beyond}. It is thus promising that such mathematical physics results can be applied to the many-body physics problem of deterministically preparing quantum states with measurement.

We note that Eq.~\eqref{eq-Aclassification} can be regarded as a classification of states which are uniform topologically gluable with abelian errors. In Theorem 3, we saw a general classification of (right-)gluable states, which says that up to physical isometry, $A \in \mathcal{C}(\{ V_{\alpha} \otimes V_{\alpha}^\dagger \})$ for the case of uniform errors. In the special case of abelian errors this commutant is generated by $V_\alpha \otimes V_\alpha^\dagger$ itself, such that we obtain the following alternative parametrized classification (also see Ref.~\onlinecite{zhang2024characterizing}):
\begin{equation} \label{eq-Aclassificationalternative}
A = \sum_\alpha \mu_\alpha V_\alpha \otimes V_\alpha^\dagger.
\end{equation}
In this expression, we interpret $A$ as a matrix from the virtual bonds (of dimension $\chi^2$) to the physical site (of dimension $\chi^2$). At face value, it is not manifestly obvious that Eq.~\eqref{eq-Aclassification} parameterizes the same space of states as Eq.~\eqref{eq-Aclassificationalternative}. It is a non-trivial consistency check that these two classifications can be shown to be equivalent; in fact in Appendix~\ref{app:connection-between-param-and-class} we derive the function which relates the parameters $\vec t$ to the parameters $\vec \mu$.

We can give an even stronger version of the above theorem for the case of bond dimension $\chi=2$. This is because the Pauli matrices are essentially the only unitary error basis for $2\times 2$ matrices \cite{klappenecker2005monomiality}. We thus have:
\begin{shaded}
    \textbf{Corollary 2} Suppose that $A$ is a uniform topologically gluable matrix product state tensor with bond dimension $\chi=2$.
    Then, up to an isometry acting at the physical level, $A$ is a is a matrix product state tensor with a physical dimension of $4$ and is given by:
    \begin{equation}
        \begin{tikzpicture} [scale = 1, baseline = {([yshift=-.5ex]current bounding box.center)}] 
        \draw[color = black] (-0.8, 0) -- (0.8, 0);
        \draw[color = black] (0, 0) -- (0, -0.8);
        \draw[fill = lightdodgerblue] (0,0) circle (0.3);
        \node at (-0.0,0) {\small $A$};
        %\node at (-0.9, 0) {\small $i$};
        %\node at (0.9, 0) {\small $j$};
        %\node at (0.25, -0.6) {\small $k$};
        %\node at (1.2, -0.05) {\small $v_i$};
        \node at (0.2, -0.75) {\small $\alpha$};
        \end{tikzpicture} = t_{\alpha}\ \  \begin{tikzpicture} [scale = 1, baseline = {([yshift=-.5ex]current bounding box.center)}]
            \draw[color = black] (-0.8, 0) -- (0.8, 0);
            \draw[fill = white] (0,0) circle (0.3);
            \node at (0,0) {\small $\sigma^\alpha$};
            \node at (0.0, -0.79) {\small $\ $};
        \end{tikzpicture}
    \end{equation}
where $\sigma^\alpha$ are the Pauli matrices $\{ \mathds{1},X,Y,Z\}$. Here $t_{\alpha=1,2,3,4}$ are arbitrary complex numbers such that $\sum_{\alpha} |t_\alpha|^2 = 1$.

%Moreover, the state is short-range entangled if and only if $|t_2t_3| + |t_3t_4| + |t_2t_4| \neq 0$.
\end{shaded}

As a particular example, the deformed GHZ state (Sec.~\ref{subsec-deformedGHZ}) corresponds to $t_1 = \frac{e^{\beta}}{\sqrt{e^{2\beta} + e^{-2\beta}}}$, $t_2=t_3=0$, and $t_4 = \sqrt{1-t_1^2}$, after a Hadamard transformation on the physical qubit. Similarly, the deformed cluster state (Sec.~\ref{subsec-deformedcluster}) corresponds to $t_\alpha \propto ( e^{2\beta},1,i e^{-2\beta},1)$ after blocking two physical qubits into one four-state qudit. Finally, let us remark that the spin-1 AKLT state \cite{AKLT_original} is also preparable with measurement \cite{smith2023aklt} and in fact is uniform topologically gluable in our framework, corresponding to $t_1=0$ and $t_2=t_3=t_4=\frac{1}{\sqrt{3}}$.

The above parameterizations reveal a rich landscape of matrix product states that are deterministically preparable using one round of measurements and given a specific measurement basis.
Indeed, in a companion paper \cite{sahay2024finite}, we explore some of the phenomenological landscape that arises as a consequence of the above parameterization (including how $t_\alpha$ relates to entanglement and correlation properties).
Below, we now formally present the ``go'' theorems mentioned in Sec.~\ref{subsec-Prelude}, making physical comments on the types of order present in the short-range entangled case of the above.

\subsection{Uniform, Short-Range Entangled Case} \label{subsec:simplygluableSPT}

In the case of short-range entangled states, there are no virtual symmetries of the matrix product state and consequently the correction unitaries in Eq.~\eqref{eq-symsimpglue} are now all non-trivial and generate a group $G_U$, which we term \textbf{index group} of the state\footnote{We borrow this terminology from the error correction literature \cite{knill1996group}.}.
From our example of the deformed cluster state (Sec.~\ref{subsec-deformedcluster}), we may naturally wonder if the gluable state is then required to be a non-trivial SPT. 
This is indeed the case! 
Specifically, we prove in Appendix~\ref{app:SPT} that for the abelian case, the measurement errors $V_{\alpha}$ must generate a non-trivial faithful irreducible projective representation of the above symmetry group.
This naturally connects uniform topologically gluable quantum states to SPT phases and results in the following theorem:
\begin{shaded}
    \textbf{Theorem 6} Every short-range entangled, uniform topologically gluable matrix product state with abelian errors is a non-trivial SPT phase protected by the symmetry group $G_U$.
\end{shaded}
A natural question is then whether the converse also holds.
In particular, does there exists a class of representatives in any abelian SPT phase that are required to be gluable?
This, in fact, is also true and the representatives in question are so-called \textit{minimally entangled} quantum states in the SPT phase.

Such quantum states are required to have a bipartite Schmidt rank that is the minimal rank permissible by the entanglement degeneracy of the SPT phase, but we remark this does not constrain the nature of the correlations of the state.
As an example, for all values of $\beta$, the deformed cluster states considered (Sec.~\ref{subsec-deformedcluster}) are all minimally entangled SPTs---there Schmidt rank is $\text{rank}(\Lambda) = \chi = 2$ as minimally required by the two-fold entanglement degeneracy required by the $\mathbb{Z}_2 \times \mathbb{Z}_2$ SPT order---but their correlation lengths for different values are drastically different and can even diverge.
The equivalence between minimally entangled abelian SPTs and uniform topologically gluable quantum states is established via the following theorem:

\begin{shaded}
    \textbf{Theorem 7} Any minimally-entangled translation-invariant SPT phase protected by a uniform abelian internal symmetry is topologically gluable.
\end{shaded}

We remark that, at a high level, this theorem shows that measurement-based preparability can go hand-in-hand with certain exotic classes of 1D orders.
% %
The starting point for proving the above theorem is rather natural: the state being in a non-trivial SPT phase for a symmetry group $G$ naturally gives virtual operators $V_\alpha$ which define a non-trivial projective representation of $G$ \cite{Pollmann_2010,Chen2011-et}. In the abelian case, there is a unique bond dimension associated to such a projective representation, and our above minimal-entanglement condition can then be used to prove that $\{V_\alpha\}$ indeed defines a unitary error basis. The full proof is provided in Appendix~\ref{app:SPT}. Here we want to highlight an interesting subtlety.
In particular, one may wonder if the group $G$ which defined the initial SPT phases necessarily coincides with the index group $G_U$ of the resulting gluable state.
Surprisingly this is not the case! In fact, $G_U$ need not even be a subgroup\footnote{However, it must be a quotient group. The simplest non-trivial example arises from the $\chi=2$ SPT phase protected by $G = \mathbb Z_4 \times \mathbb Z_4$ symmetry; the resulting index group of this gluable state will be $G_U = \mathbb Z_2 \times \mathbb Z_2$ which turns out to be identified with a quotient group of $G$ but not a subgroup.} of $G$.
Indeed, we share a corollary for the proof of the above theorem, which may be of independent interest:
\begin{shaded}
    \textbf{Corollary 3} Suppose $\ket{\Psi}$ is a minimally entangled SPT state protected by a uniform abelian internal symmetry $G$ and labeled by a factor set $\omega$.
    If $Z_{\omega} = \{g \in G | \omega(g, h) = \omega(h, g),  \forall h \in G\}$ is the projective center of the group associated with $\omega$, then $\ket{\psi}$ is also a non-trivial SPT pase protected by an abelian symmetry $G' \simeq G/Z_{\omega}$, whose factor set $\omega'$ is maximally non-commuting, i.e., its projective center is trivial $Z_{\omega'} = \{1\}$.
\end{shaded}
We refer to Refs.~\onlinecite{else_2012_maximallynon, else2012non2} for a discussion of the notion of maximally non-commuting SPT phases.

\subsection{Modulated Case: Dipole SPTs \label{subsec-modulated}}

Having addressed the uniform case in detail, we now turn to the comparatively more exotic modulated case for the case of short-range entangled states.
In such a case, the modulation of the measurement errors as they are swept through virtual bonds of the tensors is accompanied by a modulated site-to-site action of the correction unitary at the physical level.
Moreover, if the measurement error operator is inserted at the boundary (with suitable boundary conditions), the resulting correction symmetry string will generate a modulated symmetry of the system.

For arbitrary modulated symmetries, the structure of the resulting symmetry could be system-size dependent.
This sort of ``UV/IR mixing'' is a regular feature of systems with modulated symmetries and makes a completely general analysis of the modulated case challenging.
Indeed, a full theory of short-range entangled states is still lacking, though there has been recent progress in understanding several examples for certain modulated symmetries \cite{Stephen2019subsystem,dumbqc,han2024dipole} and a general classification in one-dimension for dipole modulated symmetries \cite{lam2024classification}.
As such, below we address a few examples of preparable modulated topologically gluable SPT states.
In all cases, it is in some sense the SPT properties of the state that enable the gluing of these states.
This suggests a more general connection between topologically gluable states to modulated SPTs that we leave for future work to address in generality.

\subsubsection{Review of Ref.~\onlinecite{han2024dipole}: Deformed Cluster State as Dipole SPT}

We start by recognizing that the deformed cluster state in our motivating examples (Sec.~\ref{subsec-deformedcluster}) can already be interpreted as the simplest (albeit contrived) example of a dipole SPT phase.
In particular, traditionally the cluster state, and its deformed analogs, are viewed as being protected by a $\mathbb{Z}_2^{\text{even}} \times \mathbb{Z}_2^{\text{odd}}$ symmetry (corresponding to spin flips on even and odd sites).
These can be viewed as two distinct uniform symmetries if one forgets the one-site translation invariance of the cluster state and blocks two adjacent sites on the cluster state together.
Alternatively, if one demands the single-site translation symmetry of the state, it can instead be viewed as a $\mathbb{Z}_2^Q \times \mathbb{Z}_2^D$ dipole SPT, where the former $\mathbb{Z}_2^Q$ is the uniform ``$\mathbb{Z}_2$-charge'' symmetry of the state generated by $\prod_{x} X_x$ and the latter is a modulated ``$\mathbb{Z}_2$-dipole'' symmetry generated by $\prod_{x \text{ even}} X_{x}$.
Phrased in a more tensor network-based language, the existence of a uniform symmetry of the matrix product state guarantees that:
\begin{equation}
    \begin{tikzpicture}[scale = 0.8, baseline = {([yshift=-.5ex]current bounding box.center)}]
        \draw[color = black] (-1,0) -- (-1.5 + 4, 0);
    \foreach \i in {0, ..., 1}{
        \draw[color = black] (1.5*\i,0) -- (1.5*\i, -1);
        \draw[fill = lightdodgerblue] (1.5*\i,0) circle (0.3);
        \node at (1.5*\i, 0) {\small $A$};
    }
    \node at (0,-1.2) {\small $X$};
    \end{tikzpicture} =     \begin{tikzpicture}[scale = 0.8, baseline = {([yshift=-.5ex]current bounding box.center)}]
        \draw[color = black] (-1,0) -- (-1.5 + 4, 0);
    \foreach \i in {0, ..., 1}{
        \draw[color = black] (1.5*\i,0) -- (1.5*\i, -1);
        \draw[fill = lightdodgerblue] (1.5*\i,0) circle (0.3);
        \node at (1.5*\i, 0) {\small $A$};
    }
    \node at (0, -1.2) {\small $\ $};
    \node at (-1.2,0) {\small $X$};
    \node at (2.7,0) {\small $X$};
    \end{tikzpicture}
\end{equation}
and similarly, 
\begin{equation}
    \begin{tikzpicture}[scale = 0.8, baseline = {([yshift=-.5ex]current bounding box.center)}]
        \draw[color = black] (-1,0) -- (-1.5 + 4, 0);
    \foreach \i in {0, ..., 1}{
        \draw[color = black] (1.5*\i,0) -- (1.5*\i, -1);
        \draw[fill = lightdodgerblue] (1.5*\i,0) circle (0.3);
        \node at (1.5*\i, 0) {\small $A$};
    }
    \node at (1.5,-1.2) {\small $X$};
    \end{tikzpicture} =     \begin{tikzpicture}[scale = 0.8, baseline = {([yshift=-.5ex]current bounding box.center)}]
        \draw[color = black] (-1,0) -- (-1.5 + 4, 0);
    \foreach \i in {0, ..., 1}{
        \draw[color = black] (1.5*\i,0) -- (1.5*\i, -1);
        \draw[fill = lightdodgerblue] (1.5*\i,0) circle (0.3);
        \node at (1.5*\i, 0) {\small $A$};
    }
    \node at (0, -1.2) {\small $\ $};
    \node at (-1.2,0) {\small $Z$};
    \node at (2.7,0) {\small $Z$};
    \end{tikzpicture}
\end{equation}
In contrast, the presence of the $\mathbb{Z}_2^Q \times  \mathbb{Z}_2^D$ symmetries guarantees that:
\begin{align}
        \begin{tikzpicture}[scale = 0.8, baseline = {([yshift=-.5ex]current bounding box.center)}] 
        \draw[color = black] (-1, 0) -- (1, 0);
        \draw[color = black] (0, 0) -- (0, -0.8);
        \draw[fill =  lightdodgerblue] (0,0) circle (0.3);
        \node at (-0.03,0) {\small $A$};
        %\node at (-1.2, 0) {\small $$};
        \node at (0, -1.1) {\small $X$};
 %       \node at (-1.4, 0) {\small $\{V_{\alpha} \}$ };
    \end{tikzpicture} &= -\begin{tikzpicture}[scale = 0.8, baseline = {([yshift=-.5ex]current bounding box.center)}] 
        \draw[color = black] (-1, 0) -- (1, 0);
        \draw[color = black] (0, 0) -- (0, -0.8);
        \draw[fill = lightdodgerblue] (0,0) circle (0.3);
        \node at (-0.03,0) {\small $A$};
        %\node at (-1.4, 0) {\small $\{V_{\alpha} \}$ };
        \node at (1.2, 0) {\small $Y$ };
        \node at (-1.2, 0) {\small $Y$ };
        \node at (0, -1.0) {\small $\ $};
    \end{tikzpicture} \\ 
    \begin{tikzpicture}[scale = 0.8, baseline = {([yshift=-.5ex]current bounding box.center)}] 
        \draw[color = black] (-1, 0) -- (1, 0);
        \draw[color = black] (0, 0) -- (0, -0.8);
        \draw[fill =  lightdodgerblue] (0,0) circle (0.3);
        \node at (-0.03,0) {\small $A$};
        \node at (-1.2, 0) {\small $Y$};
        \node at (0, -1.2) {\small $\ $};
 %       \node at (-1.4, 0) {\small $\{V_{\alpha} \}$ };
    \end{tikzpicture} &= \begin{tikzpicture}[scale = 0.8, baseline = {([yshift=-.5ex]current bounding box.center)}] 
        \draw[color = black] (-1, 0) -- (1, 0);
        \draw[color = black] (0, 0) -- (0, -0.8);
        \draw[fill = lightdodgerblue] (0,0) circle (0.3);
        \node at (-0.03,0) {\small $A$};
        %\node at (-1.4, 0) {\small $\{V_{\alpha} \}$ };
        \node at (1.2, 0) {\small $X$ };
        \node at (-1.2, 0) {\small $X$ };
        \node at (0, -1.2) {\small $\ $};
    \end{tikzpicture}
\end{align}
which can be checked is equivalent to Eq.~\eqref{eq-clusterpush}.
Both schemes are sufficient for guaranteeing that the deformed cluster state is gluable.

\subsubsection{Gluing $\mathbb{Z}_N^Q \times \mathbb{Z}_N^D$ Dipole SPTs}

Following our discussion of the deformed cluster state, we now discuss the gluing of a deformed variant of the $\mathbb{Z}_N^Q \times \mathbb{Z}_N^D$ dipole SPTs introduced in Ref.~\onlinecite{han2024dipole}.
To do so, it will be convenient to define the $\mathbb{Z}_N$ Pauli matrices (or clock and shift matrices) $\mathcal{X}$ and $\mathcal{Z}$.
These operators act on a local Hilbert space of dimension $N$, labeled $\ket{g \in \mathbb{Z}_N}$, and they satisfy the following algebraic relations: 
\begin{align}
 \mathcal{Z}^{N} = \mathcal{X}^N = \mathds{1}\quad\   \mathcal{Z} \mathcal{X} = \omega  \mathcal{X}\mathcal{Z} \quad\   \mathcal{Z}^{\dagger} \mathcal{X} = \bar{\omega}  \mathcal{X} \mathcal{Z}^{\dagger}
\end{align}
where $\omega = e^{2\pi i/N}$.
In terms of these, the two internal symmetry groups of these SPTs are given by $\prod_{x} \mathcal{X}_x$ and $\prod_{x} \mathcal{X}^x_x$, which generate the charge and dipole symmetry respectively. 

There are $N$ distinct variants of the dipole SPT labeled by $\eta \in \mathbb{Z}_N$.
In all such cases, a deformed variant of the fixed point wavefunction can be labeled by: 
\begin{equation}
    \begin{tikzpicture}[scale = 1, baseline = {([yshift=-.5ex]current bounding box.center)}] 
        \draw[color = black] (-0.8, 0) -- (0.8, 0);
        \draw[color = black] (0, 0) -- (0, -0.8);
        \draw[fill = lightdodgerblue] (0,0) circle (0.3);
        \node at (-0.0,0) {\small $A$};
        %\node at (-0.9, 0) {\small $i$};
        %\node at (0.9, 0) {\small $j$};
        %\node at (0.25, -0.6) {\small $k$};
        %\node at (1.2, -0.05) {\small $v_i$};
        % \node at (0.0, -1.2) {\small $\ $};
 %       \node at (-1.4, 0) {\small $\{V_{\alpha} \}$ };
    \end{tikzpicture}  = \begin{tikzpicture}[scale = 1, baseline = {([yshift=-.5ex]current bounding box.center)}] 
    \draw[color = black] (0, 0) -- (0, -1);
    \draw[color = black] (1.2, 0) -- (-0.8, 0);
    %\draw[color = black] (0, 0) -- (0, -0.8);
    \draw[fill = lightorange(ryb)] (0.6,0) circle (0.25);
    \node at (0.6, 0) {\small $H_{\eta}$};
    \draw[fill = white] (0, -0.5) circle (0.2); 
\end{tikzpicture}
\end{equation}
where the white circle is $e^{\beta (\mathcal{X} + \mathcal{X}^{\dagger})}$ and the orange circle denotes a dipolar ``Hadamard''-esque matrix:
\begin{equation}
    H_{\eta}  = \sum_{g, h = 1}^{N} \omega^{\eta g( h - g)} \ket{g} \bra{h}
\end{equation}
These tensors are in right canonical form and satisfy the following dipolar SPT rules, which can be expressed as:
\begin{align} \label{eq-dipolerules}
        \begin{tikzpicture}[scale = 0.8, baseline = {([yshift=-.5ex]current bounding box.center)}] 
        \draw[color = black] (-1, 0) -- (1, 0);
        \draw[color = black] (0, 0) -- (0, -0.8);
        \draw[fill =  lightdodgerblue] (0,0) circle (0.3);
        \node at (-0.03,0) {\small $A$};
        %\node at (-1.2, 0) {\small $$};
        \node at (0, -1.0) {\small $\mathcal{X}^{\dagger}$};
 %       \node at (-1.4, 0) {\small $\{V_{\alpha} \}$ };
    \end{tikzpicture} &= \begin{tikzpicture}[scale = 0.8, baseline = {([yshift=-.5ex]current bounding box.center)}] 
        \draw[color = black] (-1, 0) -- (1, 0);
        \draw[color = black] (0, 0) -- (0, -0.8);
        \draw[fill = lightdodgerblue] (0,0) circle (0.3);
        \node at (-0.03,0) {\small $A$};
        %\node at (-1.4, 0) {\small $\{V_{\alpha} \}$ };
        %\node at (1.2, 0) {\small $Y$ };
        %\node at (-1.2, 0) {\small $Y$ };
        \node at (-1.75, 0) {\small $(\mathcal{Z}^{\dagger})^{\eta} \mathcal{X}$};
        \node at (1.7, 0) {\small $\mathcal{X}^{\dagger}\mathcal{Z}^\eta$};
        \node at (0, -1.0) {\small $\ $};
    \end{tikzpicture} \\ 
    \begin{tikzpicture}[scale = 0.8, baseline = {([yshift=-.5ex]current bounding box.center)}] 
        \draw[color = black] (-1, 0) -- (1, 0);
        \draw[color = black] (0, 0) -- (0, -0.8);
        \draw[fill =  lightdodgerblue] (0,0) circle (0.3);
        \node at (-0.03,0) {\small $A$};
        \node at (-1.8, 0) {\small $(\mathcal{Z}^{\dagger})^{\eta} \mathcal{X}$};
        \node at (0, -1.2) {\small $\ $};
 %       \node at (-1.4, 0) {\small $\{V_{\alpha} \}$ };
    \end{tikzpicture} &= \omega^{-\eta} \begin{tikzpicture}[scale = 0.8, baseline = {([yshift=-.5ex]current bounding box.center)}] 
        \draw[color = black] (-1, 0) -- (1, 0);
        \draw[color = black] (0, 0) -- (0, -0.8);
        \draw[fill = lightdodgerblue] (0,0) circle (0.3);
        \node at (-0.03,0) {\small $A$};
        %\node at (-1.4, 0) {\small $\{V_{\alpha} \}$ };
        \node at (1.2, 0) {\small $\mathcal{X}^{\dagger}$ };
        \node at (-1.2, 0) {\small $\mathcal{X}$ };
        \node at (0, -1.2) {\small $\ $};
    \end{tikzpicture}
\end{align}
The rules above are rather generic and apply, in some form, to any minimally entangled representative of the $\mathbb{Z}_N \times \mathbb{Z}_N^D$ dipole SPT phase.
Rules in hand, we now discuss how to prepare these states with measurement.

As always, we envision preparing the tensors above as quantum states.
Subsequently, we measure the virtual qudits in the maximally entangled unitary error basis defined by the $\mathbb{Z}_N$ Pauli matrices: $\mathcal{V} = \langle \mathcal{X}, \mathcal{Z} \rangle$.
Note that the dipole SPT rules of Eq.~\eqref{eq-dipolerules}, can be rearranged to provide push through rules for $(\mathcal{Z}^{\dagger})^{\eta} \mathcal{X}^{\dagger}$ and $(\mathcal{Z}^{\dagger})^{\eta}$.
To ensure correctability, we consider the the case where $d_{\eta} = \text{gcd}(\eta, N) = 1$ (i.e. $\eta$ and $N$ are co-prime).
In such a case, the aforementioned operators generate the $\mathbb{Z}_N$ Paulis and we are guarantees that we can correct any virtual insertions just using the SPT property.
This demonstrates the preparability of the deformed $\mathbb{Z}_N^Q \times \mathbb{Z}_N^D$ dipole SPT.
The same argument applies to any minimally entangled representative of the $\mathbb{Z}_N^Q \times \mathbb{Z}_N^D$ dipole SPT for $d_{\eta} = 1$.

\section{Local and Topological Errors: Correlation versus Entanglement}\label{sec:localvtopological}

Having provided a detailed analysis of the constraints imposed on gluable quantum states with exclusively topological errors, in this section, we will now consider the case where there exist both topological and local errors in the system.
In particular, we first show that the presence of even one local error in the system forces there to exist zero correlation length operators in the state (a statement, which we will make more precise in the matrix product state language shortly).
Subsequently, we will provide some extended physical intuition on the above result before stating the main result of this section: a no-go theorem demonstrating that certain quantum many-body states are not gluable, or equivalently cannot be prepared using the ingredients informally laid out in the introduction.

\subsection{Local Errors Constrain Correlations}

As stated above, here we will consider a scenario where a subset of the measurement errors are locally correctable.
In this case, we will prove a simple, but powerful constraint on the correlations of the state.
To do so, let us first recall that correlation functions in a matrix product state are controlled by the spectrum of the transfer matrix \cite{Bridgeman_2017, Cirac2021MPSReview, tenpy, perezgarcia2006MPSrep}.
Indeed, the correlation function of a generic operator $\mathcal{O}$ (either point-like or string-like) is a sum of exponential decays given by: 
\begin{equation}
    \langle \mathcal{O}(x) \mathcal{O}(y)\rangle =  \sum_{\lambda \in \text{spec}(T)} c_{\mathcal{O}}^{\lambda} e^{-|x - y|/\xi_{\lambda}}
\end{equation}
where $\xi_{\lambda} = 1/\log(1/\lambda)$ and $c_{\mathcal{O}}^{\lambda}$ are coefficients that depend on the choice of operator.
As a consequence, we will refer to the spectrum of the transfer matrix more physically as the \textbf{correlation spectrum}.

Let us remark that, by the requirement of normalization of the state, the largest transfer matrix eigenvalue is required to be $1$ and degeneracies in this eigenvalue always signify long-range entanglement\footnote{Indeed, degeneracies in the largest transfer matrix eigenvalue imply that there exists an operator that fails to satisfy cluster decomposition: $\langle \mathcal{O}(x) \mathcal{O}(y) \rangle \to \langle \mathcal{O}(x) \rangle \langle \mathcal{O}(y) \rangle$ for large separations $|x - y|$. Consequently, the resulting state cannot even approximately be connected to a product state with a finite-depth local unitary and is hence long-range entangled.}.
Zeros in the transfer matrix spectrum are associated with operators whose correlation length is exactly zero \footnote{For fixed point states, aside from degeneracies in the largest transfer matrix eigenvalue, the remaining eigenvalues are always zero.}.
Consequently, seeing such zeros in the correlation spectrum indicates the presence of operators whose connected correlation functions, while not identically zero, are zero beyond a certain distance.
With an understanding of the correlations in the matrix product state, we are prepared to state and prove the following theorem:

\begin{shaded}
    \textbf{Theorem 8} (Local Errors Constrain Correlations) Suppose that $A$ is a matrix product state tensor and $V$ is a (non-identity) measurement error that can be corrected at the physical level using a unitary with finite local support.
    Then, the correlation spectrum of $A$ has a zero.
\end{shaded}

\textit{Proof.} Let us suppose that $V$ is an operator that that can be locally corrected.
Then, this means that the measurement-error pushes through the matrix product state onto a unitary with a support on a bounded number of continguous sites $n$.
Consequently, we have that: 
\begin{equation}
\begin{tikzpicture} [scale = 1, baseline = {([yshift=-.5ex]current bounding box.center)}] 
        \draw[color = black] (-1.4, 0) -- (1.0, 0);
        \draw[color = black] (-1.4, -1.2) -- (1.0, -1.2);
        \draw[rounded corners, fill = lightdodgerblue] (-0.4, -1.4) rectangle (0.4, 0.2) {};
        \node at (0, -0.6) { $T^n$};
        \draw[fill = white] (-0.9,0) circle (0.3);
        \node at (-0.9, 0) {\small $V$};
        \draw[fill = white] (-0.9,-1.2) circle (0.3);
        \node at (-0.9, -1.2) {\small $\bar{V}$};
     \end{tikzpicture} = 
    \begin{tikzpicture} [scale = 1, baseline = {([yshift=-.5ex]current bounding box.center)}] 
        \draw[color = black] (-1.0, 0) -- (1., 0);
        \draw[color = black] (-1.0, -1.2) -- (1., -1.2);
        \draw[rounded corners, fill = lightdodgerblue] (-0.4, -1.4) rectangle (0.4, 0.2) {};
        \node at (0, -0.6) {$T^{n}$};
     \end{tikzpicture}
\end{equation}
The above naturally implies that $(V \otimes \bar{V} - \mathds{1} \otimes \mathds{1})T = 0$.
Now, by assumption, $V\neq \mathds{1}$ and hence the above operator is non-zero.
Thus, for any vector $\bra{\phi} \in \mathbb{C}^{\chi} \otimes \mathbb{C}^{\chi}$ such that $\bra{\phi} (V \otimes \bar{V} - \mathds{1} \otimes \mathds{1}) = \bra{\psi} \neq 0$, we have that $\bra{\psi} T = 0$ \footnote{Indeed, this same line of reasoning shows that if there is a subset $\nu \subset \mathcal{V}$ that is locally correctable, then the number of zeros in the transfer matrix $\zeta$ is bounded from below by 
$$\zeta \geq \chi^2 - \text{dim} \left( \bigcap_{v \in \nu} \text{ker}(v \otimes \bar{v} - \mathds{1})\right)$$
}.
This completes the proof.

\hspace{0.44 \textwidth} $\blacksquare$

The above theorem shows that states with local errors are ``closer to their fixed points'', at least at the level of their correlations.

\subsection{Local Errors and Their Correction from Local Unitaries}

To provide some more intuition for the previous theorem, we show that in certain cases, matrix product states with local errors can be ``partially glued'' using a finite-depth local unitary circuit. 
Since such unitaries cannot increase the correlation length of a state (as they have finite light cones), this provides heuristic intuition for why the correlations of such states are more limited.

The aforementioned cases correspond to when each element of the error basis $\mathcal{V}$ can be decomposed into a product of two sets of operators $v_\alpha V_\beta$, with $i = 1, \cdots, \chi'$ and $j = 1, \cdots \chi^2/\chi'$, where the $v$'s are local errors and $V$'s are topological errors.
In other words, they satisfy:
    \begin{align}
        \begin{tikzpicture}[scale = 1, baseline = {([yshift=-.5ex]current bounding box.center)}] 
        \draw[color = black] (-1, 0) -- (1, 0);
        \draw[color = black] (0, 0) -- (0, -0.8);
        \draw[fill =  lightdodgerblue] (0,0) circle (0.3);
        \node at (-0.03,0) {\small $A$};
        \node at (-1.2, 0) {\small $V_{\alpha}$};
        \node at (0, -1.2) {\small $\ $};
 %       \node at (-1.4, 0) {\small $\{V_{\alpha} \}$ };
    \end{tikzpicture} &= \begin{tikzpicture}[scale = 1, baseline = {([yshift=-.5ex]current bounding box.center)}] 
        \draw[color = black] (-1, 0) -- (1, 0);
        \draw[color = black] (0, 0) -- (0, -0.8);
        \draw[fill = lightdodgerblue] (0,0) circle (0.3);
        \node at (-0.03,0) {\small $A$};
        %\node at (-1.4, 0) {\small $\{V_{\alpha} \}$ };
        \node at (1.3, 0) {\small $V^{[1]}_{\alpha}$ };
        \node at (0, -1.0) {\small $U^{[0]}_{\alpha}$};
    \end{tikzpicture}  \\ 
    \begin{tikzpicture}[scale = 1, baseline = {([yshift=-.5ex]current bounding box.center)}] 
        \draw[color = black] (-1, 0) -- (1, 0);
        \draw[color = black] (0, 0) -- (0, -0.8);
        \draw[fill =  lightdodgerblue] (0,0) circle (0.3);
        \node at (-0.03,0) {\small $A$};
        \node at (-1.2, 0) {\small $v_{\alpha}$};
        \node at (0, -1.2) {\small $\ $};
 %       \node at (-1.4, 0) {\small $\{V_{\alpha} \}$ };
    \end{tikzpicture} &= \begin{tikzpicture}[scale = 1, baseline = {([yshift=-.5ex]current bounding box.center)}] 
        \draw[color = black] (-1, 0) -- (1, 0);
        \draw[color = black] (0, 0) -- (0, -0.8);
        \draw[fill = lightdodgerblue] (0,0) circle (0.3);
        \node at (-0.03,0) {\small $A$};
        %\node at (-1.4, 0) {\small $\{V_{\alpha} \}$ };
        \node at (1.2, 0) {\small $\ $ };
        \node at (0, -1.0) {\small $u_{\alpha}$};
    \end{tikzpicture} 
\end{align}
where $V_{\alpha}^{[1]}$ can be iteratively pushed off to infinity and $U_{\alpha}^{[0]}$ and $u_{\alpha}$ are unitary.
Such a structure arises, for example, in the deformed trivial state (Sec.~\ref{subsec-deformedtrivial}) whose errors basis---the Pauli's---satisfy similar push-through conditions, with $Z$ being a local error and $X$ being topological.
We now demonstrate how this product structure of the errors allows to replace some of the measurements-and-feedback by conventional unitaries.
For maximal clarity we here demonstrate the key ideas using the example of the deformed trivial state, but the manipulations generalize to the more general product case structure.

The essential idea is to replace measurement and local classical post-processing with quantum processing.
To see this idea in action, let us recall that for the deformed trivial state, Bell measurement between two clusters (see Eq.~\eqref{eq-trivial_tensor_clusters} for a tensor network depiction of the clusters) are used to glue the clusters together.
Operationally, one could perform this Bell measurement by first performing a control-NOT gate between the clusters visualized as: 
\begin{equation}
    \begin{tikzpicture}[scale = 1, baseline = {([yshift=-.5ex]current bounding box.center)}] 
        \foreach \i in {0, ..., 1}{
        \draw[color = black] (-0.7 + 2.3*\i, 0) -- (0.7 + 2.3*\i, 0);
        \draw[color = black] (-0.7 + 2.3*\i, 0) -- (-0.7 + 2.3*\i, -0.8);
        \draw[color = black] (0.7 + 2.3*\i, 0) -- (0.7 + 2.3*\i, -0.8);
        \draw[color = black] (0 + 2.3*\i, 0) -- (0 + 2.3*\i, -0.8);
        \draw[fill = white] (0.7 + 2.3*\i, -0.3) circle (0.15);
        }
        \draw[color = black] (0.7, -0.8) -- (-0.7 + 2.3, -0.8);
        \draw[fill = black] (1.6, -0.8) circle (0.07);
        \draw[color = black] (0.7, -0.8) -- (0.7, -1.2);
        \draw[color = black] (1.6, -0.8) -- (1.6, -1.2);
        \node at (0.7, -0.8) {$\small \bigoplus$};
\end{tikzpicture}
\end{equation}
and subsequently, perform a $Z$-basis measurement on the target and a $X$-basis measurement on the control.
Note that if the outcome is $\ket{0}\ket{+}$, then the result will be to insert a projector onto the Bell state $\ket{\mathds{1}}$.
This constitutes the ideal measurement outcome.
In the case, where we get either the $\ket{1}$ or $\ket{0}$ state, note that: 
\begin{equation}
    \begin{tikzpicture}[scale = 0.7, baseline = {([yshift=-.5ex]current bounding box.center)}] 
        \foreach \i in {0, ..., 1}{
        \draw[color = black] (-0.7 + 2.3*\i, 0) -- (0.7 + 2.3*\i, 0);
        \draw[color = black] (-0.7 + 2.3*\i, 0) -- (-0.7 + 2.3*\i, -0.8);
        \draw[color = black] (0.7 + 2.3*\i, 0) -- (0.7 + 2.3*\i, -0.8);
        \draw[color = black] (0 + 2.3*\i, 0) -- (0 + 2.3*\i, -0.8);
        \draw[fill = white] (0.7 + 2.3*\i, -0.3) circle (0.15);
        }
        \draw[color = black] (0.7, -0.8) -- (-0.7 + 2.3, -0.8);
        \draw[fill = black] (1.6, -0.8) circle (0.07);
        \draw[color = black] (0.7, -0.8) -- (0.7, -1.4);
        \draw[color = black] (1.6, -0.8) -- (1.6, -1.4);
        \node at (0.7, -0.8) {$\small \bigoplus$};
        \draw[fill = error-red, rounded corners] (0.3, -1.25) rectangle (1.1, -1.75);
        \node at (0.7, -1.5) {\small $\bra{1}$};
        \draw[fill = lightishgray, rounded corners] (0.3 + 0.9, -1.25) rectangle (1.1 + 0.9, -1.75);
        \node at (0.7 + 0.9, -1.5) {\small $\bra{+}$};
\end{tikzpicture} =     \begin{tikzpicture}[scale = 0.7, baseline = {([yshift=-.5ex]current bounding box.center)}] 
        \foreach \i in {0, ..., 1}{
        \draw[color = black] (-0.7 + 2.3*\i, 0) -- (0.7 + 2.3*\i, 0);
        \draw[color = black] (-0.7 + 2.3*\i, 0) -- (-0.7 + 2.3*\i, -0.8);
        \draw[color = black] (0.7 + 2.3*\i, 0) -- (0.7 + 2.3*\i, -0.8);
        \draw[color = black] (0 + 2.3*\i, 0) -- (0 + 2.3*\i, -0.8);
        \draw[fill = white] (0.7 + 2.3*\i, -0.3) circle (0.15);
        }
        \draw[color = black] (0.7, -0.8 - 0.4) -- (-0.7 + 2.3, -0.8 - 0.4);
        \draw[fill = black] (1.6, -0.8- 0.4) circle (0.07);
        \draw[color = black] (0.7, -0.8) -- (0.7, -1.8);
        \draw[color = black] (1.6, -0.8) -- (1.6, -1.8);
        \node at (0.7, -0.8 - 0.4) {$\small \bigoplus$};
        \node at (0.7, -0.7) {\small $\textcolor{red}{X}$};
        \draw[fill = lightishgray, rounded corners] (0.3, -1.25 - 0.4) rectangle (1.1, -1.75 - 0.4);
        \node at (0.7, -1.5 - 0.4) {\small $\bra{0}$};
        \draw[fill = lightishgray, rounded corners] (0.3 + 0.9, -1.25 - 0.4) rectangle (1.1 + 0.9, -1.75 - 0.4);
        \node at (0.7 + 0.9, -1.5 - 0.4) {\small $\bra{+}$};
\end{tikzpicture} 
\end{equation}
and 
\begin{equation}
    \begin{tikzpicture}[scale = 0.7, baseline = {([yshift=-.5ex]current bounding box.center)}] 
        \foreach \i in {0, ..., 1}{
        \draw[color = black] (-0.7 + 2.3*\i, 0) -- (0.7 + 2.3*\i, 0);
        \draw[color = black] (-0.7 + 2.3*\i, 0) -- (-0.7 + 2.3*\i, -0.8);
        \draw[color = black] (0.7 + 2.3*\i, 0) -- (0.7 + 2.3*\i, -0.8);
        \draw[color = black] (0 + 2.3*\i, 0) -- (0 + 2.3*\i, -0.8);
        \draw[fill = white] (0.7 + 2.3*\i, -0.3) circle (0.15);
        }
        \draw[color = black] (0.7, -0.8) -- (-0.7 + 2.3, -0.8);
        \draw[fill = black] (1.6, -0.8) circle (0.07);
        \draw[color = black] (0.7, -0.8) -- (0.7, -1.4);
        \draw[color = black] (1.6, -0.8) -- (1.6, -1.4);
        \node at (0.7, -0.8) {$\small \bigoplus$};
        \draw[fill = lightishgray, rounded corners] (0.3, -1.25) rectangle (1.1, -1.75);
        \node at (0.7, -1.5) {\small $\bra{0}$};
        \draw[fill = error-red, rounded corners] (0.3 + 0.9, -1.25) rectangle (1.1 + 0.9, -1.75);
        \node at (0.7 + 0.9, -1.5) {\small $\bra{-}$};
\end{tikzpicture} =     \begin{tikzpicture}[scale = 0.7, baseline = {([yshift=-.5ex]current bounding box.center)}] 
        \foreach \i in {0, ..., 1}{
        \draw[color = black] (-0.7 + 2.3*\i, 0) -- (0.7 + 2.3*\i, 0);
        \draw[color = black] (-0.7 + 2.3*\i, 0) -- (-0.7 + 2.3*\i, -0.8);
        \draw[color = black] (0.7 + 2.3*\i, 0) -- (0.7 + 2.3*\i, -0.8);
        \draw[color = black] (0 + 2.3*\i, 0) -- (0 + 2.3*\i, -0.8);
        \draw[fill = white] (0.7 + 2.3*\i, -0.3) circle (0.15);
        }
        \draw[color = black] (0.7, -0.8 - 0.4) -- (-0.7 + 2.3, -0.8 - 0.4);
        \draw[fill = black] (1.6, -0.8- 0.4) circle (0.07);
        \draw[color = black] (0.7, -0.8) -- (0.7, -1.8);
        \draw[color = black] (1.6, -0.8) -- (1.6, -1.8);
        \node at (0.7, -0.8 - 0.4) {$\small \bigoplus$};
        \node at (1.6, -0.7) {\small $\textcolor{red}{Z}$};
        \draw[fill = lightishgray, rounded corners] (0.3, -1.25 - 0.4) rectangle (1.1, -1.75 - 0.4);
        \node at (0.7, -1.5 - 0.4) {\small $\bra{0}$};
        \draw[fill = lightishgray, rounded corners] (0.3 + 0.9, -1.25 - 0.4) rectangle (1.1 + 0.9, -1.75 - 0.4);
        \node at (0.7 + 0.9, -1.5 - 0.4) {\small $\bra{+}$};
\end{tikzpicture} 
\end{equation}
The above means that, errors on the left measurement insert an $X$ measurement error between the clusters and errors on the right measurement insert a $Z$ measurement error.
While correcting the topological $X$ error requires a non-local string as explored extensively in the previous sections, correcting the local $Z$ error only requires a \emph{local action} of $Z^a$ on the physical qubit immediately to the right of the measurement, classically conditioned on the measurement outcome of the right qubit above ($a = 0$ for $\ket{+}$ and $a = 1$ for $\ket{-}$).
This local classical post-processing can be replaced with the following quantum processing: 
\begin{equation}
    \begin{tikzpicture}[scale = 1, baseline = {([yshift=-.5ex]current bounding box.center)}] 
        \foreach \i in {0, ..., 1}{
        \draw[color = black] (-0.7 + 2.3*\i, 0) -- (0.7 + 2.3*\i, 0);
        \draw[color = black] (-0.7 + 2.3*\i, 0) -- (-0.7 + 2.3*\i, -0.8);
        \draw[color = black] (0.7 + 2.3*\i, 0) -- (0.7 + 2.3*\i, -0.8);
        \draw[color = black] (0 + 2.3*\i, 0) -- (0 + 2.3*\i, -0.8);
        \draw[fill = white] (0.7 + 2.3*\i, -0.3) circle (0.15);
        }
        \draw[color = black] (0.7, -0.7) -- (-0.7 + 2.3, -0.7);
        \draw[fill = black] (1.6, -0.7) circle (0.07);
        \draw[color = black] (0.7, -0.7) -- (0.7, -1.8);
        \draw[color = black] (1.6, -0.7) -- (1.6, -1.8);
        \draw[color = black] (2.3, -0.7) -- (2.3, -1.8);
        \node at (0.7, -0.7) {$\small \bigoplus$};
        \draw[fill = lightorange(ryb)] (1.6, -1.1) circle (0.1);
        \draw[fill = black] (1.6, -1.35) circle (0.07);
        \draw[fill = black] (2.3, -1.35) circle (0.07);
        \draw[color = black] (1.6, -1.35) -- (2.3, -1.35);
        \draw[fill = lightorange(ryb)] (1.6, -1.6) circle (0.1);
        \draw[color = red] (1.4, -1.0) rectangle (2.5, -1.7);
\end{tikzpicture}
\end{equation}
where the orange circles are once again Hadamard gates, the conditional gate is a control-$Z$ gate, and the purpose of the gate in the red box is effectively to check whether the left qubit is in the $\ket{+}$ or $\ket{-}$ state and use a quantum conditional gate to correct it on the right.
Note that the red box is simply just a control-NOT, but we have emphasized its internal structure for didactic reasons, since it can be interpreted as (quantum) feedback which checks the state of the to-be-measured ancilla.
Indeed, one can show that: 
\begin{equation}
    \begin{tikzpicture}[scale = 0.8, baseline = {([yshift=-.5ex]current bounding box.center)}] 
        \foreach \i in {0, ..., 1}{
        \draw[color = black] (-0.7 + 2.3*\i, 0) -- (0.7 + 2.3*\i, 0);
        \draw[color = black] (-0.7 + 2.3*\i, 0) -- (-0.7 + 2.3*\i, -0.8);
        \draw[color = black] (0.7 + 2.3*\i, 0) -- (0.7 + 2.3*\i, -0.8);
        \draw[color = black] (0 + 2.3*\i, 0) -- (0 + 2.3*\i, -0.8);
        \draw[fill = white] (0.7 + 2.3*\i, -0.3) circle (0.15);
        }
        \draw[color = black] (0.7, -0.7) -- (-0.7 + 2.3, -0.7);
        \draw[fill = black] (1.6, -0.7) circle (0.07);
        \draw[color = black] (0.7, -0.7) -- (0.7, -1.8);
        \draw[color = black] (1.6, -0.7) -- (1.6, -1.8);
        \draw[color = black] (2.3, -0.7) -- (2.3, -1.8);
        \node at (0.7, -0.7) {$\small \bigoplus$};
        \draw[fill = lightorange(ryb)] (1.6, -1.1) circle (0.1);
        \draw[fill = black] (1.6, -1.35) circle (0.07);
        \draw[fill = black] (2.3, -1.35) circle (0.07);
        \draw[color = black] (1.6, -1.35) -- (2.3, -1.35);
        \draw[fill = lightorange(ryb)] (1.6, -1.6) circle (0.1);
        \draw[color = red] (1.4, -1.0) rectangle (2.5, -1.7);
        \draw[fill = lightishgray, rounded corners] (0.3, -1.25 -0.1) rectangle (1.1, -1.75 - 0.1);
        \node at (0.7, -1.5 - 0.1) {\small $\bra{0}$};
        %\draw[fill = lightdodgerblue] (0,0) circle (0.3);
        %\node at (-0.0,0) {\small $A$};
        %\node at (-0.9, 0) {\small $i$};
        %\node at (0.9, 0) {\small $j$};
        %\node at (0.25, -0.6) {\small $k$};
        %\node at (1.2, -0.05) {\small $v_i$};
        % \node at (0.0, -1.2) {\small $\ $};
 %       \node at (-1.4, 0) {\small $\{V_{\alpha} \}$ };
\end{tikzpicture}\ =\     \begin{tikzpicture}[scale = 0.8, baseline = {([yshift=-.5ex]current bounding box.center)}] 
        \draw[color = black] (-0.7 + 2.3*0, 0) -- (0.7 + 2.3*1, 0);
        \draw[color = black] (-0.7 + 2.3*0, 0) -- (-0.7 + 2.3*0, -0.8);
        %\draw[color = black] (0.7 + 2.3*0, 0) -- (0.7 + 2.3*0, -0.8);
        \draw[color = black] (0 + 2.3*0, 0) -- (0 + 2.3*0, -0.8);
        \draw[fill = white] (1.15 + 2.3*0, 0) circle (0.15);
        %\draw[color = black] (-0.7 + 2.3*1, 0) -- (0.7 + 2.3*1, 0);
        %\draw[color = black] (-0.7 + 2.3*1, 0) -- (-0.7 + 2.3*1, -0.8);
        \draw[color = black] (0.7 + 2.3*1, 0) -- (0.7 + 2.3*1, -0.8);
        \draw[color = black] (0 + 2.3*1, 0) -- (0 + 2.3*1, -0.8);
        \draw[fill = white] (0.7 + 2.3*1, -0.3) circle (0.15);
        \draw[color = black] (0.7 + 1, -1.05 +0.3) -- (0.7 + 1, -1.05 - 0.2);
        \draw[fill = lightishgray, rounded corners] (0.3 + 1., -0.8 +0.3) rectangle (1.1 + 1., -1.3 +0.3);
        \node at (0.7 + 1, -1.05 +0.3) {\small $\bra{+}$};
\end{tikzpicture}
\end{equation}
What this means is that by turning our classically-conditioned on-site feedback into a quantum gate, the corresponding ancilla qubit is automatically disentangled and we thus do not need to measure it anymore. We have thus reduced the necessary measurements to a single qubit for ``gluing'' the two clusters together, with a topological error which can be corrected by a non-local string.

Since local unitary gates cannot change the correlation length of the state they are acting on, the above provides some intuition as to why local errors lead to zeros in the correlation spectrum.
Furthermore, we remark that, while unitary gates cannot change correlation lengths, they can affect the entanglement spectrum, further motivating why states with local errors can have non-flat entanglement spectrum.

\subsection{No-Go Theorem}

Given the above theorem on local errors and zeros in the correlation spectrum and intuition for this result, we are now prepared to prove our final main result of this manuscript: a powerful no-go theorem on the preparability of quantum states with a single round of measurements and left-conditioned feedback.

\begin{shaded}
    \textbf{Theorem 9} (No-Go Theorem) Suppose $\ket{\Psi}$ is a translation-invariant quantum state with a non-flat entanglement spectrum upon bipartitioning the infinite chain in two halves, and with no zeros in its correlation spectrum (i.e., its transfer matix has full rank).
    Then $\ket{\Psi}$ is not right-gluable, i.e., it cannot be prepared with the ingredients listed in Sec.~\ref{sec:setup}.
\end{shaded}

\textit{Proof.} The proof follows by establishing a contradiction.
Let us suppose that $\ket{\Psi}$ was gluable with an error basis $\mathcal{V}$.
Further, suppose that the transfer matrix of the matrix product state representation of $\ket{\Psi}$ had no zero eigenvalues in its spectrum and consequently was full rank.
Then, by Theorem~8, we know that $\ket{\Psi}$ has no locally correctable errors and its errors are exclusively topological. 
Consequently, per Theorem~4 \footnote{We remark upon a subtlety that is glossed over in this proof. Namely, the errors all being topological was not strictly sufficient for showing a flat entanglement spectrum. It was also necessary to show that each $\mathcal{V}^{[n]}$ in Eq.~\eqref{eq-pushable} was a unitary error basis. We show in Appendix~\ref{app-nogo} that, when a quantum state has no zeros in its correlation spectrum, all errors being topological implies each $\mathcal{V}^{[n]}$ is a unitary error basis.}, the entanglement spectrum is flat.
However, this is in contradiction with the assumption that $\ket{\Psi}$ had a non-flat entanglement spectrum.
Hence, $\ket{\Psi}$ is not gluable.

\hspace{0.44 \textwidth} $\blacksquare$

As an example of a state with the aforementioned properties, we can consider a \emph{combination} of the examples seen in Sec.~\ref{sec:examples_and_prelude}. For instance, $\ket{\Psi} = e^{\beta \sum_n X_n} e^{\beta' \sum_n Z_n Z_{n+1}}\ket{+}$ is a translation-invariant state which admits an exact $\chi=2$ MPS representation, and one can straightforwardly check that for $\beta\neq 0 \neq \beta'$, its entanglement spectrum is not flat and its transfer matrix has full rank. Thus, this state cannot be prepared with the ingredients explored in the present work. This highlights it is in principle possible to highlight concrete boundaries between what is and what is not possible using measurement-based state preparation.

\section{Conclusions and Outlook}

In this work, we sought to isolate the power of measurement for creating one-dimensional quantum entanglement by characterizing states accessible with a single round of measurement and left-conditioned tensor product unitary feedback.
We showed that such states are required to have an exact matrix product state description and then uncovered local criteria on the matrix product state tensors that are necessary and sufficient conditions for preparability.
These local criteria then directly informed the preparation procedure of such states and enabled a full classification of all preparable states.
Moreover, they also placed strong constraints on their physical properties---dictating which states could or could not be prepared in different settings.

In the general case, we were able to uncover an intriguing trade-off in the power of measurement for creating interesting entanglement structure and correlations.
Namely, if the correlations of the state were rich (i.e., no zero-correlation length operators present in the state), the entanglement structure was heavily constrained (i.e., forced to be flat).
Conversely, in order to have a more expressive entanglement structure, the state was required to have zero correlation length operators.
This tradeoff then naturally led to the no-go theorem in the previous section.

Moreover, in more specialized settings (i.e., for uniform topologically gluable quantum states with abelian errors), we were able to find convenient parameterizations of the complete space of preparable states, enabling a phenomenological exploration carried out as part of a companion work \cite{sahay2024finite}.
Finally, we revealed connections between preparable quantum states and order, finding that the manner in which a state was corrected, often directly informed its order.
This connection was made sharpest in the context of abelian symmetry protected topological phases where we found that all minimally entangled representatives of these phases were preparable, with the converse also holding under suitable conditions.

Our work also opens up many exciting opportunities for future research.
On the more practical side, our work opens the door to a large class of quantum many-body states that can be prepared with few quantum operations in contemporary quantum devices.
Such quantum states could be useful as initial states in quantum simulation experiments.
In such a case, they could be the starting point of some quench dynamics protocol or could be non-trivial inputs to more elaborate simulation schemes.
Moreover, they could also serve as inputs to benchmark contemporary quantum algorithms.
Such constant-depth preparation protocols are especially timely since a variety of platforms have recently demonstrated mid-circuit measurement capabilities in tuneable many-body quantum systems \cite{Singh_2023,moses2023race,deist2022midcircuit,norcia2023midcirc,graham2023midcirc,baumer2023efficient, anand2024dualspecies, baumer2024quantum, finkelstein2024universal, Bluvstein_2023}.

On the more theoretical side, our work contributes to the recent push to develop a more refined organization and understanding of the complexity of quantum entanglement and correlations.
Indeed, the states we show are non-gluable in the main text are ``more complex'' in a precise sense than those that are gluable.
Along this direction, it would be interesting to study several immediate generalizations of this work.
As an immediate question, how do our results change when incorporating feedback that is both left- and right-conditioned?
Indeed, we saw that in such cases, many basic properties (such as the measurement basis being maximally entangled) are no longer guaranteed, and it is unclear whether a rigid framework can exist.
More broadly, what states become accessible with multiple rounds of measurements\footnote{Studies of fixed point states in higher dimensions have revealed an interesting hierarchy in such scenarios \cite{LREfromSPT,verresen2021efficiently,Bravyi22,hierarchy}.}?
What if one gives up on the `measurement-only' condition and incorporates unitary evolution as part of the ingredients (either Hamiltonian- or gate-based)?
Moreover, what can be said generally in higher dimensions (of which special examples are worked out in our companion work \cite{sahay2024finite})?
In addition, if one relaxes the condition of deterministic preparation \cite{piroli2024approximate} or exact preparation \cite{piroli2021locc}, what can be achieved?
Finally, in making connections to other active fields of study, one could explore the present concepts in cases where the initial state already has interesting entanglement---which is the case for measurement-altered criticality \cite{PhysRevX.13.021026,PhysRevB.107.245132,PhysRevX.13.041042,PhysRevB.108.165120,sun2023new,ashida2023systemenvironment,ma2023exploring,PhysRevB.108.165120,paviglianiti2023enhanced,PRXQuantum.4.030317,sala2024quantum}---or where the measurement pattern forms a non-trivial pattern in time---as in Floquet codes \cite{Hastings2021dynamically,vuillot2021planar,PhysRevB.106.085122,aasen2023faulttolerant,PRXQuantum.4.020341,Townsend_Teague_2023,PhysRevB.108.195134,PhysRevB.108.205116,davydova2023quantum,zhu2023qubit,aasen2023measurement,kobayashi2024crosscap,PRXQuantum.4.010310,PhysRevLett.132.070401,PRXQuantum.5.010342,PRXQuantum.5.020305, bombin2023unifying} or in deep monitored circuits \cite{PhysRevX.9.031009,PhysRevB.98.205136,PhysRevB.101.104302,PhysRevB.103.104306,PhysRevB.100.134306,PhysRevB.99.224307,10.21468/SciPostPhys.7.2.024,PhysRevX.10.041020,PhysRevLett.125.070606,PhysRevB.101.104301,PRXQuantum.2.010352,monitored_review,Potter_2022,vijay2020measurementdriven,PhysRevB.103.174309,yoshida2021decoding,lavasani2022monitored,PhysRevB.101.060301,PhysRevResearch.2.033017,PhysRevB.108.094304,tikhanovskaya2023universality,hauser2023continuous,PRXQuantum.4.010331,suzuki2023quantum,PhysRevResearch.3.023200,Lavasani_2021,PhysRevLett.129.080501,PhysRevX.11.011030}.
A more complete understanding of any of these questions would help further foliate and organize the landscape of quantum states and provide deeper insights into the structure of accessible (i.e., physically occurring) quantum entanglement.

\vspace{5pt}

\emph{Note added:} The posting of this preprint to the arXiv was coordinated with simultaneous postings by Smith et al.~\cite{smith2024constant} and Stephen et al.~\cite{stephen2024fusing}. Both discuss the measurement-based preparation of matrix product states, and were developed independently from this work.

\emph{Note added in second version:} While preparing an updated manuscript to include Theorem 3, a preprint by Zhang et al.~\cite{zhang2024characterizing} appeared that also discusses the measurement-based preparation of tensor network states.
Our results agree where they intersect, notably our Eq.~\eqref{eq-Aclassificationalternative} (added in second version as special case of Theorem 3) appears as Theorem 6 in Ref.~\onlinecite{zhang2024characterizing}.

\section{Acknowledgements} We would like to thank Richard Allen, Soonwon Choi, Jordan Cotler, Ana Lyons, Francisco Machado, Nandagopal Manoj, Neelam Prasad, Pablo Sala, Tomohiro Soejima, Ashvin Vishwanath, and Josephine Yu for input and feedback on parts of this manuscript.
We also thank user8675309 for related discussions on Mathematics Stack Exchange.
R.S. acknowledges support from the U.S. Department of Energy, Office of Science, Office of Advanced Scientific Computing Research, Department of Energy Computational Science Graduate Fellowship under Award Number
DESC0022158.
R.V. is supported by the
Simons Collaboration on Ultra-Quantum Matter, which
is a grant from the Simons Foundation (618615, Ashvin
Vishwanath).

\bibliography{refs}

\pagebreak
\onecolumngrid
\appendix

\section{Additional Details for Motivating Examples \label{app:examples}}

In this appendix, we provide some additional details regarding the motivating examples provided in the main text.
Specifically, we provide derivations for the tensor networks of both the deformed cluster state and the deformed trivial state.
Furthemore, by computing the correlation spectrum of each state as a function of $\beta$, we justify claims made regarding their correlations in the main text.

\subsection{Additional Details for Deformed GHZ State}
In the main text we mentioned that this state has a full correlation spectrum (for $\beta \neq 0$). To be more precise, we calculate its correlation spectrum (i.e., eigenvalues of the transfer matrix) to be:
\begin{equation}
\lambda(T) = \{ 1 ,1 ,\tanh(2\beta), \tanh(2\beta)\}.
\end{equation}
The degeneracy highlights that the state is long-range entangled. The subleading values give rise to a correlation length $\xi = 1/|\ln\tanh(2\beta)|$. This is picked up, e.g., by the string correlation function of the $\prod_n X_n$ symmetry.

\subsection{Additional Details for Deformed SPT State}
To see that the protocol in Sec.~\ref{subsec-deformedcluster} leads to $e^{\beta \sum_n X_n} \ket{\textrm{cluster}}_N$ if all measurement outcomes are the ideal Bell state, let us first note that at $\beta = 0$, the state satisfies:
\begin{equation}
      \begin{tikzpicture}[scale = 1, baseline = {([yshift=-.5ex]current bounding box.center)}]
        \foreach \i in {0, ..., 2} {
        \draw[rounded corners, fill = lightdodgerblue] (\i*1.3,0) rectangle (\i*1.3 + 1, 0.4) {};
        \node at (0.5 + \i * 1.3, 0.2) {\small $\text{GHZ}_3$};
        \draw[color = black] (0.1 + 1.3*\i, 0) -- (0.1 + 1.3*\i, -0.5);
        \draw[color = black] (0.5 + 1.3*\i, 0) -- (0.5 + 1.3*\i, -0.65);
        \draw[color = black] (0.9 + 1.3*\i, 0) -- (0.9 + 1.3*\i, -0.5);
        % \draw[color = black] (1.4, 0) -- (1.4, -0.5);
        % \draw[color = black] (1.8, 0) -- (1.8, -0.65);
        % \draw[color = black] (2.2, 0) -- (2.2, -0.25);
        %\draw[fill = white] (0.5 + 1.3*\i, -0.3) circle (0.1);
        %\draw[fill = white] (1.8, -0.3) circle (0.1);
        % \node at (1.15, -0.4) { \scriptsize $\bra{\mathds{1}}$};
        };
        \foreach \i in {0, ..., 1} {
            \draw[rounded corners, fill = lightishgray] (0.8 + 1.3*\i,-0.25 - 0.25) rectangle (1.5 + 1.3*\i, -0.55 -0.25) {};
        }
        \node at (1.15, -0.65) { \scriptsize $\bra{\mathds{1}}$};
        \node at (1.15 + 1.3, -0.65) { \scriptsize $\bra{\mathds{1}}$};
        \node at (1.8, -0.65) {\small $\textcolor{red}{X}$};
        \node at (0.5, -0.65) {\small $\textcolor{red}{Z}$};
        \node at (3.1, -0.65) {\small $\textcolor{red}{Z}$};
        \draw[fill = lightorange(ryb)] (0.9, -0.3) circle (0.1);
        \draw[fill = lightorange(ryb)] (2.2, -0.3) circle (0.1);
        %\node at (0.9, -0.25) {\small $H$};
        %\node at (2.2, -0.25) {\small $H$};
        %\node at (1.15 + 2.6, -0.4) { \scriptsize $\bra{ZX}$};
     \end{tikzpicture} =       \begin{tikzpicture}[scale = 1, baseline = {([yshift=-.5ex]current bounding box.center)}]
        \foreach \i in {0, ..., 2} {
        \draw[rounded corners, fill = lightdodgerblue] (\i*1.3,0) rectangle (\i*1.3 + 1, 0.4) {};
        \node at (0.5 + \i * 1.3, 0.2) {\small $\text{GHZ}_3$};
        \draw[color = black] (0.1 + 1.3*\i, 0) -- (0.1 + 1.3*\i, -0.5);
        \draw[color = black] (0.5 + 1.3*\i, 0) -- (0.5 + 1.3*\i, -0.65);
        \draw[color = black] (0.9 + 1.3*\i, 0) -- (0.9 + 1.3*\i, -0.5);
        % \draw[color = black] (1.4, 0) -- (1.4, -0.5);
        % \draw[color = black] (1.8, 0) -- (1.8, -0.65);
        % \draw[color = black] (2.2, 0) -- (2.2, -0.25);
        %\draw[fill = white] (0.5 + 1.3*\i, -0.3) circle (0.1);
        %\draw[fill = white] (1.8, -0.3) circle (0.1);
        % \node at (1.15, -0.4) { \scriptsize $\bra{\mathds{1}}$};
        };
        % \foreach \i in {0, ..., 1} {
        \draw[rounded corners, fill = error-red] (0.8 + 1.3*0,-0.25 - 0.25) rectangle (1.5 + 1.3*0, -0.55 -0.25) {};
        \draw[rounded corners, fill = error-red] (0.8 + 1.3*1,-0.25 - 0.25) rectangle (1.5 + 1.3*1, -0.55 -0.25) {};
        % }
        %EFB2B6
        \node at (1.15, -0.65) { \scriptsize $\bra{X}$};
        \node at (1.15 + 1.3, -0.65) { \scriptsize $\bra{Z}$};
        \node at (1.8, -0.65) {\small $\ $};
        \node at (0.5, -0.65) {\small $\textcolor{red}{Z}$};
        \node at (3.1, -0.65) {\small $\textcolor{red}{Z}$};
        \draw[fill = lightorange(ryb)] (0.9, -0.3) circle (0.1);
        \draw[fill = lightorange(ryb)] (2.2, -0.3) circle (0.1);
        %\node at (0.9, -0.25) {\small $H$};
        %\node at (2.2, -0.25) {\small $H$};
        %\node at (1.15 + 2.6, -0.4) { \scriptsize $\bra{ZX}$};
     \end{tikzpicture}
\end{equation}
which followed from using the property of the GHZ that $X_{c_x} \ket{\text{GHZ}_3} = X_{l_{x}} X_{r_{x}} \ket{\text{GHZ}_3}$ and the property of the Hadamard that $XH = HZ$.
By using the property of the GHZ that $Z_{c_x} \ket{\text{GHZ}_3}_x = Z_{r_x/l_x} \ket{\text{GHZ}_3}_x$, we can thus derive that: 
\begin{equation}
     \begin{tikzpicture}[scale = 1, baseline = {([yshift=-.5ex]current bounding box.center)}]
        \foreach \i in {0, ..., 2} {
        \draw[rounded corners, fill = lightdodgerblue] (\i*1.3,0) rectangle (\i*1.3 + 1, 0.4) {};
        \node at (0.5 + \i * 1.3, 0.2) {\small $\text{GHZ}_3$};
        \draw[color = black] (0.1 + 1.3*\i, 0) -- (0.1 + 1.3*\i, -0.5);
        \draw[color = black] (0.5 + 1.3*\i, 0) -- (0.5 + 1.3*\i, -0.65);
        \draw[color = black] (0.9 + 1.3*\i, 0) -- (0.9 + 1.3*\i, -0.5);
        % \draw[color = black] (1.4, 0) -- (1.4, -0.5);
        % \draw[color = black] (1.8, 0) -- (1.8, -0.65);
        % \draw[color = black] (2.2, 0) -- (2.2, -0.25);
        %\draw[fill = white] (0.5 + 1.3*\i, -0.3) circle (0.1);
        %\draw[fill = white] (1.8, -0.3) circle (0.1);
        % \node at (1.15, -0.4) { \scriptsize $\bra{\mathds{1}}$};
        };
        % \foreach \i in {0, ..., 1} {
        %     \draw[rounded corners, fill = lightishgray] (0.8 + 1.3*\i,-0.25 - 0.25) rectangle (1.5 + 1.3*\i, -0.55 -0.25) {};
        % }
        \draw[rounded corners, fill = error-red] (0.8 + 1.3*0,-0.25 - 0.25) rectangle (1.5 + 1.3*0, -0.55 -0.25) {};
        \draw[rounded corners, fill = error-red] (0.8 + 1.3*1,-0.25 - 0.25) rectangle (1.5 + 1.3*1, -0.55 -0.25) {};
        \node at (1.15, -0.65) { \scriptsize $\bra{X}$};
        \node at (1.15 + 1.3, -0.65) { \scriptsize $\bra{Z}$};
        \node at (1.8, -0.65) {\small $\ $};
        \node at (0.5, -0.65) {\small $\textcolor{red}{Z}$};
        \node at (3.1, -0.65) {\small $\textcolor{red}{Z}$};
        \draw[fill = lightorange(ryb)] (0.9, -0.3) circle (0.1);
        \draw[fill = lightorange(ryb)] (2.2, -0.3) circle (0.1);
        %\node at (0.9, -0.25) {\small $H$};
        %\node at (2.2, -0.25) {\small $H$};
        %\node at (1.15 + 2.6, -0.4) { \scriptsize $\bra{ZX}$};
     \end{tikzpicture} =      \begin{tikzpicture}[scale = 1, baseline = {([yshift=-.5ex]current bounding box.center)}]
        \foreach \i in {0, ..., 2} {
        \draw[rounded corners, fill = lightdodgerblue] (\i*1.3,0) rectangle (\i*1.3 + 1, 0.4) {};
        \node at (0.5 + \i * 1.3, 0.2) {\small $\text{GHZ}_3$};
        \draw[color = black] (0.1 + 1.3*\i, 0) -- (0.1 + 1.3*\i, -0.5);
        \draw[color = black] (0.5 + 1.3*\i, 0) -- (0.5 + 1.3*\i, -0.65);
        \draw[color = black] (0.9 + 1.3*\i, 0) -- (0.9 + 1.3*\i, -0.5);
        % \draw[color = black] (1.4, 0) -- (1.4, -0.5);
        % \draw[color = black] (1.8, 0) -- (1.8, -0.65);
        % \draw[color = black] (2.2, 0) -- (2.2, -0.25);
        %\draw[fill = white] (0.5 + 1.3*\i, -0.3) circle (0.1);
        %\draw[fill = white] (1.8, -0.3) circle (0.1);
        % \node at (1.15, -0.4) { \scriptsize $\bra{\mathds{1}}$};
        };
        \foreach \i in {0, ..., 1} {
            \draw[rounded corners, fill = lightishgray] (0.8 + 1.3*\i,-0.25 - 0.25) rectangle (1.5 + 1.3*\i, -0.55 -0.25) {};
        }
        \node at (1.15, -0.65) { \scriptsize $\bra{\mathds{1}}$};
        \node at (1.15 + 1.3, -0.65) { \scriptsize $\bra{\mathds{1}}$};
        \node at (1.8, -0.65) {\small $\ $};
        % \node at (0.5, -0.65) {\small $Z$};
        % \node at (3.1, -0.65) {\small $Z$};
        %\node at (0.9, -0.25) {\small $H$};
        %\node at (2.2, -0.25) {\small $H$};
        \draw[fill = lightorange(ryb)] (0.9, -0.3) circle (0.1);
        \draw[fill = lightorange(ryb)] (2.2, -0.3) circle (0.1);
        %\node at (1.15 + 2.6, -0.4) { \scriptsize $\bra{ZX}$};
     \end{tikzpicture}
\end{equation}
Hence Eq.~\eqref{eq-decoupled-cluster} is the cluster state at $\beta = 0$,  and hence is the desired wavefunction for general $\beta$.

Moreover, we determine its correlation spectrum (i.e., eigenvalues of the transfer matrix) to be:
\begin{equation}
\lambda(T) = \{ 1 , \sqrt{\tanh(2\beta)}, - \sqrt{\tanh(2\beta)} , - \tanh{2\beta} \}.
\end{equation}
This state hence has several non-trivial correlation lengths, the largest of which is $\xi = 2/| \ln \tanh(2 \beta)|$.

\subsection{Additional Details for Deformed Trivial State}

Now for a simple non-SPT example, we remark that we can prepare the state $e^{\beta ZZ}\ket{+}^{\otimes N}$.
Such a state is strictly in the trivial phase for all values of $\beta< \infty$.
To see the tensor network representation of this state, we remark that: 
\begin{equation}
    \begin{tikzpicture}[scale = 1, baseline = {([yshift=-.5ex]current bounding box.center)}] 
        \draw[fill = lightorange(ryb), rounded corners] (0, 0) rectangle (1, 0.5);
        \draw[color = black] (0.1, 0) -- (0.1, -0.3);
        \draw[color = black] (0.9, 0) -- (0.9, -0.3);
        \draw[color = black] (0.1, 0.5) -- (0.1, 0.8);
        \draw[color = black] (0.9, 0.5) -- (0.9, 0.8);
        \node at (0.5, 0.25) {\small $e^{\beta ZZ}$};
    \end{tikzpicture}\quad =\quad 
    \begin{tikzpicture}[scale = 1, baseline = {([yshift=-.5ex]current bounding box.center)}] 
        %\filldraw[color = orange(ryb), rounded corners, thick] (0, 0) rectangle (1, 0.5);
        \draw[color = black] (0.1, 0.8) -- (0.1, -0.3);
        \draw[color = black] (0.9, 0.8) -- (0.9, -0.3);
        \draw[color = black] (0.1, 0.25) -- (0.9, 0.25);
        \draw[fill = white] (0.5, 0.25) circle (0.1);
        %\draw[color = black] (0.1, 0.5) -- (0.1, 0.8);
        %\draw[color = black] (0.9, 0.5) -- (0.9, 0.8);
        %\node at (0.5, 0.25) {\small $e^{\beta ZZ}$};
    \end{tikzpicture}
\end{equation}
where the red circle is the matrix:
\begin{equation}
    \begin{pmatrix}
        e^{\beta} & e^{-\beta} \\
        e^{-\beta} & e^{\beta}
    \end{pmatrix} = e^{\beta} + e^{-\beta}X \propto e^{\alpha X}
\end{equation}
where $\alpha = \text{arctanh}\left(e^{-2\beta}\right)$.
With this in mind, note that the full tensor network will look like:
\begin{equation}
    \ket{\Psi} =\ \begin{tikzpicture}[scale = 1, baseline = {([yshift=-.5ex]current bounding box.center)}] 
    \draw[color = black] (-3.5, 0) -- (3.5, 0);
    \foreach \i in {-3,...,3}{
        \draw[color = black] (\i, 0) -- (\i, -0.7);
    }
    \foreach \i in {-3,...,3}{
        \draw[fill = white] (\i + 0.3, 0) circle (0.1);
    }
    \end{tikzpicture}
\end{equation}

Moreover, we determine its correlation spectrum (i.e., eigenvalues of the transfer matrix) to be:
\begin{equation}
\lambda(T) = \{ 1 , \tanh(2\beta), 0, 0 \}.
\end{equation}
Unlike the other two examples, this state has non-trivial observables with zero correlation length. As discussed in the main text, this is intimately related to the fact that this state can be prepared where some errors are \emph{local} rather than \emph{topological}.

\section{Proof of Resource Theorem and Local Tensor Characterization \label{app:resourcetheorem}}

This appendix is devoted to proving the resource and local tensor characterizations theorems of Sec.~\ref{sec:setup} (Theorems 1 and 2).
Specifically, for the readers convenience, we restate the two theorems:

\begin{shaded}
    \textbf{Theorem 1} (Resources for Gluable Quantum States) Suppose that $\ket{\Psi}$ is a translation-invariant right-gluable quantum state.
    Then the following are true: 
    \begin{enumerate}
        \item[(1)] $\ket{\Psi}$ has an exact matrix product state description

        \item[(2)] the unentangled clusters in the state preparation protocol $\ket{\psi}$ are its matrix product state tensors in canonical form (labeled $A$) up to a tensor product of unitaries acting on each qudit of the cluster.
        In other words, 
        \begin{equation} %\label{eq-uLuPuRA}
            \begin{tikzpicture}[scale = 1, baseline = {([yshift=-.5ex]current bounding box.center)}] 
            \draw[color = black] (-1, 0) -- (1, 0);
            \draw[color = black] (0, 0) -- (0, -0.8);
            \draw[fill = lightdodgerblue] (0,0) circle (0.3);
            \node at (-0.0,0) {\small $\psi$};
            %\node at (-1.3, 0) {\small $u_L$ };
            %\node at (1.5, 0) {\small $u_R$ }; %\\
            \node at (0, -1.0) {\small $\ $};
            %\node at (-1.3, 0) {\small $V^{[n]}$};
            %\node at (0, -1.2) {\small $\ $};
     %       \node at (-1.4, 0) {\small $\{V_{\alpha} \}$ };
        \end{tikzpicture} = \begin{tikzpicture}[scale = 1, baseline = {([yshift=-.5ex]current bounding box.center)}] 
            \draw[color = black] (-1, 0) -- (1, 0);
            \draw[color = black] (0, 0) -- (0, -0.8);
            \draw[fill = lightdodgerblue] (0,0) circle (0.3);
            \node at (-0.03,0) {\small $A$};
            \node at (-1.3, 0) {\small $u_L$ };
            \node at (1.5, 0) {\small $u_R$ };
            \node at (0, -1.0) {\small $u_P$};
        \end{tikzpicture}
    \end{equation}
    where $A$ is the MPS in right-canonical form and $u_L, u_R,$ and $u_P$ are unitaries (which are determined by the feedback protocol and measurement basis).

    \item[(3)] The measurement basis $\mathcal{V}$ used for the preparation is maximally entangled.
    Equivalently, viewed as operators, the measurement basis is a unitary error basis.
    \end{enumerate}
\end{shaded}

\begin{shaded}
    \textbf{Theorem 2} (Tensor Characterization for Gluable States) 
    A translation-invariant state $\ket{\Psi}$ is right-gluable if and only if its matrix product state representation $A$ admits an error basis of right-pushable operators.
    
    This condition is equivalent to the existence of $\chi^2$ trace-orthonormal operators $\{V_\alpha\}$ such that
\begin{equation}
        \begin{tikzpicture} [scale = 1, baseline = {([yshift=-.5ex]current bounding box.center)}] 
    
    \draw[color = black] (-0.6, 0) -- (0.6, 0); %  top horizontal bar
    \draw[color = black] (-0.6, -1.2) -- (0.6, -1.2); %  bottom horizontal bar
    \draw[color = black] (0, 0) -- (0, -1.2); % vertical bar
     %\draw[color = black] (-0.8, 0) -- (-0.8, -1.2);
    % \draw[color = black] (-1, 0) -- (-1, -1.2);
    \draw[fill = lightdodgerblue] (0,0) circle (0.3); % circle
    \draw[fill = lightdodgerblue] (0,-1.2) circle (0.3); % circle
    %% Text
    \node at (-0.0,0) {\small $A$};
    \node at (-0.0,-1.2) {\small $\bar{A}$};
    %\draw[fill = lightishgray] (-1,0) circle (0.4);
    \node at (-1, 0) {\small $V_{\alpha}^{[n]}$};
    %\draw[fill = lightishgray] (-1,-1.2) circle (0.4);
    \node at (-1, -1.2) {\small $\bar{V}_{\alpha}^{[n]}$};
    \end{tikzpicture}  \; \; = \; \;
        \begin{tikzpicture} [scale = 1, baseline = {([yshift=-.5ex]current bounding box.center)}] 
    
    \draw[color = black] (-0.6, 0) -- (0.6, 0); %  top horizontal bar
    \draw[color = black] (-0.6, -1.2) -- (0.6, -1.2); 
    \draw[color = black] (0, 0) -- (0, -1.2); % vertical bar
     %\draw[color = black] (-0.8, 0) -- (-0.8, -1.2);
    % \draw[color = black] (-1, 0) -- (-1, -1.2);
    \draw[fill = lightdodgerblue] (0,0) circle (0.3); % circle
    \draw[fill = lightdodgerblue] (0,-1.2) circle (0.3); % circle
    %\draw[fill = lightishgray] (1.05,0) circle (0.5);
    \node at (1.05, 0) {\small $V^{[n+1]}_{\alpha}$};
    %\draw[fill = lightishgray] (1.05,-1.2) circle (0.5);
    \node at (1.05, -1.2) {\small $\bar{V}^{[n+1]}_{\alpha}$};
    %% Text
    \node at (-0.0,0) {\small $A$};
    \node at (-0.0,-1.2) {\small $\bar{A}$};
    \end{tikzpicture} 
\end{equation}
where $\chi$ is the bond dimension of the MPS tensor, $n$ is any non-negative integer, and $V_\alpha^{[0]} = V_\alpha$.
\end{shaded}

\noindent
Since the proof is rather lengthy, we organize the following logical sections
\begin{description}
    \item[Subsection 1] We start by providing a proof of claim (1) of Theorem 1---every gluable quantum state is a matrix product state.

    \item[Subsection 2] Subsequently, it will be necessary to prove Theorem 2, which shows that gluable is equivalent to a local criteria on the MPS tensors.

    \item[Subsection 3] We conclude by claims (2) and then (3), given the ingredients in the previous subsections.
\end{description}

\subsection{Every Gluable Quantum State is a Matrix Product State \label{app:reduction}}

The proof of claim (1) of Theorem 1 proceeds in two stages.
First, we prove that if a state is gluable, then its measurement preparation protocol can be viewed as performing nearest-neighbor measurements between disentangled clusters.
We subsequently show that this setting naturally produces matrix product states.

\subsubsection{Reduction to Nearest-Neighbor Measurements}

To reduce a more general measurement scheme to a nearest-neighbor measurement scheme, let us recall that definition, measurements between clusters can only occur with some maximal range $R$, measured in units of the distance between clusters (e.g. in Fig.~\ref{fig:reduction_to_MPS} (a, top row), $R = 2$).
By then blocking $R$ clusters together, we can convert this to a protocol with at most nearest-neighbor measurements [in units of the new clusters, cf. Fig.~\ref{fig:reduction_to_MPS}(a, second row)].
However, at this stage, recognize that there will be measurements that occur within clusters, unlike the setting considered in our examples.
To mitigate this, we can perform two tricks.
\begin{enumerate}
    \item[(1)] First, we re-order the measured qudits such that all inter-cluster and right-neighbor measured qudits are to the right of all the unmeasured qudits, which are to the right of all left-neighbor measured qudits [ as shown in  Fig.~\ref{fig:reduction_to_MPS}(a, third row)]. 
    Note that this inter-cluster re-ordering of the measured qudits has no physical effect on the resulting state as we discard the measured qudits in our measurement protocol.

    \item[(2)] Second, we can ``group'' measurements together to eliminate all inter-cluster measurements [shown in the last panel of Fig.~\ref{fig:reduction_to_MPS}(a)].
\end{enumerate}
The result is a set of disentangled clusters with nearest-neighbor measurements with non-overlapping geometric support, identical to all the examples that we considered [shown in Fig.~\ref{fig:reduction_to_MPS}(b)].
With this simplification, left conditioning is equivalent to saying that the correction unitary for a given unmeasured qudit, depends on the measurement outcome for qudits states to the left of it.

% %

\subsubsection{Gluable Quantum States are Matrix Product States}

Given the reduction to nearest neighbor measurements, we are prepared to prove claim (1) of Theorem 1.
\begin{shaded}
    \textbf{Theorem B.1} [Claim (1) of Theorem 1] Any gluable quantum state $\ket{\Psi}$ has an exact matrix product state description.
\end{shaded}

\textit{Proof.} Suppose that a global state $\ket{\Psi}$ is gluable.
Then this means that there exists a measurement basis $|\widetilde{\mathcal{V}}\rangle = \{|\widetilde{V}_{\alpha}\rangle\}$ and an associated set of correction unitaries $U^{\mathbf{m}}_x$ (with $\mathbf{m} = \{\alpha_x = 1, \cdots \chi^2\}$ denoting the measurement record) that enable deterministically preparing $\ket{\Psi}$.
As a remark, for notational convenience, we take $x \in \mathbb{Z}$, i.e. the chain is infinite, though the analysis below is identical for the open boundary condition case.
To set notation, let us denote the initial state as: 
\begin{equation}
    \ket{\Psi_0} = \bigotimes_{x \in \mathbb{Z}} \ket{\psi}_{(x_L) x (x_R)} = \cdots \begin{tikzpicture}[scale = 1, baseline = {([yshift=-.5ex]current bounding box.center)}]
        \foreach \i in {0, ..., 2} {
        \draw[rounded corners, fill = lightdodgerblue] (\i*1.3,0) rectangle (\i*1.3 + 1, 0.4) {};
        \node at (0.5 + \i * 1.3, 0.2) {\small $\psi_\i$};
        \draw[color = black] (0.1 + 1.3*\i, 0) -- (0.1 + 1.3*\i, -0.4);
        \draw[color = black] (0.5 + 1.3*\i, 0) -- (0.5 + 1.3*\i, -0.4);
        \draw[color = black] (0.9 + 1.3*\i, 0) -- (0.9 + 1.3*\i, -0.4);
        % \draw[color = black] (1.4, 0) -- (1.4, -0.5);
        % \draw[color = black] (1.8, 0) -- (1.8, -0.65);
        % \draw[color = black] (2.2, 0) -- (2.2, -0.25);
        %\draw[fill = white] (0.5 + 1.3*\i, -0.2) circle (0.1);
        %\draw[fill = white] (1.8, -0.3) circle (0.1);
        % \node at (1.15, -0.4) { \scriptsize $\bra{\mathds{1}}$};
        };
        %\node at (2.7, -0.3) {\small $Z$};
        %\node at (1.15 + 2.6, -0.4) { \scriptsize $\bra{ZX}$};
     \end{tikzpicture} \cdots
\end{equation}
Here, $x_L$ and $x_R$ label the qubits to be measured to the left and the right respectively.
Note that the clusters $\ket{\psi_x}$ can in principle depend on position $x$.
Once again, for sake of notation we choose to drop this label but it can be restored easily.
With the above, we have that for every measurement outcome $\mathbf{m}$, we have that:
\
\begin{align}
    \ket{\Psi} = \bigotimes_{x \in \mathbb{Z}} (U^{\mathbf{m}}_x)^{\dagger} \langle \widetilde{V}^{\alpha_x}|_{(x-1)_R\, x_L}  \left[\bigotimes_{x \in \mathbb{Z}} \ket{\psi}_{(x_L)\, x\, (x_R)} \right] &= \sum_{\mathbf{s}} \prod_{x \in \mathbb{Z}} [(U^{\mathbf{m}}_{x})^{\dagger}]_{s_x s'_x}  \widetilde{V}_{i_{(x-1)_R} i_{x_{L}}}^{\alpha_x} \psi_{i_{x_L} \, i_{x_R}}^{s'_x}  \ket{\mathbf{s}} \\
    &= \begin{tikzpicture}[scale = 0.8, baseline = {([yshift=-.5ex]current bounding box.center)}] 
    \draw[color = black] (-1.5,0) -- (7.5 -1.7, 0);
    \foreach \i in {0, ..., 2}{
        \draw[color = black] (2*\i,0) -- (2*\i, -1.3);
        \draw[fill = lightdodgerblue] (2*\i,0) circle (0.3);
        \node at (2*\i, 0) {\small $\psi$};
        \draw[fill = lightishgray] (2*\i+ 1,0) circle (0.3);
        \draw[fill = lightorange(ryb)] (2*\i,-0.8) circle (0.25);
    }
    \node at (2*0 + 1, 0) {\scriptsize  $\widetilde{V}^{\alpha}$};
    \node at (2*1 + 1, 0) {\scriptsize  $\widetilde{V}^{\beta}$};
    \node at (2*2 + 1, 0) {\scriptsize  $\widetilde{V}^{\gamma}$};
    \node at (-1.8, -0.02) {\small $\cdots$};
    \node at (2*0, -0.8) {\tiny  $U^{\mathbf{m}}_0$};
    \node at (2*1, -0.8) {\tiny  $U^{\mathbf{m}}_1$};
    \node at (2*2, -0.8) {\tiny  $U^{\mathbf{m}}_2$};
    \node at (2*0 + 0.3, -0.65) {\tiny  $\dagger$};
    \node at (2*1 + 0.3, -0.65) {\tiny  $\dagger$};
    \node at (2*2 + 0.3, -0.65) {\tiny  $\dagger$};    
    \node at (6.2, -0.02) {\small $\cdots$};
\end{tikzpicture}
\end{align}
where in the last line, we implicitly sum over repeated indices and $\mathbf{s} = \{s_x\}$ labels the local Hilbert space dimension of the unmeasured qubits, with $s_x = 0, \cdots d-1$ and the use of upper and lower indices having no meaning.
Note that the above automatically implies a matrix product state representation of the state.
In particular, let the tensor at location $x$ be equal to 
\begin{equation} \label{eq-Afrompsi}
    [A^\mathbf{m}_x]^{s_x}_{i_{x_L} i_{(x + 1)_L}} = [U^{\mathbf{m}}_{x}]^{\dagger}_{s_x s'_x}  \widetilde{V}_{i_{(x-1)_R} i_{x_{L}}}^{\alpha_x} \psi_{i_{x_L} \, i_{x_R}}^{s'_x} = \begin{tikzpicture}[scale = 0.8, baseline = {([yshift=-.5ex]current bounding box.center)}] 
    \draw[color = black] (-1.5,0) -- (1.0, 0);
    \foreach \i in {0, ..., 0}{
        \draw[color = black] (2*\i,0) -- (2*\i, -1.3);
        \draw[fill = lightdodgerblue] (2*\i,0) circle (0.3);
        \node at (2*\i, 0) {\small $\psi$};
        \draw[fill = lightishgray] (2*\i- 1,0) circle (0.3);
        \draw[fill = lightorange(ryb)] (2*\i,-0.8) circle (0.25);
    }
    \node at (2*0 - 1, 0) {\scriptsize  $\widetilde{V}^{\alpha}$};
    \node at (2*0, -0.8) {\tiny  $U^{\mathbf{m}}_x$};
    \node at (2*0 + 0.3, -0.65) {\tiny  $\dagger$}; 
\end{tikzpicture}
\end{equation}
Note that, in the above representation, the tensor may depend on $\mathbf{m}$ but, by design, the global state is not.

\hspace{0.95\textwidth} $\blacksquare$ 

\vspace{5 mm}

As explained in Sec.~\ref{subsec:setup} of the main text, we can always presume that there exists a measurement outcome $\mathbf{m}_0$ such the post-measurement state is the desired state, i.e., there is no need for correction in that particular case. Indeed, Eq.~\eqref{eq-Afrompsi} confirms this point, since one can perform a basis transformation on our initial $\ket{\psi}$ to absorb the correcting unitary for a given outcome. Without loss of generality, we can take this `ideal' measurement outcome to be $\mathbf{m}_0 = \mathbf{1} = (\cdots, 1, 1, 1, \cdots )$. The MPS tensor is thus equal to
\begin{equation}%\label{eq-Ainit}
    [A]^{s_x}_{i_{(x - 1)_R} i_{x_R}} =  \widetilde{V}_{i_{(x-1)_R} i_{x_{L}}}^{1} (\psi_x)_{i_{x_L} \, i_{x_R}}^{s_x'}
= \begin{tikzpicture}[scale = 0.8, baseline = {([yshift=-.5ex]current bounding box.center)}] 
    \draw[color = black] (-1.5,0) -- (0.7, 0);
    \foreach \i in {0, ..., 0}{
        \draw[color = black] (2*\i,0) -- (2*\i, -0.8);
        \draw[fill = lightdodgerblue] (2*\i,0) circle (0.3);
        \node at (2*\i, 0) {\small $\psi$};
        \draw[fill = lightishgray] (2*\i- 1,0) circle (0.3);
    };
    \node at (2*0 - 1, 0) {\scriptsize  $\widetilde{V}^{1}$};
\end{tikzpicture}
\end{equation}

Let us further remark that $\widetilde{V}^1$ must be invertible because if it were not, we could find smaller dimensional ``virtual qudits'', with which to create our initial product state clusters, contradicting the minimality condition in Sec.~\ref{sec:setup}.
Then, without loss of generality, we will write the measurement basis as $\widetilde{\mathcal{V}} = \mathcal{V}B = \{V_{\alpha}B\}$ with the property that $V_1 = \mathds{1}$ (achieved by setting $B = \widetilde{V}_1$). In summary, we have
\begin{equation} \label{eq-A}
 \begin{tikzpicture}[scale = 0.8, baseline = {([yshift=-.5ex]current bounding box.center)}] 
    \draw[color = black] (-0.7,0) -- (0.7, 0);
    \foreach \i in {0, ..., 0}{
        \draw[color = black] (2*\i,0) -- (2*\i, -0.8);
        \draw[fill = lightdodgerblue] (2*\i,0) circle (0.3);
        \node at (2*\i, 0) {\small $A$};;
    };
\end{tikzpicture}
=
 \begin{tikzpicture}[scale = 0.8, baseline = {([yshift=-.5ex]current bounding box.center)}] 
    \draw[color = black] (-1.5,0) -- (0.7, 0);
    \foreach \i in {0, ..., 0}{
        \draw[color = black] (2*\i,0) -- (2*\i, -0.8);
        \draw[fill = lightdodgerblue] (2*\i,0) circle (0.3);
        \node at (2*\i, 0) {\small $\psi$};
        \draw[fill = lightishgray] (2*\i- 0.9,0) circle (0.3);
    };
    \node at (2*0 - 0.9, 0) {\small  $B$};
\end{tikzpicture}
\end{equation}
with measurement outcomes leading to random insertions of $V_\alpha$ on the bonds of the matrix product state.

Finally, we remind the reader that in Sec.~\ref{sec:setup} we presume translation-invariance for convenience. More precisely, this means we will take $A$ in Eq.~\eqref{eq-A} to be site-independent. However, many of our proofs can be extended beyond this.

\subsection{Proof of Theorem 2: Local Criteria on Matrix Product State Tensors \label{app:pushable}}

We now turn to proving our local criteria for matrix product states.
To do so, our strategy will be to first prove that:
\begin{equation}
\begin{tikzpicture}[scale = 0.8, baseline = {([yshift=-.5ex]current bounding box.center)}] 
    \node at (-1.25, 0) {\small $V_{\alpha}$};
    \draw[color = black] (-1.0,0) -- (4, 0);
    \node at (4.4, 0) {\small $\cdots$};
    \foreach \i in {0, ..., 2}{
        \draw[color = black] (1.5*\i,0) -- (1.5*\i, -1);
        \draw[fill = lightdodgerblue] (1.5*\i,0) circle (0.3);
        \node at (1.5*\i, 0) {\small $A$};
        \node at (1.5*\i, -1.6) {$\ $};
    } 
\end{tikzpicture} = \begin{tikzpicture}[scale = 0.8, baseline = {([yshift=-.5ex]current bounding box.center)}] 
    %\node at (-1.25, 0) {\small $V_{\alpha}$};
    \draw[color = black] (-1.0,0) -- (4, 0);
    \node at (4.4, 0) {\small $\cdots$};
    \foreach \i in {0, ..., 2}{
        \draw[color = black] (1.5*\i,0) -- (1.5*\i, -1);
        \draw[fill = lightdodgerblue] (1.5*\i,0) circle (0.3);
        \node at (1.5*\i, 0) {\small $A$};
        \node at (1.5*\i, -1.3) {\small $U^{[\i]}_{\alpha}$};
    } 
\end{tikzpicture}
\end{equation}
for a single $V_{\alpha}$.
While the above might seem manifestly true given the assumption of left-conditioning, there is an important subtlety.
In particular, any measurement error with a zero probability, need not push through to a unitary (since we will never be required to correct such an error).
As such, we first prove that a single measurement error $V_{\alpha}$ has a non-zero probability (on an open chain geometry with generic boundary conditions) before proving the condition on the local tensor criteria.

\begin{shaded}
    \textbf{Lemma B.1.} For any translationally invariant gluable wavefunction $\ket{\Psi}$ defined on a finite system of length $L$ (assumed to be much larger than the correlation length\footnotemark) with particular open boundary conditions, the probability of measuring a single virtual error is non-zero.%
\end{shaded}
\footnotetext[\value{footnote}]{To be precise, by correlation length here, $\xi_{\lambda} = 1/\log(1/\lambda)$ where $\lambda$ is the largest non-identity eigenvalue of the transfer matrix.}

\textit{Proof.} Let us start by showing that, on a finite chain with open boundary conditions, the measurement outcome corresponding to a single virtual insertion $V_{\alpha}B$ in the center of the chain has a non-zero probability density of being measured.
Since we are predominantly interested in bulk properties, we let our target wavefunction have the following boundary conditions: 
\begin{equation}
    \ket{\Psi} = \frac{1}{\mathcal{N}}\begin{tikzpicture}[scale = 0.8, baseline = {([yshift=-.5ex]current bounding box.center)}] 
    \draw[color = black] (-1.5,0) -- (7.5, 0);
    \foreach \i in {0, ..., 3}{
        \draw[color = black] (2*\i,0) -- (2*\i, -1);
        \draw[fill = lightdodgerblue] (2*\i,0) circle (0.3);
        \node at (2*\i, 0) {\small $A$};
    }
    % \node at (7.5, 0) {$\cdots$};
    % \node at (-1.5, 0) {$\cdots$};
    %\node at (3, 0) {$V_{\alpha}$};
    \draw[color = black] (-1.5, 0) -- (-1.5, -1);
    \draw[color = black] (7.5, 0) -- (7.5, -1);
    \draw[fill = white] (-1.5, -0.5) circle (0.35);
    \draw[fill = white] (7.5, -0.5) circle (0.35);
    \node at (-1.5, -0.5) {\scriptsize $v_L^{1/2}$};
    \node at (7.5, -0.5) {\small $v_R^{1/2}$};
\end{tikzpicture}
\end{equation}
where $v_L$ and $v_R$ define vectors with non-zero overlap on the largest positive right and left eigenvectors of the transfer matrix, $R$ and $L$ and with no overlap on any other potential dominant eigenvectors. Note that this is generic in the case of short-range entangled phases, since they have a unique dominant eigenvector of the transfer matrix. There is thus only a non-trivial condition in the case of long-range entangled phases, in which case it exactly corresponds to the physical requirement that we obtain the appropriate bulk state in the thermodynamic limit.
For simplicity, we proceed by taking $v_L = L$ and $v_R = R$, though the proof carries through more generally under these conditions.
With this target state in mind, the post-measurement state with a single bulk measurement error is:
\begin{equation}
    p_{\alpha} = \frac{1}{\chi^{N} \mathcal{N}}     \begin{tikzpicture}[scale = 0.8, baseline = {([yshift=-.5ex]current bounding box.center)}] 
        \draw[color = black] (-1,0) -- (7, 0);
        \draw[color = black] (-1,-1) -- (7, -1);
        \foreach \i in {0, ..., 3}{
            \draw[color = black] (2*\i,0) -- (2*\i, -1);
            \draw[fill = lightdodgerblue] (2*\i,0) circle (0.3);
            \draw[fill = lightdodgerblue] (2*\i,-1) circle (0.3);
            \node at (2*\i, 0) {\small $A$};
            \node at (2*\i, -1) {\small $\bar{A}$};
        }
        \node at (7.5, 0) {$\cdots$};
        \node at (-1.5, 0) {$\cdots$};
        \node at (3, 0) {$V_{\alpha}$};
        \node at (7.5, -1) {$\cdots$};
        \node at (-1.5, -1) {$\cdots$};
        \node at (3, -1) {$\bar{V}_{\alpha}$};
    \end{tikzpicture}% \qquad \mathcal{N} =     \begin{tikzpicture} [scale = 0.75, baseline = {([yshift=-.5ex]current bounding box.center)}] 
%     \draw[color = black] (-0.8, 0) -- (0.8, 0); %  top horizontal bar
%     \draw[color = black] (-0.8, -1.2) -- (0.8, -1.2); %  bottom horizontal bar
%     \draw[color = black] (0, 0) -- (0, -1.2); % vertical bar
%     \draw[fill = lightorange(ryb)] (0,0) circle (0.3); % circle
%     \draw[fill = lightorange(ryb)] (0,-1.2) circle (0.3); % circle
%     \draw[color = black] (-0.8, 0) -- (-0.8, -1.2);
%     \draw[color = black] (0.8, 0) -- (0.8, -1.2);
%     \node at (-0.0,0) {\small $\psi$};
%     \node at (-0.0,-1.2) {\small $\bar{\psi}$};
%     \end{tikzpicture}
 \end{equation}
where $\mathcal{N}$ is the normalization for the initial state of decoupled clusters and $\chi$ arose from the normalization in Eq.~\eqref{eq-V}.
Now, provided that system size $N$ is sufficiently large, we have, by the power method:
\begin{equation}
    p_{\alpha} = \frac{1}{\chi^{N} \mathcal{N}} \begin{tikzpicture} [scale = 0.75, baseline = {([yshift=-.5ex]current bounding box.center)}] 
    \draw[color = black] (-0.8, 0) -- (0.8, 0); %  top horizontal bar
    \draw[color = black] (-0.8, -1.2) -- (0.8, -1.2); %  bottom horizontal bar
    %\draw[color = black] (0, 0) -- (0, -1.2); % vertical bar
    %\draw[fill = lightorange(ryb)] (0,0) circle (0.3); % circle
    %\draw[fill = lightorange(ryb)] (0,-1.2) circle (0.3); % circle
    \draw[color = black] (-0.8, 0) -- (-0.8, -1.2);
    \draw[color = black] (0.8, 0) -- (0.8, -1.2);
    \node at (-0.0,0) {\small $V_{\alpha}$};
    \node at (-0.0,-1.2) {\small $\bar{V}_{\alpha}$};
    \draw[fill = white] (-0.8,-0.6) circle (0.2); % circle
    \draw[fill = white] (0.8,-0.6) circle (0.2);
    \node at (0.8, -0.6) {\scriptsize $R$};
    \node at (-0.8, -0.6) {\scriptsize $L$};
    \end{tikzpicture} = \frac{1}{(\chi \mathcal{N})^{N}} \text{tr}\left\{[M^{-1}V_{\alpha} M] [M^{\dagger} V_{\alpha}^{\dagger} (M^{-1})^{\dagger}] \Lambda^2 \right\} = \frac{1}{(\chi\mathcal{N})^{N}} \text{tr}(W^{\dagger}_{\alpha} W_{\alpha} \Lambda^2) \neq 0 .
\end{equation}
Here, we used our assumption that $\text{rank}(\Lambda) = \chi$ and the subsequent existence of the canonical form to set  $L = M^{-1} \Lambda^2 (M^{-1})^{\dagger}$ and $R = M M^{\dagger}$, for some non-singular matrix $M$.
Note that in the last step, we used the fact that $W_{\alpha}^{\dagger} W_{\alpha} \Lambda^2$ is positive semi-definite and non-zero.
Hence the probability density of the outcome above is non-zero.

\hspace{0.95\textwidth} $\blacksquare$

Given this proof on the probabilities, we now prove Theorem 2: our local tensor criteria for gluable quantum states.

\begin{shaded}
    \textbf{Theorem B.2} (First Part of Theorem 2  of Main Text) 
    A translationally invariant matrix product state is gluable if and only if there exists an error basis of pushable operators.
\end{shaded}

\textit{Proof.} Let us remark that if there existed an error basis of pushable operators, then it is clear why the matrix product state is gluable.
As such, the only non-trivial direction is the ``only if'' direction.
If $\ket{\Psi}$ is preparable with left conditioned feedback, then we know that the unitaries used for correction $U_x^{\mathbf{m}} = U_x^{\mathbf{m}_{<x}}$, where $\mathbf{m}_{<x} = \{\alpha_{y < x}\}$.
Note that this naturally implies that $U^{\mathbf{1}_{<x}} = \mathds{1}$ (i.e. if there are only $1$\footnote{Recall that ``1'' was the measurement outcome that we took, without loss of generality, to be the identity.} outcomes appearing to the left of $x$, then the correction unitary at $x$ is the identity).
With this in mind, let us consider the measurement outcome $\mathbf{m} = \mathbf{1}_{< x}\alpha \mathbf{1}_{>x}$ (that is we measured $1$ everywhere except for at $x$ where we measured $\alpha$), which we know can happen according to the above lemma.
Then, we have that: 
\begin{equation}
    \begin{tikzpicture}[scale = 0.8, baseline = {([yshift=-.5ex]current bounding box.center)}] 
        \draw[color = black] (-1,0) -- (7, 0);
        \foreach \i in {0, ..., 3}{
            \draw[color = black] (2*\i,0) -- (2*\i, -1);
            \draw[fill = lightdodgerblue] (2*\i,0) circle (0.3);
            \node at (2*\i, 0) {\small $A$};
        }
        \node at (7.5, 0) {$\cdots$};
        \node at (-1.5, 0) {$\cdots$};
        \node at (3, 0) {$V_{\alpha}$};
    \end{tikzpicture}
    =     \begin{tikzpicture}[scale = 0.8, baseline = {([yshift=-.5ex]current bounding box.center)}] 
        \draw[color = black] (-1,0) -- (7, 0);
        \foreach \i in {0, ..., 3}{
            \draw[color = black] (2*\i,0) -- (2*\i, -1);
            \draw[fill = lightdodgerblue] (2*\i,0) circle (0.3);
            \node at (2*\i, 0) {\small $A$};
        }
        \node at (7.5, 0) {$\cdots$};
        \node at (-1.5, 0) {$\cdots$};
        \node at (4.1, -1.2) {$U_0({\alpha})$};
        \node at (6.1, -1.2) {$U_1(\alpha)$};
    \end{tikzpicture}
\end{equation}
where we have emphasized that $U_n(\alpha)$ is a function of $\alpha$ and a function of $n$, the distance away from where the first virtual defect occurred.
Now, note that, by the minimality assumption of gluable quantum states, the bonds of the matrix product state are full-rank.
In other words, the following map (which takes into account the boundary):
\begin{equation} \label{eq-GammaL}
    \Gamma_L^x: \textcolor{red}{\mathcal{H}_{\text{phys}}^{L}} \to \textcolor{orange}{\mathcal{H}_{\text{virtual}}^{L}} \qquad \qquad   \Gamma_L^x =\  \begin{tikzpicture}[scale = 0.75, baseline = {([yshift=-.5ex]current bounding box.center)}] 
        \draw[color = black] (-1.5,0) -- (7, 0);
        \foreach \i in {0, ..., 3}{
            \draw[color = black] (2*\i,0) -- (2*\i, -1);
            \draw[fill = lightdodgerblue] (2*\i,0) circle (0.3);
            \node at (2*\i, 0) {\small $A$};
            \draw[color = red, -stealth, thick] (2*\i + 0.3, -1) -- (2*\i + 0.3, -0.5);
        }
        \draw[color = red, -stealth, thick] (-1.3, -1) -- (-1.3, -0.5);
        \draw[color = orange, -stealth, thick] (6.8, -0.2) -- (7.3, -0.2);
        %\node at (7.5, 0) {$\cdots$};
        \draw[color = black] (-1.5, 0) -- (-1.5, -1);
        %\node at (3, 0) {$V_{\alpha}$};
    \end{tikzpicture}
\end{equation}
is surjective for all $x$. \footnote{Note that, if it was not, we could insert a projector into the virtual bond without affecting the state. Consequently, its bond dimension $\chi \geq \text{rank}(\Lambda)$ violating the minimality assumption. Further note this is completely distinct from the concept of MPS injectivity, which we are not assuming.}.
A similar condition holds for $\Gamma_R$.
Consequently, we have that:
\begin{equation}
\begin{tikzpicture}[scale = 0.8, baseline = {([yshift=-.5ex]current bounding box.center)}] 
    \node at (-1.25, 0) {\small $V_{\alpha}$};
    \draw[color = black] (-1.0,0) -- (4, 0);
    \node at (4.4, 0) {\small $\cdots$};
    \foreach \i in {0, ..., 2}{
        \draw[color = black] (1.5*\i,0) -- (1.5*\i, -1);
        \draw[fill = lightdodgerblue] (1.5*\i,0) circle (0.3);
        \node at (1.5*\i, 0) {\small $A$};
        \node at (1.5*\i, -1.6) {$\ $};
    } 
\end{tikzpicture} = \begin{tikzpicture}[scale = 0.8, baseline = {([yshift=-.5ex]current bounding box.center)}] 
    %\node at (-1.25, 0) {\small $V_{\alpha}$};
    \draw[color = black] (-1.0,0) -- (4, 0);
    \node at (4.4, 0) {\small $\cdots$};
    \foreach \i in {0, ..., 2}{
        \draw[color = black] (1.5*\i,0) -- (1.5*\i, -1);
        \draw[fill = lightdodgerblue] (1.5*\i,0) circle (0.3);
        \node at (1.5*\i, 0) {\small $A$};
        \node at (1.5*\i, -1.3) {\small $U^{[\i]}_{\alpha}$};
    } 
\end{tikzpicture}
\end{equation}
Now, note that because $\Gamma_R$ (the right analog of Eq.~\eqref{eq-GammaL}) is surjective, there exists a map $(\Gamma^x_R)^+$ such that $(\Gamma^x_R) (\Gamma^x_R)^+ = \mathds{1}_{\mathcal{H}_{\text{virtual}}^R}$ (i.e. it is a right-inverse of $\Gamma^x_R$).
This map is the so-called Moore-Penrose pseudoinverse of $\Gamma^x_R$ \footnote{In canonical form, the Moore-Penrose inverse of $\Gamma_x^R$ is simply $(\Gamma_x^R)^{\dagger}$.
Since we are not assuming canonical form (indeed, we later derive it), we work with the more general object.
}.
Now, we apply $(\Gamma_R^{x+1})^+$ to both sides of the above equation to get: 
\begin{equation} \label{eq-MoorePenrose}
        \begin{tikzpicture}[scale = 1, baseline = {([yshift=-.5ex]current bounding box.center)}] 
        \draw[color = black] (-1, 0) -- (1, 0);
        \draw[color = black] (0, 0) -- (0, -0.8);
        \draw[fill =  lightdodgerblue] (0,0) circle (0.3);
        \node at (-0.03,0) {\small $A$};
        \node at (-1.3, 0) {\small $V_{\alpha}$};
        \node at (0, -1.2) {\small $\ $};
 %       \node at (-1.4, 0) {\small $\{V_{\alpha} \}$ };
    \end{tikzpicture} = \begin{tikzpicture}[scale = 0.8, baseline = {([yshift=-.5ex]current bounding box.center)}] 
    \node at (-1.25, 0) {\small $V_{\alpha}$};
    \draw[color = black] (-1.0,0) -- (4, 0);
    \draw[color = black] (0.5,-1.0) -- (4, -1.0);
    %\node at (4.4, 0) {\small $\cdots$};
    \foreach \i in {0, ..., 2}{
        \draw[color = black] (1.5*\i,0) -- (1.5*\i, -1);
        \draw[fill = lightdodgerblue] (1.5*\i,0) circle (0.3);
        \node at (1.5*\i, 0) {\small $A$};
        \node at (1.5*\i, -1.6) {$\ $};
    } 
    \draw[color = black] (4, -1.0) -- (4, 0);
    \draw[fill = lightorange(ryb), rounded corners] (-0.5 + 1.5, -1.3) rectangle (4.2, -0.8);
    \node at  (2.7, -1) {\scriptsize $(\Gamma_R^{x + 1})^+$};
\end{tikzpicture} = \begin{tikzpicture}[scale = 0.8, baseline = {([yshift=-.5ex]current bounding box.center)}] 
    \node at (-1.25, 0) {\small $\ $};
    \draw[color = black] (-1.0,0) -- (4, 0);
    \draw[color = black] (0.5,-1.8) -- (4, -1.8);
    %\node at (4.4, 0) {\small $\cdots$};
    \foreach \i in {0, ..., 2}{
        \draw[color = black] (1.5*\i,0) -- (1.5*\i, -0.8);
        %\draw[color = black] (1.5*\i,-1.8) -- (1.5*\i, -1.4);
        \draw[fill = lightdodgerblue] (1.5*\i,0) circle (0.3);
        \node at (1.5*\i, 0) {\small $A$};
        % \node at (1.5*\i, -1.6) {$\ $};
        \node at (1.5*\i, -1.1) {\small $U^{[\i]}_{\alpha}$};
    } 
    \draw[color = black] (1.5*1,-1.8) -- (1.5*1, -1.4);
    \draw[color = black] (1.5*2,-1.8) -- (1.5*2, -1.4);
    \draw[color = black] (4, -1.0 - 0.8) -- (4, 0);
    \draw[fill = lightorange(ryb), rounded corners] (-0.5 + 1.5, -1.3 - 0.8) rectangle (4.2, -0.8 - 0.8);
    \node at  (2.7, -1 - 0.8) {\scriptsize $(\Gamma_R^{x + 1})^+$};
    \draw[color = red, dashed] (0.4, -2.4) rectangle (4.3, 0.5);
    \node[red] at (5.2, 0) {\small $= V_{\alpha}^{[1]}$}; 
\end{tikzpicture} 
\end{equation}
Hence, we have that: 
    \begin{equation} 
        \begin{tikzpicture}[scale = 1, baseline = {([yshift=-.5ex]current bounding box.center)}] 
        \draw[color = black] (-1, 0) -- (1, 0);
        \draw[color = black] (0, 0) -- (0, -0.8);
        \draw[fill =  lightdodgerblue] (0,0) circle (0.3);
        \node at (-0.03,0) {\small $A$};
        \node at (-1.3, 0) {\small $V_{\alpha}$};
        \node at (0, -1.2) {\small $\ $};
 %       \node at (-1.4, 0) {\small $\{V_{\alpha} \}$ };
    \end{tikzpicture} = \begin{tikzpicture}[scale = 1, baseline = {([yshift=-.5ex]current bounding box.center)}] 
        \draw[color = black] (-1, 0) -- (1, 0);
        \draw[color = black] (0, 0) -- (0, -0.8);
        \draw[fill = lightdodgerblue] (0,0) circle (0.3);
        \node at (-0.03,0) {\small $A$};
        %\node at (-1.4, 0) {\small $\{V_{\alpha} \}$ };
        \node at (1.5, 0) {\small $V^{[1]}_\alpha$ }; \\
        \node at (0.1, -1.0) {\small $U^{[0]}_{\alpha}$};
    \end{tikzpicture}
    \end{equation}
The above argument can be repeated arbitrarily and hence, we have shown that $V_{\alpha}$ is pushable for all $\alpha$, proving the claim.

\hspace{0.95 \textwidth} $\blacksquare$

\begin{shaded}
    \textbf{Corollary B.1} [Second Part of Theorem 2] A matrix product state having $\chi^2$ trace-orthonormal pushable operators is equivalent to it having $\chi^2$ such operators that satisfy:
    \begin{equation} \label{eq-TV=VT}
        \begin{tikzpicture} [scale = 1, baseline = {([yshift=-.5ex]current bounding box.center)}] 
    
    \draw[color = black] (-0.6, 0) -- (0.6, 0); %  top horizontal bar
    \draw[color = black] (-0.6, -1.2) -- (0.6, -1.2); %  bottom horizontal bar
    \draw[color = black] (0, 0) -- (0, -1.2); % vertical bar
     %\draw[color = black] (-0.8, 0) -- (-0.8, -1.2);
    % \draw[color = black] (-1, 0) -- (-1, -1.2);
    \draw[fill = lightdodgerblue] (0,0) circle (0.3); % circle
    \draw[fill = lightdodgerblue] (0,-1.2) circle (0.3); % circle
    %% Text
    \node at (-0.0,0) {\small $A$};
    \node at (-0.0,-1.2) {\small $\bar{A}$};
    %\draw[fill = lightishgray] (-1,0) circle (0.4);
    \node at (-1, 0) {\small $V_{\alpha}^{[n]}$};
    %\draw[fill = lightishgray] (-1,-1.2) circle (0.4);
    \node at (-1, -1.2) {\small $\bar{V}_{\alpha}^{[n]}$};
    \end{tikzpicture}  \; \; = \; \;
        \begin{tikzpicture} [scale = 1, baseline = {([yshift=-.5ex]current bounding box.center)}] 
    
    \draw[color = black] (-0.6, 0) -- (0.6, 0); %  top horizontal bar
    \draw[color = black] (-0.6, -1.2) -- (0.6, -1.2); 
    \draw[color = black] (0, 0) -- (0, -1.2); % vertical bar
     %\draw[color = black] (-0.8, 0) -- (-0.8, -1.2);
    % \draw[color = black] (-1, 0) -- (-1, -1.2);
    \draw[fill = lightdodgerblue] (0,0) circle (0.3); % circle
    \draw[fill = lightdodgerblue] (0,-1.2) circle (0.3); % circle
    %\draw[fill = lightishgray] (1.05,0) circle (0.5);
    \node at (1.05, 0) {\small $V^{[n+1]}_{\alpha}$};
    %\draw[fill = lightishgray] (1.05,-1.2) circle (0.5);
    \node at (1.05, -1.2) {\small $\bar{V}^{[n+1]}_{\alpha}$};
    %% Text
    \node at (-0.0,0) {\small $A$};
    \node at (-0.0,-1.2) {\small $\bar{A}$};
    \end{tikzpicture} 
\end{equation}
\end{shaded}

\textit{Proof.} We remark that if a matrix product state has a pushable operator, then Eq.~\eqref{eq-TV=VT} automatically holds.
It suffices to show therefore that the converse is true.
Note that the transfer matrix of a matrix product state uniquely determines its tensors up to isometries acting at the physical level\footnote{This follows from the fact that the transfer matrix defines a complete positive channel whose Kraus operators are the matrix product state tensors \cite{Cirac2021MPSReview}.}.
Since Eq.~\eqref{eq-TV=VT} implies that $(V^{[n]}_{\alpha} \otimes \bar{V}^{[n]}_{\alpha})T(V^{[n]}_{\alpha} \otimes \bar{V}^{[n]}_{\alpha})^{-1} = T$,%T(V^{[n+1]}_{\alpha} \otimes \bar{V^{[n+1]}_{\alpha})^{-1} = T$$, 
 we have that:
\begin{equation} %\label{eq-pushable}
    \begin{tikzpicture}[scale = 1, baseline = {([yshift=-.5ex]current bounding box.center)}] 
    \draw[color = black] (-1, 0) -- (1, 0);
    \draw[color = black] (0, 0) -- (0, -0.8);
    \draw[fill =  lightdodgerblue] (0,0) circle (0.3);
    \node at (-0.03,0) {\small $A$};
    \node at (-1.4, 0) {\small $V^{[n]}_{\alpha}$};
    \node at (1.8, 0) {\small $(V^{[n + 1]}_{\alpha})^{-1}$};
    \node at (0, -1.2) {\small $\ $};
%       \node at (-1.4, 0) {\small $\{V_{\alpha} \}$ };
\end{tikzpicture} = \begin{tikzpicture}[scale = 1, baseline = {([yshift=-.5ex]current bounding box.center)}] 
    \draw[color = black] (-1, 0) -- (1, 0);
    \draw[color = black] (0, 0) -- (0, -0.8);
    \draw[fill = lightdodgerblue] (0,0) circle (0.3);
    \node at (-0.03,0) {\small $A$};
    %\node at (-1.4, 0) {\small $\{V_{\alpha} \}$ };
    \node at (1.5, 0) {\small $\ $ }; \\
    \node at (0.1, -1.0) {\small $U^{[n]}_{\alpha}$};
\end{tikzpicture}
\end{equation}
where $U_{\alpha}^{[n]}$ is a square isometric matrix, which implies that it is unitary.
This completes the proof.

\hspace{0.95 \textwidth} $\blacksquare$

Theorem 2 in hand, we can move onto proving the rest of Theorem 1.

\subsection{Proof of Claims (2) and (3) in Theorem 1}

We now aim to prove claims $(2)$ and $(3)$.
We do so in parallel, first showing that $A$ (as defined in Eq.~\eqref{eq-A}) is in canonical form, then proving that $V_{\alpha}$  (which appear in the error basis $V_{\alpha} B$ [See discussion above Eq.~\eqref{eq-A}]) is unitary, and then finally proving that $B$ is unitary.
These three facts together simultaneously prove the two claims.

$\ $

$\ $

\begin{shaded}
    \textbf{Theorem B.3} Suppose that $\ket{\Psi}$ is a translation-invariant gluable quantum state.
    Then the matrix product state representation of $\ket{\Psi}$, as given by the tensor in Eq.~\eqref{eq-A}, is in right canonical form up to a unitary gauge transformation.% and the $V_{\alpha}$'s appearing in the measurement basis are unitary.
\end{shaded}

\textit{Proof.} To show that we are in right canonical form, it suffices to show that the identity is in the dominant right eigenspace of the transfer matrix.
To do so, recall that since  $\ket{\Psi}$ is a gluable matrix product state, for all $\alpha$ in the error basis, we have from Theorem 2, that:
\begin{equation}
        \begin{tikzpicture} [scale = 1, baseline = {([yshift=-.5ex]current bounding box.center)}] 
    
    \draw[color = black] (-0.6, 0) -- (0.6, 0); %  top horizontal bar
    \draw[color = black] (-0.6, -1.2) -- (0.6, -1.2); %  bottom horizontal bar
    \draw[color = black] (0, 0) -- (0, -1.2); % vertical bar
     %\draw[color = black] (-0.8, 0) -- (-0.8, -1.2);
    % \draw[color = black] (-1, 0) -- (-1, -1.2);
    \draw[fill = lightdodgerblue] (0,0) circle (0.3); % circle
    \draw[fill = lightdodgerblue] (0,-1.2) circle (0.3); % circle
    %% Text
    \node at (-0.0,0) {\small $\psi$};
    \node at (-0.0,-1.2) {\small $\bar{\psi}$};
    %\draw[fill = lightishgray] (-1,0) circle (0.4);
    \node at (-1.25, 0) {\small $V_{\alpha}B$};
    %\draw[fill = lightishgray] (-1,-1.2) circle (0.4);
    \node at (-1.25, -1.2) {\small $\bar{V}_{\alpha} \bar{B}$};
    \end{tikzpicture}  \; \; = \; \;
        \begin{tikzpicture} [scale = 1, baseline = {([yshift=-.5ex]current bounding box.center)}] 
    
    \draw[color = black] (-0.6, 0) -- (0.6, 0); %  top horizontal bar
    \draw[color = black] (-0.6, -1.2) -- (0.6, -1.2); 
    \draw[color = black] (0, 0) -- (0, -1.2); % vertical bar
     %\draw[color = black] (-0.8, 0) -- (-0.8, -1.2);
    % \draw[color = black] (-1, 0) -- (-1, -1.2);
    \draw[fill = lightdodgerblue] (0,0) circle (0.3); % circle
    \draw[fill = lightdodgerblue] (0,-1.2) circle (0.3); % circle
    %\draw[fill = lightishgray] (1.05,0) circle (0.5);
    \node at (1.05, 0) {\small $V^{[1]}_{\alpha}$};
    %\draw[fill = lightishgray] (1.05,-1.2) circle (0.5);
    \node at (1.05, -1.2) {\small $\bar{V}^{[1]}_{\alpha}$};
    %% Text
    \node at (-0.0,0) {\small $A$};
    \node at (-0.0,-1.2) {\small $\bar{A}$};
    \end{tikzpicture} 
\end{equation}
In other words, $V_{\alpha}$ pushes through $A$.
Now, since $V_{\alpha}B$ is an error basis, we can sum both sides of the equation by $\alpha$ yielding (using Eq.~\eqref{eq-niceidentity}): 
\begin{equation}
       \chi \begin{tikzpicture} [scale = 1, baseline = {([yshift=-.5ex]current bounding box.center)}] 
    
    \draw[color = black] (-0.8, 0) -- (0.8, 0); %  top horizontal bar
    \draw[color = black] (-0.8, -1.2) -- (0.8, -1.2); %  bottom horizontal bar
    \draw[color = black] (-0.8, 0) -- (-0.8, -1.2);
    \draw[color = black] (-1, 0) -- (-1, -1.2);
    \draw[color = black] (-1, 0) -- (-1.4, 0);
    \draw[color = black] (-1, -1.2) -- (-1.4, -1.2);
    %\draw[color = black] (-1, 0) -- (-1.4, 0); %  top horizontal bar
    %\draw[color = black] (-1, -1.2) -- (-1.4, -1.2); %  bottom horizontal bar
    \draw[color = black] (0, 0) -- (0, -1.2); % vertical bar
     %\draw[color = black] (-0.8, 0) -- (-0.8, -1.2);
    % \draw[color = black] (-1, 0) -- (-1, -1.2);
    \draw[fill = lightdodgerblue] (0,0) circle (0.3); % circle
    \draw[fill = lightdodgerblue] (0,-1.2) circle (0.3); % circle
    %% Text
    \node at (-0.0,0) {\small $\psi$};
    \node at (-0.0,-1.2) {\small $\bar{\psi}$};
    %
    % \node at (-1, -1.2) {\small $\bar{V}_{\alpha}$};
    % \node at (1.05, -1.2) {\small $\bar{W}_{\alpha} $ };
    % \node at (-1, 0) {\small ${V}_{\alpha}$};
    % \node at (1.05, 0) {\small $W_{\alpha} $ };
    \end{tikzpicture} =       \chi  \begin{tikzpicture} [scale = 1, baseline = {([yshift=-.5ex]current bounding box.center)}] 
    
    \draw[color = black] (-1.0, 0) -- (0.8, 0); %  top horizontal bar
    \draw[color = black] (-1.0, -1.2) -- (0.8, -1.2); %  bottom horizontal bar
    \draw[color = black] (-1.0, 0) -- (-1.0, -1.2);
    \draw[color = black] (-1.2, 0) -- (-1.2, -1.2);
    \draw[color = black] (-1.2, 0) -- (-1.6, 0);
    \draw[color = black] (-1.2, -1.2) -- (-1.6, -1.2);
    %\draw[color = black] (-1, 0) -- (-1.4, 0); %  top horizontal bar
    %\draw[color = black] (-1, -1.2) -- (-1.4, -1.2); %  bottom horizontal bar
    \draw[color = black] (0, 0) -- (0, -1.2); % vertical bar
     %\draw[color = black] (-0.8, 0) -- (-0.8, -1.2);
    % \draw[color = black] (-1, 0) -- (-1, -1.2);
    \draw[fill = lightdodgerblue] (0,0) circle (0.3); % circle
    \draw[fill = lightdodgerblue] (0,-1.2) circle (0.3); % circle
    %% Text
    \node at (-0.0,0) {\small $A$};
    \node at (-0.0,-1.2) {\small $\bar{A}$};
    \draw[fill = white] (-0.7, 0) circle (0.25);
    \draw[fill = white] (-0.7, -1.2) circle (0.25);
    \node at (-0.7, 0) {\tiny $B^{-1}$};
    \node at (-0.7, -1.2) {\tiny $\bar{B}^{-1}$};
    %
    % \node at (-1, -1.2) {\small $\bar{V}_{\alpha}$};
    % \node at (1.05, -1.2) {\small $\bar{W}_{\alpha} $ };
    % \node at (-1, 0) {\small ${V}_{\alpha}$};
    % \node at (1.05, 0) {\small $W_{\alpha} $ };
    \end{tikzpicture}  = \sum_{\alpha} 
        \begin{tikzpicture} [scale = 1, baseline = {([yshift=-.5ex]current bounding box.center)}] 
    
    \draw[color = black] (-0.8, 0) -- (1.4, 0); %  top horizontal bar
    \draw[color = black] (-0.8, -1.2) -- (1.4, -1.2); %  bottom horizontal bar

    %\draw[color = black] (-1, 0) -- (-1.4, 0); %  top horizontal bar
    %\draw[color = black] (-1, -1.2) -- (-1.4, -1.2); %  bottom horizontal bar
    \draw[color = black] (0, 0) -- (0, -1.2); % vertical bar
     %\draw[color = black] (-0.8, 0) -- (-0.8, -1.2);
    % \draw[color = black] (-1, 0) -- (-1, -1.2);
    \draw[fill = lightdodgerblue] (0,0) circle (0.3); % circle
    \draw[fill = lightdodgerblue] (0,-1.2) circle (0.3); % circle
    \draw[fill = lightishgray] (0.9,0) circle (0.3);
    \node at (0.9, 0) {\small $V^{[1]}_{\alpha}$};
    \draw[fill = lightishgray] (0.9,-1.2) circle (0.3);
    \node at (0.9, -1.2) {\small $\bar{V}^{[1]}_{\alpha}$};
    %% Text
    \node at (-0.0,0) {\small $A$};
    \node at (-0.0,-1.2) {\small $\bar{A}$};
    %
    % \node at (-1, -1.2) {\small $\bar{V}_{\alpha}$};
    % \node at (1.05, -1.2) {\small $\bar{W}_{\alpha} $ };
    % \node at (-1, 0) {\small ${V}_{\alpha}$};
    % \node at (1.05, 0) {\small $W_{\alpha} $ };
    \end{tikzpicture} 
\end{equation}
Now, since $V_{\alpha}^{[1]}$ is pushable, we know that we can iterate the action of the transfer matrix above to get: 
\begin{equation} \label{eq-fundamental_theorem1}
      \chi \begin{tikzpicture} [scale = 1, baseline = {([yshift=-.5ex]current bounding box.center)}] 
    
    \draw[color = black] (-1.0, 0) -- (0.8, 0); %  top horizontal bar
    \draw[color = black] (-1.0, -1.2) -- (0.8, -1.2); %  bottom horizontal bar
    \draw[color = black] (-1.0, 0) -- (-1.0, -1.2);
    \draw[color = black] (-1.2, 0) -- (-1.2, -1.2);
    \draw[color = black] (-1.2, 0) -- (-1.6, 0);
    \draw[color = black] (-1.2, -1.2) -- (-1.6, -1.2);
    %\draw[color = black] (-1, 0) -- (-1.4, 0); %  top horizontal bar
    %\draw[color = black] (-1, -1.2) -- (-1.4, -1.2); %  bottom horizontal bar
    \draw[color = black] (0, 0) -- (0, -1.2); % vertical bar
     %\draw[color = black] (-0.8, 0) -- (-0.8, -1.2);
    % \draw[color = black] (-1, 0) -- (-1, -1.2);
    %\draw[fill = lightdodgerblue] (0,0) circle (0.3); % circle
    %\draw[fill = lightdodgerblue] (0,-1.2) circle (0.3); % circle
    %% Text
    %\node at (-0.0,0) {\small $A$};
    %\node at (-0.0,-1.2) {\small $\bar{A}$};
    \draw[fill = white] (-0.7, 0) circle (0.25);
    \draw[fill = white] (-0.7, -1.2) circle (0.25);
    \node at (-0.7, 0) {\tiny $B^{-1}$};
    \node at (-0.7, -1.2) {\tiny $\bar{B}^{-1}$};
    \draw[rounded corners, fill = lightdodgerblue] (-0.4, -1.4) rectangle (0.4, 0.2) {};
    \node at (0, -0.6) { $T^n$};
    %
    % \node at (-1, -1.2) {\small $\bar{V}_{\alpha}$};
    % \node at (1.05, -1.2) {\small $\bar{W}_{\alpha} $ };
    % \node at (-1, 0) {\small ${V}_{\alpha}$};
    % \node at (1.05, 0) {\small $W_{\alpha} $ };
    \end{tikzpicture} = 
    \begin{tikzpicture} [scale = 1, baseline = {([yshift=-.5ex]current bounding box.center)}] 
        \draw[color = black] (-1.0, 0) -- (1.4, 0);
        \draw[color = black] (-1.0, -1.2) -- (1.4, -1.2);
        \draw[rounded corners, fill = lightdodgerblue] (-0.4, -1.4) rectangle (0.4, 0.2) {};
        \node at (0, -0.6) { $T^n$};
        \draw[color = black] (0.9, 0) -- (0.9, -1.2);
        \draw[fill = lightishgray] (0.9,0) circle (0.3);
        \node at (0.9, 0) {\small $\mathcal{V}^{[n]}$};
        \draw[fill = lightishgray] (0.9,-1.2) circle (0.3);
        \node at (0.9, -1.2) {\small $\bar{\mathcal{V}}^{[n]}$};
     \end{tikzpicture}
\end{equation}
Now, we take $n$ to be arbitrarily large yielding:

\begin{align}
    \begin{tikzpicture}[scale = 1, baseline = {([yshift=-.5ex]current bounding box.center)}]
        \draw[color = black] (-1.2, 0) -- (-1.2, -1.2);
        \draw[color = black] (-1.2, 0) -- (-1.6, 0);
        \draw[color = black] (-1.2, -1.2) -- (-1.6, -1.2);
    \end{tikzpicture}\ \ &\left\{
    \sum_{\beta} \text{tr}(B^{-1}R_{\beta}(B^{-1})^{\dagger})\ \ \begin{tikzpicture}[scale = 1, baseline = {([yshift=-.5ex]current bounding box.center)}]
        \draw[color = black] (-0.5, 0) -- (0,0);
        \draw[color = black] (-0.5, 0) -- (-0.5,-1.2);
        \draw[color = black] (-0.5, -1.2) -- (0,-1.2);
        \draw[fill = white] (-0.5, -0.6) circle (0.25);
        \node at (-0.5, -0.6) {\small $L_{\beta}$};
    \end{tikzpicture}\ \ \  \right\} =\frac{1}{\chi}  \sum_{\gamma} \begin{tikzpicture}[scale = 1, baseline = {([yshift=-.5ex]current bounding box.center)}]
        \draw[color = black] (-0.5, 0) -- (0,0);
        \draw[color = black] (0, 0) -- (0,-1.2);
        \draw[color = black] (-0.5, -1.2) -- (0,-1.2);
        \draw[fill = white] (0, -0.6) circle (0.25);
        \node at (0, -0.6) {\small $R_{\gamma}$};
    \end{tikzpicture}\ \begin{tikzpicture}[scale = 1, baseline = {([yshift=-.5ex]current bounding box.center)}]
        \draw[color = black] (-0.5, 0) -- (0.9,0);
        \draw[color = black] (-0.5, 0) -- (-0.5,-1.2);
        \draw[color = black] (-0.5, -1.2) -- (0.9,-1.2);
        \draw[fill = white] (-0.5, -0.6) circle (0.25);
        \node at (-0.5, -0.6) {\small $L_{\gamma}$};
        \draw[color = black] (0.4, 0) -- (0.4, -1.2);
        \draw[fill = lightishgray] (0.4,0) circle (0.3);
        \node at (0.4, 0) {\small $\mathcal{V}^{[n]}$};
        \draw[fill = lightishgray] (0.4,-1.2) circle (0.3);
        \node at (0.4, -1.2) {\small $\bar{\mathcal{V}}^[n]$};
    \end{tikzpicture}\\
    &\implies  \text{tr}(B^{-1}R_{\beta}(B^{-1})^{\dagger}) \begin{tikzpicture}[scale = 1, baseline = {([yshift=-.5ex]current bounding box.center)}]
        \draw[color = black] (-1.2, 0) -- (-1.2, -1.2);
        \draw[color = black] (-1.2, 0) -- (-1.6, 0);
        \draw[color = black] (-1.2, -1.2) -- (-1.6, -1.2);
    \end{tikzpicture} = \frac{1}{\chi } \sum_{\gamma} \begin{tikzpicture}[scale = 1, baseline = {([yshift=-.5ex]current bounding box.center)}]
        \draw[color = black] (-0.5, 0) -- (0,0);
        \draw[color = black] (0, 0) -- (0,-1.2);
        \draw[color = black] (-0.5, -1.2) -- (0,-1.2);
        \draw[fill = white] (0, -0.6) circle (0.25);
        \node at (0, -0.6) {\small $R_{\gamma}$};
    \end{tikzpicture}\ \begin{tikzpicture}[scale = 1, baseline = {([yshift=-.5ex]current bounding box.center)}]
        \draw[color = black] (-0.5, 0) -- (1.4,0);
        \draw[color = black] (-0.5, 0) -- (-0.5,-1.2);
        \draw[color = black] (-0.5, -1.2) -- (1.4,-1.2);
        \draw[color = black] (1.4, 0) -- (1.4, -1.2);
        \draw[fill = white] (-0.5, -0.6) circle (0.25);
        \node at (-0.5, -0.6) {\small $L_{\gamma}$};
        \draw[color = black] (0.4, 0) -- (0.4, -1.2);
        \draw[fill = white] (0.4,0) circle (0.3);
        \draw[fill = white] (1.4,-0.6) circle (0.25);
        \node at (1.4, -0.6) {\small $R_{\beta}$};
        \node at (0.4, 0) {\small $\mathcal{V}^{[n]}$};
        \draw[fill = white] (0.4,-1.2) circle (0.3);
        \node at (0.4, -1.2) {\small $\bar{\mathcal{V}}^{[n]}$};
    \end{tikzpicture}
\end{align}
Note that there exists a $\beta$ for which $R_{\beta}$ is full rank and positive \cite{Cirac2021MPSReview} and hence the left-hand side is non-zero.
Consequently, the above naturally implies that $\ket{\mathds{1}}$ lives in the span of dominant right eigenvector $\ket{R_\gamma}$.
As a consequence, the matrix product state is in canonical form up to a unitary gauge transformation (to make the corresponding left eigenvector diagonal).

\hspace{0.95 \textwidth} $\blacksquare$

We now prove Claim (3).

\begin{shaded}
    \textbf{Theorem B.4} Let $A$ be a translation-invariant gluable matrix product state. Then, the $V_{\alpha}$ operators that appear in the measurement basis $\mathcal{V}B$ are unitary.
\end{shaded}

\textit{Proof.} To prove this, let us note that by Theorem B.2 and B.3, we have that $A$ is in right canonical form.
As a consequence, the Moore-Penrose pseudoinverse of the map $\Gamma^R_x$ defined near Eq.~\eqref{eq-GammaL}, $(\Gamma^R_x)^+ = (\Gamma^R_x)^{\dagger}$.
Consequently, using methods similar to Eq.~\eqref{eq-MoorePenrose} we have that:
\begin{equation}
    V_{\alpha} = \Gamma^R_x \prod_{n} U_{\alpha} (\Gamma^R_x)^{\dagger}
\end{equation}
Now, because 
\begin{equation}
\begin{tikzpicture}[scale = 0.8, baseline = {([yshift=-.5ex]current bounding box.center)}] 
    \node at (-1.25, 0) {\small $V^{-1}_{\alpha}$};
    \draw[color = black] (-1.0,0) -- (4, 0);
    \node at (4.4, 0) {\small $\cdots$};
    \foreach \i in {0, ..., 2}{
        \draw[color = black] (1.5*\i,0) -- (1.5*\i, -1);
        \draw[fill = lightdodgerblue] (1.5*\i,0) circle (0.3);
        \node at (1.5*\i, 0) {\small $A$};
        \node at (1.5*\i, -1.6) {$\ $};
    } 
\end{tikzpicture} = \begin{tikzpicture}[scale = 0.8, baseline = {([yshift=-.5ex]current bounding box.center)}] 
    %\node at (-1.25, 0) {\small $V_{\alpha}$};
    \draw[color = black] (-1.0,0) -- (4, 0);
    \node at (4.4, 0) {\small $\cdots$};
    \foreach \i in {0, ..., 2}{
        \draw[color = black] (1.5*\i,0) -- (1.5*\i, -1);
        \draw[fill = lightdodgerblue] (1.5*\i,0) circle (0.3);
        \node at (1.5*\i, 0) {\small $A$};
        \node at (1.5*\i, -1.3) {\small $(U^{[\i]}_{\alpha})^{\dagger}$};
    } 
\end{tikzpicture}
\end{equation}
Therefore, we also have that: $V^{-1}_{\alpha} = \Gamma^R_x \prod_{n} U_{\alpha}^{\dagger} (\Gamma^R_x)^{\dagger}$.
Finally, note that $V_{\alpha}^{\dagger} = \Gamma^R_x \prod_{n} U^{\dagger}_{\alpha} (\Gamma^R_x)^{\dagger}$.
But then, this implies that $V_{\alpha}^{\dagger} = V_{\alpha}^{-1}$.
This completes the proof.

\hspace{0.95\textwidth} $\blacksquare$ 

As an important remark, the same argument shows that \textit{any} $V_{\alpha}^{[n]}$ that $V_{\alpha}$ pushes through to, is required to be unitary.
We now turn to proving the main theorem of this subsection.

\begin{shaded}
    \textbf{Theorem B.5} [Claims (1) and (2) of Theorem 1] Suppose that $\ket{\Psi}$ is a translation-invariant gluable quantum state.
    Then, the unentangled clusters in the state preparation protocol $\ket{\psi}$ are the matrix product state tensor in canonical form up to a tensor product of unitaries acting on each qudit of the cluster.
    In other words,    
    \begin{equation}
        \begin{tikzpicture}[scale = 1, baseline = {([yshift=-.5ex]current bounding box.center)}] 
        \draw[color = black] (-1, 0) -- (1, 0);
        \draw[color = black] (0, 0) -- (0, -0.8);
        \draw[fill = lightdodgerblue] (0,0) circle (0.3);
        \node at (-0.03,0) {\small $\psi$};
        %\node at (-1.3, 0) {\small $V^{[n]}$};
        %\node at (0, -1.2) {\small $\ $};
 %       \node at (-1.4, 0) {\small $\{V_{\alpha} \}$ };
    \end{tikzpicture} = \begin{tikzpicture}[scale = 1, baseline = {([yshift=-.5ex]current bounding box.center)}] 
        \draw[color = black] (-1, 0) -- (1, 0);
        \draw[color = black] (0, 0) -- (0, -0.8);
        \draw[fill = lightdodgerblue] (0,0) circle (0.3);
        \node at (-0.03,0) {\small $A$};
        \node at (-1.3, 0) {\small $u_L$ };
        \node at (1.5, 0) {\small $u_R$ }; \\
        \node at (0, -1.0) {\small $u_P$};
    \end{tikzpicture}
\end{equation}
    where $A$ is the MPS in right canonical form and $u_L, u_R,$ and $u_P$ are unitaries.
    Moreover, the measurement basis is maximally entangled or, when viewed as operators, form a unitary error basis.
\end{shaded}

\textit{Proof.} If $\ket{\Psi}$ is right gluable, then we know that $\ket{\Psi}$ can be prepared from clusters $\ket{\psi}$ along with entangling measurements in an error basis $\mathcal{V}B$ with $\mathcal{V}$ unitary (Theorem B.4).
Furthermore, we know from Eq.~\eqref{eq-A} that the matrix product state tensor associated with $\ket{\Psi}$ is:
\begin{equation}
    [A]^{s_x}_{i_{(x - 1)_R} i_{x_R}} =  B_{i_{(x-1)_R} i_{x_{L}}}\psi_{i_{x_L}\, i_{x_R}}^{s_x'} [u_P]_{s_x' s_x} 
\end{equation}
which is in right canonical form up to conjugating unitaries acting at the virtual level.
Therefore, it suffices to show that $B$ is unitary.
To do so, let us start by noting that since $\mathcal{V}B$ is an error basis, a nice graphical notation for this error basis is:
\begin{equation}
    V_{\alpha}B = \begin{tikzpicture}[scale = 1, baseline = {([yshift=-.5ex]current bounding box.center)}] 
        \draw[color = black] (-0.8, 0) -- (0.8, 0);
        \draw[color = black] (0, 0) -- (0, -0.8);
        \draw[fill = lightishgray] (0,0) circle (0.3);
        \node at (-0.0,0) {\small $\mathcal{V}B$};
        \node at (-0.9, 0) {\small $\chi$};
        \node at (0.9, 0) {\small $\chi$};
        \node at (0, -0.9) {\small $\chi^2$};
        %\node at (1.2, -0.05) {\small $v_i$};
        % \node at (0.0, -1.2) {\small $\ $};
 %       \node at (-1.4, 0) {\small $\{V_{\alpha} \}$ };
    \end{tikzpicture}
\end{equation}
where the leg labels indicate dimensionality of the leg and the downwards facing leg is the ``$\alpha$'' leg.
Note that trace orthonormality implies that:
\begin{equation}
    \begin{tikzpicture} [scale = 1, baseline = {([yshift=-.5ex]current bounding box.center)}] 
    \draw[color = black] (-0.8, 0) -- (0.8, 0); %  top horizontal bar
    \draw[color = black] (0, 0) -- (0, 0.5); % vertical bar
    \draw[fill = lightishgray] (0,0) circle (0.3); % circle
    % \draw[fill = white] (0,0.5) circle (0.14);
    %% Text
    \node at (0.0,0) {\small $\overline{\mathcal{V}B}$};
    % \node at (0.0 ,0.5) {\small $\hat{s}$};
    %
    %
    \draw[color = black] (-0.8, 0) -- (-0.8, -1);
    \draw[color = black] (0.8, 0) -- (0.8, -1);
    \draw[color = black] (-0.8, -1) -- (0.8, -1); %  top horizontal bar
    \draw[color = black] (0, -1) -- (0, -0.5-1); % vertical bar
    \draw[fill = lightishgray] (0,-1) circle (0.3); % circle
    % \draw[fill = white] (0,-0.5-1) circle (0.14);
    % \draw[rounded corners, fill = lightdodgerblue] (-1, -0.7) rectangle (1, -0.3);
    %% Text
    \node at (0.,-1) {\small $\mathcal{V}B$};
    % \node at (0.0 ,-0.5-1) {\small $s$};
    %\node at (-0.55, -1) {\small $V_{\alpha}$};
    %\node at (0.55, -1) {\small $W_{\alpha}$};
    % \node at (-1, 0) {\small ${V}_{\alpha}$};
    % \node at (1.05, 0) {\small $W_{\alpha} $ }; 
    \end{tikzpicture}\ \ =\  \chi\ \  \begin{tikzpicture}[scale = 1, baseline =  {([yshift=-.5ex]current bounding box.center)}] 
        \draw[color = black] (0,0.4) -- (0, -1.6);
    \end{tikzpicture} \qquad \qquad   \begin{tikzpicture} [scale = 1, baseline = {([yshift=-.5ex]current bounding box.center)}] 
    \draw[color = black] (-0.8, 0) -- (0.8, 0); %  top horizontal bar
    \draw[color = black] (-0.8, -1.2) -- (0.8, -1.2); %  bottom horizontal bar
    \draw[color = black] (0, 0) -- (0, -1.2); % vertical bar
    \draw[fill = lightishgray] (0,0) circle (0.3); % circle
    \draw[fill = lightishgray] (0,-1.2) circle (0.3); % circle
    \draw[color = black] (-0.8, 0) -- (-0.8, 0.2);
    \draw[color = black] (0.8, 0) -- (0.8, 0.2);
    \draw[color = black] (-0.8, -1.2) -- (-0.8, -1.4);
    \draw[color = black] (0.8, -1.2) -- (0.8, -1.4);
    %% Text
    \node at (-0.0,0) {\small $\mathcal{V}B$};
    \node at (-0.0,-1.2) {\small $\overline{\mathcal{V}B}$};
    %
    % \node at (-1, -1.2) {\small $\bar{V}_{\alpha}$};
    % \node at (1.05, -1.2) {\small $\bar{W}_{\alpha} $ };
    % \node at (-1, 0) {\small ${V}_{\alpha}$};
    % \node at (1.05, 0) {\small $W_{\alpha} $ };
    \end{tikzpicture}\ \ \  =\ \ \chi \ \  \begin{tikzpicture} [scale = 1, baseline = {([yshift=-.5ex]current bounding box.center)}] 
        \draw[color = black] (-0.8, 0.4) -- (-0.8, -1.6);
        \draw[color = black] (0.8, 0.4) -- (0.8, -1.6);
    \end{tikzpicture}
\end{equation}
Now, note that the second equality automatically implies that: 
\begin{equation}
    \begin{tikzpicture} [scale = 1, baseline = {([yshift=-.5ex]current bounding box.center)}] 
    \draw[color = black] (-0.8, 0) -- (0.8, 0); %  top horizontal bar
    \draw[color = black] (-0.8, -1.2) -- (0.8, -1.2); %  bottom horizontal bar
    \draw[color = black] (0, 0) -- (0, -1.2); % vertical bar
    \draw[fill = lightishgray] (0,0) circle (0.3); % circle
    \draw[fill = lightishgray] (0,-1.2) circle (0.3); % circle
    \draw[color = black] (-0.8, 0) -- (-0.8, -1.2);
    %\draw[color = black] (0.8, 0) -- (0.8, 0.4);
    %\draw[color = black] (-0.8, -1.2) -- (-0.8, -1.6);
    %\draw[color = black] (0.8, -1.2) -- (0.8, -1.6);
    %% Text
    \node at (-0.0,0) {\small $\mathcal{V}B$};
    \node at (-0.0,-1.2) {\small $\overline{\mathcal{V}}B$};
    %
    % \node at (-1, -1.2) {\small $\bar{V}_{\alpha}$};
    % \node at (1.05, -1.2) {\small $\bar{W}_{\alpha} $ };
    % \node at (-1, 0) {\small ${V}_{\alpha}$};
    % \node at (1.05, 0) {\small $W_{\alpha} $ };
    \end{tikzpicture} = \sum_{\alpha} BV_{\alpha} V^{\dagger}_{\alpha}B^{\dagger} = \sum_{\alpha} B B^{\dagger} =  \chi^2  \mathds{1},
\end{equation}
where we used our earlier result that $V_\alpha$ is unitary.
Thus, we have that $BB^{\dagger} = 1$, implying that $B$ is unitary.
Consequently, $\psi$ is related to the matrix product state tensor in canonical form up to a tensor product of unitaries acting on each qudit of the cluster.
Moreover, $\mathcal{V}B$ is a unitary error basis (or equivalently, the measurement basis is maximally entangled).

\hspace{0.95\textwidth} $\blacksquare$ 

\subsection{An Aside on Measurement Probabilities \label{app-equalprob}}

\begin{shaded}
    \textbf{Theorem B.6} For any gluable wavefunction $\ket{\Psi}$ defined on a finite-size system with particular open boundary conditions, every measurement outcome is equally likely.
\end{shaded}

\textit{Proof.} We do this by computing the probability explicitly.
Once again, we work with open boundary conditions and since we are predominantly interested in bulk properties, we let our target wavefunction have the following boundaries: 
\begin{equation}
    \ket{\Psi} = \begin{tikzpicture}[scale = 1, baseline = {([yshift=-.5ex]current bounding box.center)}] 
    \draw[color = black] (-1.5,0) -- (7.5, 0);
    \foreach \i in {0, ..., 3}{
        \draw[color = black] (2*\i,0) -- (2*\i, -1);
        \draw[fill = lightdodgerblue] (2*\i,0) circle (0.3);
        \node at (2*\i, 0) {\small $A$};
    }
    % \node at (7.5, 0) {$\cdots$};
    % \node at (-1.5, 0) {$\cdots$};
    %\node at (3, 0) {$V_{\alpha}$};
    \draw[color = black] (-1.5, 0) -- (-1.5, -1);
    \draw[color = black] (7.5, 0) -- (7.5, -1);
    \draw[fill = white] (-1.5, -0.5) circle (0.35);
    \draw[fill = white] (7.5, -0.5) circle (0.35);
    \node at (-1.5, -0.5) {\scriptsize $v_L^{1/2}$};
    \node at (7.5, -0.5) {\small $v_R^{1/2}$};
\end{tikzpicture}
\end{equation}
where $v_L$ and $v_R$ define vectors with non-zero overlap on the largest positive right and left eigenvectors of the transfer matrix, $R$ and $L$ (with no overlap on any other potential dominant eigenvectors).
For convenience, we choose $v_L = L$ and $v_R = R$.
We have that:
\begin{equation}
    p_{\mathbf{m}} = \frac{1}{\chi^N\mathcal{N}}     \begin{tikzpicture}[scale = 1, baseline = {([yshift=-.5ex]current bounding box.center)}] 
        \draw[color = black] (-1,0) -- (7, 0);
        \draw[color = black] (-1,-1) -- (7, -1);
        \foreach \i in {0, ..., 3}{
            \draw[color = black] (2*\i,0) -- (2*\i, -1);
            \draw[fill = lightdodgerblue] (2*\i,0) circle (0.3);
            \draw[fill = lightdodgerblue] (2*\i,-1) circle (0.3);
            \node at (2*\i, 0) {\small $A$};
            \node at (2*\i, -1) {\small $\bar{A}$};
        }
        \node at (7.5, 0) {$\cdots$};
        \node at (-1.5, 0) {$\cdots$};
        \node at (3, 0) {$V_{\alpha_x}$};
        \node at (5, 0) {$V_{\alpha_{x+1}}$};
        \node at (1, 0) {$V_{\alpha_{x-1}}$};
        \node at (7.5, -1) {$\cdots$};
        \node at (-1.5, -1) {$\cdots$};
        \node at (3, -1) {$\bar{V}_{\alpha_x}$};
        \node at (5, -1) {$\bar{V}_{\alpha_{x + 1}}$};
        \node at (1, -1) {$\bar{V}_{\alpha_{x - 1}}$};
    \end{tikzpicture}
\end{equation}
Now, from the above theorem, each of these are pushable operators, indicating that the can be swept to the right to unitaries.
Consequently, we have that the overlap above is equal to $1$.
Hence, 
\begin{equation}
    p_{\mathbf{m}} = \frac{1}{\chi^N \mathcal{N}} 
\end{equation}
for all measurement outcomes.

\hspace{0.95\textwidth} $\blacksquare$

\section{Proof of Parameterization and Go Theorems} \label{app:parameterization-and-go}

This appendix is devoted to providing proofs for the parameterization and ``go'' theorems in the main text (Theorems 5 and 6, 7 respectively), which were proved in the context of uniform topological gluable quantum states.
To do so, we start by providing a more formal treatise on the relationship between uniform topological errors and symmetry.
This will rely heavily on the concept of an index group, introduced in Ref.~\onlinecite{knill1996group}.
We will then show that the measurement errors form a faithful irreducible projective representation of this group and will rely heavily on the theory of abelian projective representations to prove our parameterization theorem.
Moreover, the connection with projective representations will make the connection with SPT phases rather natural.
This will enable proving our two go theorems.

\subsection{Uniform Topologically Gluable, Symmetry, and Parameterization \label{app-classification}}

Let us suppose that we have a uniform topologically gluable matrix product state $A$, with error basis $\mathcal{V}$.
Then it is useful to have a name for the group generated by $\mathcal{V} \otimes \mathcal{V}^*$:

\begin{shaded}
    \textbf{Definition} (Index Group) Suppose $A$ is a uniform topologically gluable matrix product state tensor with an error basis $\mathcal{V} = \{V_{\alpha}\}$.
    We call the group  $G_U \simeq \langle \{{V}_{\alpha} \otimes \bar{V}_{\alpha}\}\rangle$ the index group of the matrix product state.
\end{shaded}
\noindent
Several useful lemmas immediately follow from this definition.
In particular:
\begin{shaded}
    \textbf{Lemma C.1} If $A$ is a short-range entangled uniformly topologically gluable matrix product state tensor, then the index group is a symmetry group of the state.
\end{shaded} 

\textit{Proof.} To prove this, it suffices to show that the index group is a symmetry group of the matrix product state.
Suppose that $A$ is uniform topologically gluable with error basis $\mathcal{V}$ and index group $G_U \simeq \langle \{V_{\alpha} \otimes \bar{V}_{\alpha} \} \rangle$, then:
\begin{equation} \label{eq-VVU}
    \begin{tikzpicture}[scale = 1, baseline = {([yshift=-.5ex]current bounding box.center)}] 
    \draw[color = black] (-1, 0) -- (1, 0);
    \draw[color = black] (0, 0) -- (0, -0.8);
    \draw[fill = lightdodgerblue] (0,0) circle (0.3);
    \node at (-0.03,0) {\small $A$};
    \node at (-1.3, 0) {\small ${V}_{\alpha}$};
    \node at (0, -1.2) {\small $\ $};
    \node at (1.3, 0) {\small $V^{\dagger}_{\alpha} $ };
\end{tikzpicture} = \begin{tikzpicture}[scale = 1, baseline = {([yshift=-.5ex]current bounding box.center)}] 
    \draw[color = black] (-1, 0) -- (1, 0);
    \draw[color = black] (0, 0) -- (0, -0.8);
    \draw[fill = lightdodgerblue] (0,0) circle (0.3);
    \node at (-0.03,0) {\small $A$};
    %\node at (-1.4, 0) {\small $\{V_{\alpha} \}$ };
   % \node at (1.3, 0) {\small ${V}_{\alpha}$ }; \\
    \node at (0, -1.0) {\small $U_{\alpha}$};
\end{tikzpicture}
\end{equation}
Hence, the global matrix product state is symmetric under $U_{\alpha}$.
Let us further note that the matrix product state tensor $A$, viewed as a map from virtual degrees of freedom to physical degrees of freedom, defines a group homomorphism from $\langle \{V_{\alpha} \otimes \bar{V}_{\alpha} \}\rangle$ to $\langle \{U_{\alpha} \}\rangle$.
Consequently, the $U_{\alpha}$'s define a linear representation of the index group $G_U$ and the matrix product state is symmetric under $\langle\{ U_{\alpha}\} \rangle$.
Finally, since $A$ is injective after blocking \cite{Cirac2021MPSReview}, this representation is faithful and hence $\langle \{U_{\alpha}\} \rangle \simeq G_U$.
Hence, the matrix product state is symmetric under $G_U$.

\hspace{0.95 \textwidth} $\blacksquare$

\begin{shaded}
    \textbf{Lemma C.2.} Suppose $A$ is a topologically gluable matrix product state with an error basis $\mathcal{V} = \{V_{\alpha} \}$.
    Then, the canonical isomorphism between $G_U$ and $\langle \{V_{\alpha} \otimes \bar{V}_{\alpha}\} \rangle$ defines a faithful linear representation of $G_U$.
    Moreover, the canonical map between the generating set of $G_U$ and $\{V_{\alpha}\}$ induces a faithful irreducible projective representation of $G_U$.
\end{shaded}

\textit{Proof.} Let us note that, by definition, $G_U \simeq \langle \{V_{\alpha} \otimes \bar{V}_{\alpha} \} \rangle$ and denote the isomorphism by $\rho: G \to  \langle \{V_{\alpha} \otimes \bar{V}_{\alpha}\} \rangle \subseteq \text{GL}_{\mathbb{C}}(\mathbb{C}^{\chi} \otimes \mathbb{C}^{\chi})$.
It immediately follows that this defines a faithful linear representation of $G_U$ because the map is injective, proving the first part of the theorem.

We now prove that there exists a map between $G_U$ and $\langle \{V_{\alpha}\} \rangle$ that defines a faithful and irreducible projective representation of $G_U$.
Note that for any element $g \in G_U$: 
\begin{equation}
    \rho(g) = \prod_{\alpha \in S_g} V_{\alpha} \otimes \prod_{\alpha \in S_g} \bar{V}_{\alpha} = \nu(g) \otimes \bar{\nu}(g)\qquad \text{ where }\qquad \nu(g) \equiv \prod_{\alpha \in S_g} V_{\alpha}
\end{equation}
for some list $S_g$ of numbers between $1, \cdots, \chi^2$ where $\nu: G \to \langle \{V_{\alpha}\} \rangle \subseteq \text{GL}_{\mathbb{C}}(\mathbb{C}^{\chi})$
Note that since $\rho(g) \rho(h) = \rho(gh)$, we have that $\nu(g) \nu(h) = \omega(g, h) \nu(gh)$ where $\omega(g, h): G \times G \to U(1)$.
Thus, there exists a map between $G_U$ and $\langle \{V_{\alpha}\} \rangle$ that defines a projective representation of $G_U$.

We now show that this representation is faithful and irreducible.
To show that the representation is irreducible, let us remark that, by assumption that $\{V_{\alpha}\}$ are trace orthonormal and span the set of $\chi \times \chi$ matrices.
As a consequence, any subspace of $\mathbb{C}^{\chi}$ that is invariant under the action of all the $V_{\alpha}$'s must be invariant under any $\chi \times \chi$ matrix.
Hence, the only invariant subspaces for the $\langle \{V_{\alpha}\} \rangle$ are $\{0\}$ and $\mathbb{C}^{\chi}$ and hence the representation is irreducible.
We now show that the representation is faithful.
We do so by contradiction.
Suppose there existed a $g \neq 1$ (the identity) such that $\nu(g) = e^{i \theta} \mathds{1}$.
Then, 
\begin{equation}
    \rho(g) = \nu(g) \otimes \bar{\nu}(g) = \mathds{1} \otimes \mathds{1}
\end{equation}
But since $\rho(g) = \rho(1)$ and $g \neq 1$ this contradicts the fact that $\rho$ is a faithful representation.
Hence, if $\nu(g) = e^{i\theta} \mathds{1}$, $g = 1$ implying that it is faithful.

\hspace{0.95 \textwidth} $\blacksquare$

\subsubsection{A Technical Foray Into Nice Error Bases}

Having built up some of the technology for understanding the interplay between topological errors and symmetry, it will be helpful to introduce the concept of nice error basis and a convenient result on these error bases from literature.
This section can largely be skipped if one is only interested in proving the parameterization.
However, the results here are invaluable for demonstrating the connection with SPT.
In particular, we start by defining nice error basis as \cite{knill1996group, klappenecker2002beyond}:

\begin{shaded}
    \textbf{Definition} (Nice Error Basis) Let $G$ be a group of order $\chi^2$.
    A nice error basis in $\chi$ dimensions is given by a set $\{\rho(g) \in \mathcal{U}(n) | g \in G \}$ of unitary matrices that satisfy: 
    \begin{enumerate}
        \item[(i)] $\rho(1) = \mathds{1}$

        \item[(ii)] $\rho(g) \rho(h) = \omega(g, h) \rho(gh)$ for all $g, h \in G$, where $\omega: G \times G \to U(1)$

        \item[(iii)] $\text{tr}(\rho(g)) = 0$ for all $g \neq 1$.
    \end{enumerate}
    In the following Lemma, we will show that these are in fact unitary error bases with index groups given by $G$.
\end{shaded}

Conditions $(i)$ and $(ii)$ above gaurentee that nice error bases form some projective representation of a group.
We also quote a result by Knill~\cite{knill1996group} that shows that nice error bases are unitary error bases.

\begin{shaded}
    \textbf{Lemma C.3} Nice error bases are unitary error bases 
\end{shaded}

Finally, we remark upon one more useful lemma and then a theorem from literature (Theorem 1 of Ref.~\cite{klappenecker2002beyond}):

\begin{shaded}
    \textbf{Lemma C.4} A nice error basis is necessarily a non-trivial projective representation of the group $G$.  
\end{shaded}

\textit{Proof.} This proof follows from the theory of characters.
Suppose that one has a group of order $|G| = \chi^2$ and $\rho$ be a regular (i.e. trivial projective) representation of the group.
Then, $\chi(g) = \text{Tr}(\rho(g))$ is traditionally called the character function of the group representation.
Now, note that if $\chi(g) = 0$ for all $g \neq 1$, then $\chi(1)$ must be an integer multiple of $|G|$ \cite{isaacs2006character}.
This immediately implies that nice error bases form a non-trivial projective representation.
In particular, note that, by property $(i)$ of nice error basis, $\rho(1) = \mathds{1}$ and hence $\chi(1) = \chi$.
But this is not an integer multiple of $\chi^2$ for any $\chi > 1$.
Therefore, $\rho$ cannot be a linear representation of $G$.

\hspace{0.95\textwidth} $\blacksquare$

\begin{shaded}
    \textbf{Theorem C.1} Let $\{\rho(g)\}$ be a set of unitary matrices parameterized by the elements of a finite group $G$. 
    The set is a nice error basis with index group $G$ if and only if $\rho$ is a unitary faithful irreducible projective representation of $G$ of degree $|G|^{1/2}$.
\end{shaded}

With these results about nice error bases, we can prove a set of results that will naturally lead to our parameterization.
First,
\begin{shaded}
    \textbf{Theorem C.2} Let $G_U$ be the index group associated to the unitary error basis $\mathcal{V}$. If $G_U$ is abelian, then the degree of the projective representation defined by $\{V_{\alpha}\}$ is $\sqrt{|G_U|}$.
    Equivalently, any unitary error basis with abelian index group is equivalent to a nice error basis.
\end{shaded}

\textit{Proof.}
From the corollary above, we know that the canonical map between the generating set of $G_U$ and $\{V_{\alpha}\}$ defines a faithful and irreducible projective representation of $G_U$:
\begin{equation}
    \nu(g) \equiv \prod_{\alpha \in S_g} V_{\alpha} \qquad \text{such that} \qquad \nu(g) \nu(h) = \omega(g, h) \nu(gh)
\end{equation}
for some list $S_g$ of numbers between $1, \cdots, \chi^2$.
We start by proving that $Z_{\omega} = \{g \in G_U | \omega(g, h) = \omega(h, g),  \forall h \in G_U\} = \{1\}$ (i.e. the projective center of the group is trivial).
To do so, suppose that $g \in Z_{\omega}$.
Then, for all $h \in G_U$, particularly those $h$ associated with $\nu(h) = V_{\alpha}$, we have that:
\begin{equation}
    \nu(g) \nu(h) = \omega(g, h) \nu(gh) = \omega(h, g) \nu(hg) = \nu(h) \nu(g)
\end{equation}
But since the $V_{\alpha}$'s form a complete basis for the set of all $\chi \times \chi$ matrices by assumption, $\nu(g)$ commutes with all $\chi \times \chi$ matrices.
Hence, $\nu(g) = e^{i \theta} \mathds{1}$.
Since $\nu(g)$ is a faithful representation, $g = 1$.
Thus, $Z_{\omega} = \{1\}$.
Now, by Ref.~\cite{berkovich2018yakov, frucht}, we have that: 
\begin{equation}
    \text{deg}(\nu) = \sqrt{|G_U|/|Z_{\omega}|} = \sqrt{|G_U|}
\end{equation}
proving the first claim of the theorem.
Finally, since the projective representation defined by $\{V_{\alpha}\}$ is an irreducible and faithful projective representation of degree $\sqrt{|G_U|}$, $\{V_{\alpha}\}$ is equivalent to a nice error basis by Theorem C.1.

\hspace{0.95 \textwidth} $\blacksquare$

\begin{shaded}
    \textbf{Corollary C.1} The error basis $\mathcal{V}$ of every topologically gluable matrix product state with an abelian index group is equivalent to a nice error basis.
\end{shaded}

\textit{Proof.} The error basis $\mathcal{V}$ of a topologically gluable matrix product state is a unitary error basis. Since by assumption, the index group is abelian, the error basis is then nice.

\hspace{0.95\textwidth} $\blacksquare$

\subsubsection{Proof of Parameterization Theorem}

We are now prepared to prove our parameterization.

\begin{shaded}
    \textbf{Theorem C.3} (Theorem 5 of the Main Text) Suppose that $A$ is a uniform topologically gluable matrix product state tensor with error basis $\mathcal{V} = \{V_g\}$ and an abelian index group $G_U$.
    Then, the transfer matrix associated with matrix product state tensor can be written as:
    \begin{equation}
        \begin{tikzpicture} [scale = 1, baseline = {([yshift=-.5ex]current bounding box.center)}] 
    \draw[color = black] (-0.8, 0) -- (0.8, 0); %  top horizontal bar
    \draw[color = black] (-0.8, -1.2) -- (0.8, -1.2); %  bottom horizontal bar
    \draw[color = black] (0, 0) -- (0, -1.2); % vertical bar
    \draw[fill = lightdodgerblue] (0,0) circle (0.3); % circle
    \draw[fill = lightdodgerblue] (0,-1.2) circle (0.3); % circle
    %% Text
    \node at (-0.03,0) {\small $A$};
    \node at (-0.03,-1.2) {\small $\bar{A}$};
    \end{tikzpicture} = \sum_{g \in G_U} |t_g|^2 \begin{tikzpicture} [scale = 1, baseline = {([yshift=-.5ex]current bounding box.center)}] 
    \draw[color = black] (-0.8, 0) -- (0.8, 0); %  top horizontal bar
    \draw[color = black] (-0.8, -1.2) -- (0.8, -1.2); %  bottom horizontal bar
    %\draw[color = black] (0, 0) -- (0, -1.2); % vertical bar
    \draw[fill = white] (0,0) circle (0.25); % circle
    \draw[fill = white] (0,-1.2) circle (0.25); % circle
    %% Text
    \node at (0,0) {\small $V_g$};
    \node at (0,-1.2) {\small $\bar{V}_g$};
    \end{tikzpicture}
    \end{equation}
where $t_g$ are arbitrary complex numbers such that $\sum_{g \in G_U} |t_g|^2 = 1$.
Since the transfer matrix uniquely determines the matrix product state tensors up to a local isometry, this parameterizes (and can be considered to classify) all topologically gluable matrix product state tensors with abelian index groups.
\end{shaded}

\textit{Proof.} Since $\mathcal{V} = \{V_g\}$ forms a complete basis of $\chi \times \chi$ matrices, $\{V_{g} \otimes \bar{V}_{h}\}$ forms a complete basis for all $\chi^2 \times \chi^2$ matrices.
Hence, 
\begin{equation}
    \begin{tikzpicture} [scale = 1, baseline = {([yshift=-.5ex]current bounding box.center)}] 
    \draw[color = black] (-0.8, 0) -- (0.8, 0); %  top horizontal bar
    \draw[color = black] (-0.8, -1.2) -- (0.8, -1.2); %  bottom horizontal bar
    \draw[color = black] (0, 0) -- (0, -1.2); % vertical bar
    \draw[fill = lightdodgerblue] (0,0) circle (0.3); % circle
    \draw[fill = lightdodgerblue] (0,-1.2) circle (0.3); % circle
    %% Text
    \node at (-0.03,0) {\small $A$};
    \node at (-0.03,-1.2) {\small $\bar{A}$};
    \end{tikzpicture} = \sum_{g, h \in G_U} t_{g, h} \begin{tikzpicture} [scale = 1, baseline = {([yshift=-.5ex]current bounding box.center)}] 
    \draw[color = black] (-0.8, 0) -- (0.8, 0); %  top horizontal bar
    \draw[color = black] (-0.8, -1.2) -- (0.8, -1.2); %  bottom horizontal bar
    %\draw[color = black] (0, 0) -- (0, -1.2); % vertical bar
    \draw[fill = white] (0,0) circle (0.25); % circle
    \draw[fill = white] (0,-1.2) circle (0.25); % circle
    %% Text
    \node at (0,0) {\small $V_g$};
    \node at (0,-1.2) {\small $\bar{V}_h$};
    \end{tikzpicture}
\end{equation}
Now, since $A$ is topologically gluable, for all $a \in G_U$
\begin{equation}
     \begin{tikzpicture} [scale = 1, baseline = {([yshift=-.5ex]current bounding box.center)}] 
    \draw[color = black] (-0.8, 0) -- (0.8, 0); %  top horizontal bar
    \draw[color = black] (-0.8, -1.2) -- (0.8, -1.2); %  bottom horizontal bar
    \draw[color = black] (0, 0) -- (0, -1.2); % vertical bar
    \draw[fill = lightdodgerblue] (0,0) circle (0.3); % circle
    \draw[fill = lightdodgerblue] (0,-1.2) circle (0.3); % circle
    %% Text
    \node at (-0.03,0) {\small $A$};
    \node at (-0.03,-1.2) {\small $\bar{A}$};
    \end{tikzpicture} = \begin{tikzpicture} [scale = 1, baseline = {([yshift=-.5ex]current bounding box.center)}] 
    \draw[color = black] (-0.8, 0) -- (0.8, 0); %  top horizontal bar
    \draw[color = black] (-0.8, -1.2) -- (0.8, -1.2); %  bottom horizontal bar
    \draw[color = black] (0, 0) -- (0, -1.2); % vertical bar
    \draw[fill = lightdodgerblue] (0,0) circle (0.3); % circle
    \draw[fill = lightdodgerblue] (0,-1.2) circle (0.3); % circle
    %% Text
    \node at (-0.03,0) {\small $A$};
    \node at (-1,0) {\small $V_a$};
    \node at (1,0) {\small $V^{\dagger}_a$};
    \node at (-1,-1.2) {\small $\bar{V}_a$};
    \node at (1,-1.2) {\small $V^{\mathsf{T}}_a$};
    \node at (-0.03,-1.2) {\small $\bar{A}$};
    \end{tikzpicture} = \sum_{g, h \in G_U} t_{g, h} \begin{tikzpicture} [scale = 1, baseline = {([yshift=-.5ex]current bounding box.center)}] 
    \draw[color = black] (-0.8, 0) -- (0.8, 0); %  top horizontal bar
    \draw[color = black] (-0.8, -1.2) -- (0.8, -1.2); %  bottom horizontal bar
    %\draw[color = black] (0, 0) -- (0, -1.2); % vertical bar
    \draw[fill = white] (0,0) circle (0.25); % circle
    \draw[fill = white] (0,-1.2) circle (0.25); % circle
    %% Text
    \node at (0,0) {\small $V_g$};
    \node at (0,-1.2) {\small $\bar{V}_h$};
    \node at (-1,0) {\small $V_a$};
    \node at (1,0) {\small $V^{\dagger}_a$};
    \node at (-1,-1.2) {\small $\bar{V}_a$};
    \node at (1,-1.2) {\small $V^{\mathsf{T}}_a$};
\end{tikzpicture}    
\end{equation}
Now, since $\{V_{\alpha}\}$ forms a projective representation of $G_U$ (Lemma C.2): 
\begin{equation}
    V_{a} V_g V_{a}^{\dagger} = \frac{\omega(a, g) \omega(ag, a^{-1})}{\omega(a, a^{-1})} V_{aga^{-1}} \equiv \chi_{g, a} V_{aga^{-1}} = \chi_{g, a} V_{g} 
\end{equation}
where in the last step we used that $G_U$ was abelian.
Then: 
\begin{equation}
     \sum_{g, h \in G_U} t_{g, h} \begin{tikzpicture} [scale = 1, baseline = {([yshift=-.5ex]current bounding box.center)}] 
    \draw[color = black] (-0.8, 0) -- (0.8, 0); %  top horizontal bar
    \draw[color = black] (-0.8, -1.2) -- (0.8, -1.2); %  bottom horizontal bar
    %\draw[color = black] (0, 0) -- (0, -1.2); % vertical bar
    \draw[fill = white] (0,0) circle (0.25); % circle
    \draw[fill = white] (0,-1.2) circle (0.25); % circle
    %% Text
    \node at (0,0) {\small $V_g$};
    \node at (0,-1.2) {\small $\bar{V}_h$};
    \node at (-1,0) {\small $V_a$};
    \node at (1,0) {\small $V^{\dagger}_a$};
    \node at (-1,-1.2) {\small $\bar{V}_a$};
    \node at (1,-1.2) {\small $V^{\mathsf{T}}_a$};
    \end{tikzpicture} =  \sum_{g, h \in G_U} \frac{t_{g, h} \chi_{g, a}}{\chi_{ h, a}}\begin{tikzpicture} [scale = 1, baseline = {([yshift=-.5ex]current bounding box.center)}] 
    \draw[color = black] (-0.8, 0) -- (0.8, 0); %  top horizontal bar
    \draw[color = black] (-0.8, -1.2) -- (0.8, -1.2); %  bottom horizontal bar
    %\draw[color = black] (0, 0) -- (0, -1.2); % vertical bar
    \draw[fill = white] (0,0) circle (0.25); % circle
    \draw[fill = white] (0,-1.2) circle (0.25); % circle
    %% Text
    \node at (0,0) {\small $V_g$};
    \node at (0,-1.2) {\small $\bar{V}_h$};
    \end{tikzpicture}  \implies \frac{t_{g, h} \chi_{g, a}}{\chi_{ h, a}} = t_{g, h}
\end{equation}
Thus, for all non-zero $t_{g, h}$, $\chi_{g, a}/\chi_{h, a} = 1$.
As a consequence of this, note that, for all $V_a \in \mathcal{V}$
\begin{equation}
    V_a V_{h}^{\dagger} V_{g} V_{a}^{\dagger} = \frac{\chi_{g, a}}{\chi_{h, a}} V_{h}^{\dagger} V_{g} = V_{h}^{\dagger} V_{g} \implies [V_{a}, V_{h}^{\dagger} V_g] = 0 \quad \forall\ V_a \in \mathcal{V}  
\end{equation}
But because $\mathcal{V}$ forms a complete basis of all $\chi \times \chi$ matrices, the above implies that $V_{h}^{\dagger} V_g = e^{i \theta} \mathds{1}$.
Hence, $V_h = e^{-i \theta}V_{g}$.
Since $\text{tr}(V_g^{\dagger}V_h) = \chi \delta_{g, h}$, we have that $\theta = 0$ and $g = h$ for non-zero $t_{g, h}$.
Hence:
\begin{equation}
       \begin{tikzpicture} [scale = 1, baseline = {([yshift=-.5ex]current bounding box.center)}] 
    \draw[color = black] (-0.8, 0) -- (0.8, 0); %  top horizontal bar
    \draw[color = black] (-0.8, -1.2) -- (0.8, -1.2); %  bottom horizontal bar
    \draw[color = black] (0, 0) -- (0, -1.2); % vertical bar
    \draw[fill = lightdodgerblue] (0,0) circle (0.3); % circle
    \draw[fill = lightdodgerblue] (0,-1.2) circle (0.3); % circle
    %% Text
    \node at (-0.03,0) {\small $A$};
    \node at (-0.03,-1.2) {\small $\bar{A}$};
    \end{tikzpicture} = \sum_{g \in G_U} t_{g, g} \begin{tikzpicture} [scale = 1, baseline = {([yshift=-.5ex]current bounding box.center)}] 
    \draw[color = black] (-0.8, 0) -- (0.8, 0); %  top horizontal bar
    \draw[color = black] (-0.8, -1.2) -- (0.8, -1.2); %  bottom horizontal bar
    %\draw[color = black] (0, 0) -- (0, -1.2); % vertical bar
    \draw[fill = white] (0,0) circle (0.25); % circle
    \draw[fill = white] (0,-1.2) circle (0.25); % circle
    %% Text
    \node at (0,0) {\small $V_g$};
    \node at (0,-1.2) {\small $\bar{V}_g$};
    \end{tikzpicture}
    \end{equation}
What is left is to show that $t_{g, g}$ are real, positive, and sum to $1$.
The fact that $\sum_{g \in G_U} t_{g, g} = 1$ follows from the fact that the identity is the dominant right (and left) eigenvector of the transfer matrix with eigenvalue $1$.
The real and positive condition follows from the fact that the transfer matrix, when viewed in the vertical direction, is a density matrix.
In this language, $t_{g, g}$ are the eigenvalues of the density matrix and thus must be real and positive.
As such:
    \begin{equation}
        \begin{tikzpicture} [scale = 1, baseline = {([yshift=-.5ex]current bounding box.center)}] 
    \draw[color = black] (-0.8, 0) -- (0.8, 0); %  top horizontal bar
    \draw[color = black] (-0.8, -1.2) -- (0.8, -1.2); %  bottom horizontal bar
    \draw[color = black] (0, 0) -- (0, -1.2); % vertical bar
    \draw[fill = lightdodgerblue] (0,0) circle (0.3); % circle
    \draw[fill = lightdodgerblue] (0,-1.2) circle (0.3); % circle
    %% Text
    \node at (-0.03,0) {\small $A$};
    \node at (-0.03,-1.2) {\small $\bar{A}$};
    \end{tikzpicture} = \sum_{g \in G_U} |t_g|^2 \begin{tikzpicture} [scale = 1, baseline = {([yshift=-.5ex]current bounding box.center)}] 
    \draw[color = black] (-0.8, 0) -- (0.8, 0); %  top horizontal bar
    \draw[color = black] (-0.8, -1.2) -- (0.8, -1.2); %  bottom horizontal bar
    %\draw[color = black] (0, 0) -- (0, -1.2); % vertical bar
    \draw[fill = white] (0,0) circle (0.25); % circle
    \draw[fill = white] (0,-1.2) circle (0.25); % circle
    %% Text
    \node at (0,0) {\small $V_g$};
    \node at (0,-1.2) {\small $\bar{V}_g$};
    \end{tikzpicture}
    \end{equation}
where $|t_g|^2 = t_{g, g}$ are arbitrary numbers such that $\sum_{g \in G_U} |t_g|^2 = 1$.

\hspace{0.95 \textwidth} $\blacksquare$

Note that in the main text, we phrase the statement in terms of the MPS tensor directly.
As stated in the theorem above, this is equivalent to the transfer matrix statement above because the transfer matrix uniquely determines the MPS tensor up to an isometry at the physical level \cite{perezgarcia2006MPSrep, Cirac2021MPSReview}.

\subsubsection{Connection Between Parameterization and Classification} \label{app:connection-between-param-and-class}

We now offer a connection between our classification (Theorem 3 of the main text) and the parameterization for abelian and uniform topologically gluable matrix product states in Theorem 5 of the main text (Theorem C.3 above).
%
%In particular, we will see that the two classification results make manifest distinct properties of the state.
%
Prior to stating the precise connection, let us remind ourselves that the commutant of an abelian nice error basis (i.e. the measurement basis of any abelian uniform topologically gluable matrix product state) is the algebra generated by $V_{\alpha} \otimes V_{\alpha}^{\dagger}$.
Since the set $\{V_{\alpha} \otimes V_{\alpha}^{\dagger}\}$ is closed under multiplication, such gluable matrix product states in definite form \eqref{eq-definiteform} are given:
\begin{equation} \label{eq-AHHHHH}
    \A = \sum_{\alpha} \mu_{\alpha} V_{\alpha} \otimes V_{\alpha}^{\dagger} 
\end{equation}
This formula also appeared in Eq.~\eqref{eq-Aclassificationalternative} of the main text.
We now relate these parameters $\mu$ to the parameters $t$ appearing in Theorem 5 (equiv. Theorem C.3).

\begin{shaded}
    \textbf{Theorem C.4} \textit{(Equivalence of Abelian Classification)}     Suppose $A$ is an abelian and uniform topologically gluable matrix product state.
    If $t_{\alpha}$ are the parameters that appear in Theorem 5 (or equivalently Theorem C.3 of the Appendix) and $\mu_{\alpha}$ are the parameters that appear in Eq.~\eqref{eq-AHHHHH}, then:
    \begin{equation}
    \begin{tikzpicture} [scale = 1, baseline = {([yshift=-.5ex]current bounding box.center)}] 
    \draw[color = black] (-0.8, 0) -- (0.8, 0); %  top horizontal bar
    \draw[color = black] (-0.8, -1.2) -- (0.8, -1.2); %  bottom horizontal bar
    \draw[color = black] (0, 0) -- (0, -1.2); % vertical bar
    \draw[fill = lightishgray] (0,0) circle (0.3); % circle
    \draw[fill = lightishgray] (0,-1.2) circle (0.3); % circle
    \draw[fill = lightdodgerblue] (0, -0.6) circle (0.2);
    \node at (0, -0.6) {\small $t$};
    %% Text
    \node at (-0.03,0) {\small $\mathcal{V}$};
    \node at (-0.03,-1.2) {\small $\bar{\mathcal{V}}$};
    %\draw[color = black] (-0.8, -1.2) -- (-0.8, -1.5);
    %\draw[color = black] (0.8, -1.2) -- (0.8, -1.5);
    \end{tikzpicture} = \sum_{\alpha} \mu_{\alpha} V_{\alpha} \otimes V_{\alpha}^{\dagger} 
    \end{equation}
where note that the left hand side is the MPS form in Theorem 5 up to a physical isometry of $\bar{\mathcal{V}}$ and the right hand side is the MPS form in Eq.~\eqref{eq-AHHHHH}, read as a matrix in the vertical direction.
\end{shaded}

\textit{Proof.} Note that it is sufficient to prove that $t$ is diagonal in the formula below:
    \begin{equation}
         \begin{tikzpicture} [scale = 1, baseline = {([yshift=-.5ex]current bounding box.center)}] 
    \draw[color = black] (0, 0.5) -- (0, -1.0); 
    \draw[fill = lightdodgerblue] (0, -0.25) circle (0.25);
    \node at (0, 0.7) {\small $\beta$};
    \node at (0, -1.2) {\small $\gamma$};
    \node at (0, -0.25) {\small $t$};
    \end{tikzpicture} =  \sum_{\alpha} \mu_{\alpha} \begin{tikzpicture} [scale = 1, baseline = {([yshift=-.5ex]current bounding box.center)}] 
    \draw[color = black] (-0.8, 0) -- (0.8, 0); %  top horizontal bar
    \draw[color = black] (0, 0) -- (0, 0.5); % vertical bar
    \draw[fill = lightishgray] (0,0) circle (0.24); % circle
    % \draw[fill = white] (0,0.5) circle (0.14);
    %% Text
    \node at (0.0,0) {\small $\bar{\mathcal{V}}$};
    % \node at (0.0 ,0.5) {\small $\hat{s}$};
    %
    %
    \draw[color = black] (-0.8, 0) -- (-0.8, -1.5);
    \draw[color = black] (0.8, 0) -- (0.8, -1.5);
    \draw[color = black] (-0.8, -1.5) -- (0.8, -1.5); %  top horizontal bar
    \draw[color = black] (0, -1.5) -- (0, -0.5-1.5); % vertical bar
    \draw[fill = lightishgray] (0,-1.5) circle (0.24); % circle
    \node at (0.,-1.5) {\small $\mathcal{V}$};
    %\draw[color = black] (-0.8, -0.75) -- (0.8, -0.75);
    \draw[fill = lightishgray] (-0.8, -0.75) circle  (0.25);
    \draw[fill = lightishgray] (0.8, -0.75) circle  (0.25);
    \node at (-0.8, -0.75) {\small $V_{\alpha}$};
     \node at (0.8, -0.75) {\small $V_{\alpha}^{\dagger}$};
    % \draw[fill = lightdodgerblue] (0.0, -0.75) circle  (0.25);
    % \node at (0,-0.75) {\small $\mu$};
    \node at (0, 0.7) {\small $\beta$};
    \node at (0, -2.2) {\small $\gamma$};
    \end{tikzpicture}
    \end{equation}
This can be verified through explicit calculation.
We have that the right hand side is: 
\begin{equation}
    t_{\beta \gamma} = \sum_{\alpha} \mu_{\alpha} \text{tr}(V_{\alpha} V_{\beta}^{\dagger} V_{\alpha}^{\dagger} V_{\gamma}) = \sum_{\alpha} \mu_{\alpha}\, \bar{\chi}_{\beta, \alpha} \text{tr}(V_{\beta}^{\dagger} V_{\gamma}) = \chi \delta_{\beta \gamma}\left[\sum_{\alpha}\mu_{\alpha} \bar{\chi}_{\beta, \alpha} \right] .
\end{equation}
Crucially, the above is diagonal.

\hspace{0.95 \textwidth} $\blacksquare$

\subsection{Connection Between Uniform Topologically Gluable and SPT Phases \label{app:SPT}}

Having provided a proof for the parameterization theorem and further connected it to the classification of Theorem 3, we now discuss the connection between such states and symmetry-protected topological phases.
In particular, we start with Theorem 6:
\begin{shaded}
    \textbf{Theorem C.5} (Theorem 6 of the Main Text)  Every short-range entangled, uniform topologically gluable matrix product state with abelian errors is a non-trivial SPT phase protected by the symmetry group $G_U$.
\end{shaded}

\textit{Proof.} Recall that if the matrix product state is short-range entangled and has abelian errors, let us note a couple of facts.
First, its index group $G_U$ is abelian and a symmetry of the state (Lemma C.1). 
Moreover, its errors $\mathcal{V} = \{V_g\}$ form a nice error basis and necessarily form a non-trivial projective representation of the group  (Theorem C.2).
Consequently, we have from the uniform topologically gluable condition that:
\begin{equation}
      \begin{tikzpicture}[scale = 1, baseline = {([yshift=-.5ex]current bounding box.center)}] 
    \draw[color = black] (-1, 0) -- (1, 0);
    \draw[color = black] (0, 0) -- (0, -0.8);
    \draw[fill = lightdodgerblue] (0,0) circle (0.3);
    \node at (-0.03,0) {\small $A$};
    %\node at (-1.4, 0) {\small $\{V_{\alpha} \}$ };
    \node at (0, -1.0) {\small $U_g$};
\end{tikzpicture} = \begin{tikzpicture}[scale = 1, baseline = {([yshift=-.5ex]current bounding box.center)}] 
    \draw[color = black] (-1, 0) -- (1, 0);
    \draw[color = black] (0, 0) -- (0, -0.8);
    \draw[fill = lightdodgerblue] (0,0) circle (0.3);
    \node at (-0.03,0) {\small $A$};
    \node at (-1.3, 0) {\small $V_g$};
    \node at (1.35, 0) {\small $V_g^{\dagger}$};
    \node at (0, -1.2) {\small $\ $};
%       \node at (-1.4, 0) {\small $\{V_{\alpha} \}$ };
\end{tikzpicture}
\end{equation}
This implies that physical symmetries fractionalize into virtual operators that form a non-trivial projective representation of $G_U$ labeled by some $2$-cocyle $\omega$.
Since this co-cycle defines a discrete invariant of the state, robust provided that the $G$-symmetry remains unbroken, the matrix product state is in a non-trivial SPT phase.

\hspace{0.95\textwidth} $\blacksquare$

Finally, we provide a proof for Theorem 7.

\begin{shaded}
    \textbf{Theorem C.6} (Theorem 7 of the Main Text) Any minimally entangled, translation-invariant, SPT protected by an abelian internal symmetry is topologically gluable.
\end{shaded}

\textit{Proof.} Suppose that $\ket{\psi}$ is an infinite translationally-invariant SPT wavefunction protected by an abelian group $G = \{\prod_{x} U_{g, x}\}$.
From Ref.~\cite{Pollmann_2010}, it is known that semi-infinite string operators of the symmetry $\prod_{x < x_0} U_{g, x}$ act on the Schmidt vectors to the left and right of their endpoints as a projective representation of the group $G$ with factor set $\omega \in H^{2}(G, U(1))$, which uniquely characterizes the SPT phase.
Consequently, the Schmidt spectrum $\Lambda$ across any cut of the state can be organized into degenerate blocks, each labeled by an irreducible projective representation of the group $G$ with factor set $\omega$.% 

\vspace{2 mm}

Since $G$ is abelian, it follows that the degree of these irreducible projective representations are all the same and are given by $\text{deg}(\nu) = \sqrt{|G|/|Z_{\omega}|}$, where $Z_{\omega}$ is the projective center of the factor set (see Theorem 2.21 of Ref.~\cite{karpilovsky1994group}).
Thus, an SPT has minimal entanglement if its Schmidt spectrum is only non-zero for a single block with the entanglement across a cut being: $S = \log(\sqrt{|G|/|Z_{\omega}|})$.
Consequently, translationally-invariant minimal entanglement SPT's are (1) captured by translationally-invariant matrix product states with bond dimension $\chi = \sqrt{|G|/|Z_{\omega}|}$, (2) have flat entanglement spectrum, and (3) have virtual symmetry operators $\nu(g)$ satisfying:
\begin{equation} \label{eq-nupush}
    \begin{tikzpicture}[scale = 1, baseline = {([yshift=-.5ex]current bounding box.center)}] 
    \draw[color = black] (-1, 0) -- (1, 0);
    \draw[color = black] (0, 0) -- (0, -0.8);
    \draw[fill = lightdodgerblue] (0,0) circle (0.3);
    \node at (-0.03,0) {\small $A$};
    %\node at (-1.4, 0) {\small $\{V_{\alpha} \}$ };
   % \node at (1.3, 0) {\small ${V}_{\alpha}$ }; \\
    \node at (0, -1.0) {\small $U_{g}$};
\end{tikzpicture} = e^{i \theta_g} \begin{tikzpicture}[scale = 1, baseline = {([yshift=-.5ex]current bounding box.center)}] 
    \draw[color = black] (-1, 0) -- (1, 0);
    \draw[color = black] (0, 0) -- (0, -0.8);
    \draw[fill = lightdodgerblue] (0,0) circle (0.3);
    \node at (-0.03,0) {\small $A$};
    \node at (-1.3, 0) {\small $\nu(g)$};
    \node at (0, -1.2) {\small $\ $};
    \node at (1.4, 0) {\small $\nu^{\dagger}(g) $ };
\end{tikzpicture}
\end{equation}
and which form an irreducible projective representation of $G$ with factor set $\omega$.

\vspace{2 mm}

We now wish to show that such states are topologically gluable by showing that some subset of the virtual symmetry operators form a nice error basis for this matrix product state.
To do so, let us note that, from the proof of Theorem 2.21 in Ref.~\cite{karpilovsky1994group}, if $\nu(g) \in Z_{\omega}$, then $\nu(g) = e^{i \phii_g} \mathds{1}$.
With this in mind, let us consider the map $\rho: G \to \text{GL}_{\mathbb{C}}(\mathbb{C}^{\chi})$ such that $\rho(g) = \nu(g) \otimes \bar{\nu}(g)$ with $\text{ker}(\rho) = Z_{\omega}$.
Note that $\rho$ forms a linear representation of $G$ but is also a group homomorphism from $G$ to $G_U = \langle \{ \nu(g) \otimes \bar{\nu}(g) \} \rangle \simeq G/Z_{\omega}$ with size $|G_U| = \sqrt{|G|/|Z_{\omega}|}$.
By picking representatives $g$ of the cosets $[g] \in G/Z_{\omega}$, we can define a closely related map $\rho': G_U \to \text{GL}_{\mathbb{C}}(\mathbb{C}^{\chi})$ with $\rho'([g]) = \nu(g) \otimes \bar{\nu}(g)$, which defines a faithful linear representation of $G_U$.

The map $\nu'([g]) = \nu(g)$ then defines a faithful projective representation of $G_U$ with factor set $\omega': G_U \times G_U \to U(1)$ with $\omega([g], [h]) = \omega(g, h)$.
The representation is further irreducible since the set of matrices $\{\nu'([g])\}$ consist of all elements of $\{\nu(g) | g \in G\}$ that are distinct up to phases.
Thus, since the former has no invariant subspaces, the latter also has no invariant subspaces.
Thus, $\nu([g])$ is a faithful irreducible projective representation of degree $\sqrt{|G_U|} = \chi$.
It follows from Theorem C.1 that it forms a nice error basis.
This, with the property in Eq.~\eqref{eq-nupush}, cements $\ket{\psi}$ as topologically gluable.

\hspace{0.95 \textwidth} $\blacksquare$

An independently interesting results follows from this.

\begin{shaded}
    \textbf{Corollary C.2} Suppose $\ket{\Psi}$ is a minimally entangled SPT state protected by an abelian internal symmetry $G$ and labeled by a factor set $\omega$.
    Then $\ket{\psi}$ is also an SPT protected by abelian symmetry $G' \simeq G/Z_{\omega}$ labeled by a maximally non-commuting factor set $\omega'$.
\end{shaded}

\textit{Proof.} The proof of this immediately follows from the proof of the theorem above.

\section{Some Additional Details on No-Go Theorem Proof \label{app-nogo}}

In the main text, we proved a no-go theorem demonstrating that a state with no zeros in its correlation spectrum and a non-flat entanglement spectrum cannot be right-gluable.
We proved this by contradiction: if the state was gluable and had no zeros in its correlation spectrum, it could not have any locally correctable errors.
Consequently, all of its errors were topological.
To show indeed that this implies that the entanglement spectrum is flat (hence, deriving the contradiction required), we need the following Lemma.
\begin{shaded}
    \textbf{Lemma D.1} Suppose that $\ket{\Psi}$ is a gluable matrix product state with an error basis $\mathcal{V}$ of (exclusively) topological errors that push through to $\mathcal{V}^{[n]}$.
    If $\ket{\Psi}$ has a full correlation spectrum, then $\mathcal{V}^{[n]}$ is a unitary error basis for all $n$. 
\end{shaded}

\textit{Proof.} We proceed inductively.
First, let us note that $\mathcal{V}^{[0]} \equiv \mathcal{V}$ is a unitary error basis.
Now, assume that $\mathcal{V}^{[n]}$ is a unitary error basis.
Subsequently, if $\ket{\Psi}$ is gluable, then the transfer matrix satisfies:
\begin{equation}
\begin{tikzpicture} [scale = 1, baseline = {([yshift=-.5ex]current bounding box.center)}] 
    \draw[color = black] (-0.8, 0) -- (0.8, 0); %  top horizontal bar
    \draw[color = black] (-0.8, -1.2) -- (0.8, -1.2); %  bottom horizontal bar
    \draw[color = black] (0, 0) -- (0, -1.2); % vertical bar
    \draw[fill = lightdodgerblue] (0,0) circle (0.3); % circle
    \draw[fill = lightdodgerblue] (0,-1.2) circle (0.3); % circle
    %% Text
    \node at (-0.03,0) {\small $A$};
    \node at (-0.03,-1.2) {\small $\bar{A}$};
    \end{tikzpicture} = \begin{tikzpicture} [scale = 1, baseline = {([yshift=-.5ex]current bounding box.center)}] 
    \draw[color = black] (-0.8, 0) -- (0.8, 0); %  top horizontal bar
    \draw[color = black] (-0.8, -1.2) -- (0.8, -1.2); %  bottom horizontal bar
    \draw[color = black] (0, 0) -- (0, -1.2); % vertical bar
    \draw[fill = lightdodgerblue] (0,0) circle (0.3); % circle
    \draw[fill = lightdodgerblue] (0,-1.2) circle (0.3); % circle
    %% Text
    \node at (-0.03,0) {\small $A$};
    \node at (-1.3,0) {\small $(V^{[n]}_\alpha)$};
    \node at (1.4,0) {\small $(V^{[n + 1]}_{\alpha})^{\dagger}$};
    \node at (-1.3,-1.2) {\small $(\bar{V}_\alpha^{[n]})$};
    \node at (1.4,-1.2) {\small $(V^{[n + 1]}_\alpha)^{\mathsf{T}}$};
    \node at (-0.03,-1.2) {\small $\bar{A}$};
    \end{tikzpicture}
\end{equation}
where $V_{\alpha}^{[n]}$ is unitary (see remark below the Theorem B.4) for all $n$ and $\alpha$.
To show that $V_{\alpha}^{[n]}$ is in fact an error basis,  first write the above equation as $T = \mathbb{V}_{\alpha} T \mathbb{W}_{\alpha}^{\dagger} $ where  $\mathbb{V}_{\alpha} = V^{[n]}_{\alpha} \otimes \bar{V}^{[n]}_{\alpha}$ and $\mathbb{W}_{\alpha} = V_{\alpha}^{[n + 1]} \otimes \bar{V}^{[n + 1]}_{\alpha}$.
Now, if we perform a singular value decomposition on $T$, we have that:
\begin{equation}
    T = X S Y^{\dagger} = \mathbb{V}_{\alpha} X S Y^{\dagger} \mathbb{W}_{\alpha} \implies S = (X^{\dagger} \mathbb{V}_{\alpha} X) S(Y^{\dagger} \mathbb{W}^{\dagger}_{\alpha} Y)
\end{equation}
where $X$ and $Y$ are unitary.
But note that both the left and right hand side of the implied equation defines a valid singular value decompositon for $S$.
Since $\ket{\Psi}$ has a full correlation spectrum, $S$ is non-singular.
Hence, since the singular vectors are unique up to unitary transformations $U$ in degenerate singular subspaces, we have that $(X^{\dagger} \mathbb{V}_{\alpha} X) = U$ and $(Y^{\dagger} \mathbb{W}_{\alpha} Y) = U$.
Consequently, 
\begin{equation}
    (X^{\dagger} \mathbb{V}_{\alpha} X)  = (Y^{\dagger} \mathbb{W}_{\alpha} Y) \implies \mathbb{W}_{\alpha} = YX^{\dagger} \mathbb{V}_{\alpha} (YX^{\dagger})^{\dagger} 
\end{equation}
Thus, $\mathbb{W}_{\alpha}$ is related to $\mathbb{V}_{\alpha}$ by unitary conjugation.
Thus, we have that: 
\begin{equation}
    \left|\text{tr}\left[(V_{\alpha}^{[n + 1]})^{\dagger} V_{\beta}^{[n + 1]}\right] \right|^2 = \text{tr}(\mathbb{W}_{\alpha}^{\dagger} \mathbb{W}_{\beta}) = \text{tr}(\mathbb{V}_{\alpha}^{\dagger} \mathbb{V}_{\beta}) = \left|\text{tr}\left[(V_{\alpha}^{[n]})^{\dagger} V_{\beta}^{[n]}\right] \right|^2 = \chi^2 \delta_{\alpha \beta}
\end{equation}
Thus, $\text{tr}\left[(V_{\alpha}^{[n + 1]})^{\dagger} V_{\beta}^{[n + 1]}\right] = \chi \delta_{\alpha \beta}$ (note that when $\alpha = \beta$, the argument of the trace is positive semi-definite picking out the phase).
Hence, it is a unitary error basis.
This proves the above assertion.

\hspace{0.95 \textwidth} $\blacksquare$

\end{document}